\documentclass[11pt,twoside,german]{scrreprt-pro}
\usepackage{amsmath,latexsym,cite,subfigure,longtable}
\usepackage[english,german,swedish,italian]{babel}
\usepackage{epsfig}
\typearea[10mm]{10}
\begin{document}
\selectlanguage{english}
%
\vspace*{0.36\textheight}
\begin{flushright}\large
Andreas E. Klinkm{\"u}ller $\bullet$ Doctorate thesis 1997
\end{flushright}
\vspace*{0.23\textheight}
\begin{center}
Akademisk avhandling, som f{\"o}r avl{\"a}ggande av filosofie
doktorsexamen \linebreak
vid G{\"o}teborgs Universitet, f{\"o}rsvaras vid offentlig disputation
\linebreak 
fredagen den 10 oktober 1997 kl 10.10, sal HA2, H{\"o}rsalsv{\"a}gen 4,
\linebreak 
Chalmers Tekniska H{\"o}gskola, G{\"o}teborg.
\end{center}
\begin{center}\small
Avhandlingen f{\"o}rsvaras p{\aa} engelska
\end{center}
\vfill
\begin{center}\small
Department of Physics\linebreak
Gothenburg University and Chalmers University of Technology \linebreak
Fysikgr{\"a}nd 3 \linebreak
SE-412\,96 G{\"o}teborg, Sweden
\end{center}
\clearpage
\vspace*{0.9\textheight}
\vfill
\begin{flushright}
Electronically available from \texttt{http://xxx.lanl.gov}
\end{flushright}
\clearpage

%
\thispagestyle{empty}
\begin{center}
\vfill
\vspace*{45mm}
\textbf{\sffamily\LARGE On doubly excited states in negative
ions}\linebreak\par
\textbf{\sffamily\large Dubbelt exciterade tillst{\aa}nd i negativa
joner}\linebreak
\par\vspace*{70mm}
\textsc{\large Andreas E. Klinkm{\"u}ller}\linebreak
\vspace*{3mm}
G{\"o}teborg 1997
\par\vspace*{12mm}
\parbox{0.60\textwidth}{
\epsfig{file=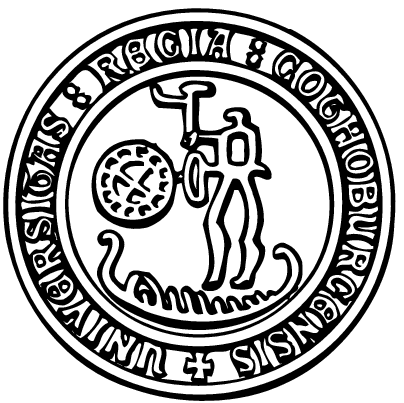, width=35mm}\hfill
\epsfig{file=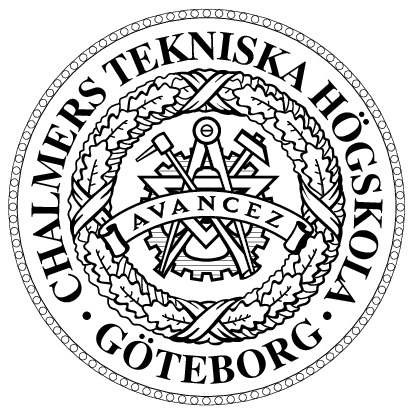, width=35mm}
}\par\bigskip
{Fysiska Institutionen, G{\"o}teborgs Universitet \linebreak
och Chalmers Tekniska H{\"o}gskola}
\vfill
\end{center}
\clearpage
\vspace*{\fill}
\begin{center}
\textsf{ISBN 91-7197-547-0}\linebreak
Chalmers bibliotekets tryckeri
\end{center}
\begin{center}
\copyright This copy of the thesis has been supplied on condition that
anyone \linebreak
who consults it is understood to recognise that its copyright rests with
the \linebreak author and 
that no quotation from the thesis, nor any information derived
\linebreak therefrom, may be 
published without the author's prior written consent.
\end{center}
\clearpage

%
\setlength{\abovecaptionskip}{0pt}
\setlength{\belowcaptionskip}{0pt}
%
\setlength{\LTcapwidth}{\textwidth}
%
\renewcommand{\subfigcapskip}{-2pt}
\renewcommand{\subfigtopskip}{0pt}
\renewcommand{\subfigbottomskip}{3pt}
\renewcommand{\capfont}{\normalfont\slshape}
%
%
\selectlanguage{english}
\minisec{Abstract} 
Atomic negative ions are fragile quantum systems with
a binding energy one order of magnitude smaller than for atoms. Due to
this weak binding, correlation among the valence electrons plays a very
important role.  For doubly excited states, where two electrons are
simultaneously promoted, correlation becomes dominant and is thus no
longer a small perturbation to the independent particle behaviour.

Photodetachment is the process where a negative ion absorbs a photon and
as a consquence ejects the outermost electron.  Investigations of this
process have been performed with the laser and ion beams merged over
0.5~m in a collinear geometry. This simultaneously enables high
resolution and high sensitivity.  By combining measurements with co- and
counter-propagating laser and ion beams the Doppler shift can be
eliminated to \emph{all} orders.  The ion beam facility was equipped
with a sputter ion source and a plasma ion source, which together allow
the production of atomic negative ions of almost any element. The laser
light for the investigations was generated by two excimer pumped dye
lasers.

A new detection scheme has been developed.  The residual atom of a
photodetachment is state selectively ionised by resonance ionisation
spectroscopy. This novel detection scheme retains all advantages of the
collinear geometry and, at the same time, offers the possibility of
measuring partial photodetachment cross sections.

Investigations of the photodetachment threshold of tellurium with
neutral particle detection have been performed and yielded an electron
affinity of $1\,589\,618(5)\,\text{m}^{-1}$. With state selective
detection the Li(2p) photodetachment threshold was investigated and the
electron affinity was determined to be 498\,490(17)~m$^{-1}$.

Doubly excited states of both Li$^{-}$ and He$^{-}$ have been
studied. Resonance structure in the Li$^{-}$ photodetachment cross
section near the Li(3p) threshold has been investigated.  The structure
arises from the autodetaching decay of doubly excited
$^{1}$P$^{\text{o}}$ states of Li$^{-}$ that are bound with respect to
the Li(3p) state. According to an assisting calculation this structure
is analogous to the symmetrically excited intra-shell
$(_{3}\{0\}_{3}^{+})$\,$^{1}$P$^{\text{o}}$ state in H$^{-}$. Higher
lying asymmetrically excited inter-shell states were observed to
converge on the Li(3p) limit.

In He$^{-}$ the three doubly excited states, 1s3s4s\,$^{4}$S,
1s3p$^{2}$\,$^{4}$P and 1s3p4p\,$^{4}$P were investigated. The Feschbach
resonance associated with the 1s3s4s\,$^{4}$S state was observed in
both the He(1s2s\,$^{3}$S) + e$^{-}$($\epsilon$s) and the
He(1s2p\,$^{3}$P$^{\text{o}}$) + e$^{-}$($\epsilon$p) partial
photodetachment cross section, whereas the 1s3p$^{2}$\,$^{4}$P and
1s3p4p\,$^{4}$P resonances were observed in the
He(1s2p\,$^{3}$P$^{\text{o}}$) + e$^{-}$($\epsilon$p) partial
photodetachment cross section.

The energy positions of all these resonances [1s3s4s\,$^{4}$S: $E_{0}=
(2\,386\,803.1\pm4.2)~\text{m}^{-1}$, $\Gamma = 160(16)~\text{m}^{-1}$;
1s3p$^{2}$\,$^{4}$P: $E_{0}= 2.478\,2(55)\times 10^{6}~\text{m}^{-1}$,
$\Gamma = 40(3)\times 10^{3}~\text{m}^{-1}$; 1s3p4p\,$^{4}$P: $E_{0}=
2\,633\,297(40)~\text{m}^{-1}$, $\Gamma = 492(35)~\text{m}^{-1}$] agree
with a recent ab initio calculation by Xi and Froese Fischer
[Phys. Rev. A,
\textbf{53}(5), 3169, (1996)] to an accuracy of up to $10^{-4}$. 
The widths, however, are less accurate both in experiment and theory:
The width of the 1s3s4s\,$^{4}$S resonance agrees with this calculation,
but for the 1s3p$^{2}$\,$^{4}$P resonance there is a slight discrepancy
and the 1s3p4p\,$^{4}$P resonance is only half as broad as predicted.
\clearpage

%
\minisec{Sammanfattning}
{\small
\selectlanguage{swedish}
Atom"ara negativa joner "ar br"ackliga kvantsystem med
bindingsenergier som "ar typiskt  en storleksordning l"agre "an f"or
atomer. P{\aa} grund av denna svaga bindning spelar korrelationen mellan
valenselektronerna en mycket viktig roll. I dubbelt exciterade
tillst{\aa}nd, d"ar tv{\aa} elektroner "ar samtidigt exciterade, kan
korrelationen dominera och den utg"or f"oljdaktligen inte l"angre endast
en
liten st"orning till den oberoende partikelmodellen.

I fotoneutralisationsprocessen absorberar en negativ jon en foton varvid
den yttersta elektronen emitteras.  Unders"okningar av denna process har
genomf"orts i en kollinj"ar geometri d"ar laser- och jonstr{\aa}larna
"overlappade varandra "over 0.5~m. Detta m"ojliggjorde samtidigt
b{\aa}de h"og uppl"osning och h"og k"anslighet i experimenten. Genom att
kombinera m"atningar med parallella och anti-parallella laser och
jonstr{\aa}lar kan Dopplerf"orskjutning elemineras till alla
ordningar. Jonstr{\aa}leanl"aggningen "ar utrustad med en sputter
jonk"alla och en plasma jonk"alla, vilka tillsammans m"ojligg"or
produktionen av atom"arer negativa joner av n"astan alla
grund"amnen. Laserljuset genereras av tv{\aa} excimerpumpade
f"arg"amneslasrar.

En ny detektionsmetod har utveklats. De i fotoneutralisationsprocessen
bildade atomerna detekterades h"arvid tillst{\aa}ndsselektivt med hj"alp
av resonansjonisationsspektroskopi. Med denna nya detektionsmetod kan
f"ordelarna med en kolinj"ar geometri bevaras, samtidigt som det blir
m"ojligt att m"ata partiella fotoneutralisationstv"arsnitt.

Unders"okningar av fotoneutralisationstr"oskeln hos tellur med
neutralpartikeldetektion har genomf"orts och gav en elektronaffinitet
p{\aa} 1\,589\,618(5)~m$^{-1}$. Med den tillst{\aa}ndsselektiva
detektionsmetoden unders"oktes Li(2p)-tr"oskeln och elektronaffiniteten
bet"amdes i detta fall till 498\,490(17)~m$^{-1}$.

Dubbelt exciterade tillst{\aa}nd hos b{\aa}de Li$^{-}$ och He$^{-}$ har
studerats.  I Li$^{-}$ har resonansstrukturer i
fotoneutralisationstv"arsnittet n"ara Li(3p)-tr"oskeln
studerats. Strukturen h"arstammar fr{\aa}n autoneutraliserande dubbelt
exciterat $^{1}$P$^{\text{o}}$-tillst{\aa}nd hos Li$^{-}$ som "ar bundet
till 3p-tillst{\aa}ndet i den neutrala atomen. Teoretiska studier har
visat att denna struktur "ar analog till det symmetriskt exciterade
tillst{\aa}ndet $(_{1}\{0\}^{+}_{3})\,^{1}\text{P}^{\text{o}}$ i
H$^{-}$. H"ogre liggande assymmetriskt exciterade tillst{\aa}nd som
konvergera till Li(3p) gr"ansen har ocks{\aa} observerats.

I He$^{-}$ har de tre dubbelt exciterade tillst{\aa}nden,
1s3s4s\,$^{4}$S, 1s3p$^{2}$\,$^{4}$P och 1s3p4p\,$^{4}$P, oberverats. En
Feschbachresonans associerad med 1s3s4s\,$^{4}$S tillst{\aa}ndet har
studerats i de partiella fotoneutralisationstv"arsnitten
He(1s2s\,$^{3}$S) + e$^{-}$($\epsilon$s) och
He(1s2p\,$^{3}$P$^{\text{o}}$) + e$^{-}$($\epsilon$p),  och
1s3p$^{2}$\,$^{4}$P och 1s3p4p\,$^{4}$P resonanserna har observerats i
det partiella fotoneutralisationstv"arsnittet
He(1s2p\,$^{3}$P$^{\text{o}}$) + e$^{-}$($\epsilon$p).

Energierna hos alla resonanserna [1s3s4s\,$^{4}$S: $E_{0}=
(2\,386\,803.1\pm4.2)~\text{m}^{-1}$, $\Gamma = 160(16)~\text{m}^{-1}$;
1s3p$^{2}$\,$^{4}$P: $E_{0}= 2.478\,2(55)\times 10^{6}~\text{m}^{-1}$,
$\Gamma = 40(3)\times 10^{3}~\text{m}^{-1}$; 1s3p4p\,$^{4}$P: $E_{0}=
2\,633\,297(40)~\text{m}^{-1}$, $\Gamma = 492(35)~\text{m}^{-1}$]
"overensst"ammer med en nyligen publicerad ab initio ber"akning av Xi
och Froese Fischer [Phys. Rev. A, \textbf{53}(5), 3169, (1996)] upp till
en noggrannhet av $10^{-4}$. Vidderna hos resonansen "ar dock mindre
noggrannt best"amda i s{\aa}v"al experiment som teori och
"overensst"ammelsen "ar h"ar mindre god. F"or 1s3s4s\,$^{4}$S resonansen
"ar skillnad mellan teori och experiment mindre "an den experimentella
noggrannheten, medan det f"or 1s3p$^{2}$\,$^{4}$P resonansen finns en
liten avvikelse. F"or 1s3p4p\,$^{4}$P resonansen, slutligen, "ar det
experimentella v"ardet bara h"alften av den teoteriska f"oruts\"agelsen.
}
\clearpage

%
\minisec{Zusammenfassung}
{\small
\selectlanguage{german}
Atomare negative Ionen sind zerbrechliche Quantensysteme mit einer um
eine Gr"o\-"sen\-ordnung niedrigeren Bindungsenergie als Atome. Aufgrund
dieser schwachen Bindung spielt Korrelation unter den Valenzelektronen
eine sehr wichtige Rolle. In doppelt angeregten Zust"anden, wo zwei
Elektronen gleichzeitig angeregt sind, kann Korrelation dominierend
werden und ist folglich keine kleine St"orung des unabh"angigen
Teilchenverhaltens mehr.

Photoneutralisation ist der Prozess bei dem ein negatives Ion ein Photon
absorbiert und infolgedessen das "au"serste Elektron abgibt.  Die
Untersuchungen sind in kollinearer Geometrie mit einem Laser und
Ionen\-strahl\-"uber\-lapp von 0.5~m L"ange durchgef"uhrt. Dieses
erm"oglicht gleichzeitig hohe Aufl"osung und hohe Empfindlichkeit. Durch
Kombinieren von Messungen mit gleich- und gegenl"aufigen Laser- und
Ionenstrahlen kann die Dopplerverschiebung in \emph{allen} Ordnungen
beseitigt werden. Die Ionenstrahlanlage ist mit einer Zerst"auberquelle
und einer Plasmaquelle ausger"ustet, die zusammen die Produktion
negativer Ionen nahezu aller Elemente erlauben. Das Laserlicht f"ur die
Untersuchungen wird von zwei excimergepumpten Farbstoff\/lasern erzeugt.

Eine neue Nachweismethode ist entwickelt worden. Das von einer
Photoneutralisation verbleibende Atom wird mit
Resonanzionisationsspektroskopie zustandsempfindlich ionisiert. Diese
neuartige Nachweismethode bewahrt alle Vorteile der kollinearen
Geometrie und bietet gleichzeitig die M"oglichkeit partielle
Photo\-neutralisations\-quer\-schnitte zu messen.

Untersuchungen der Photoneutralisationsschwelle von Tellur mit
Neutralteilchennachweis wurden durchgef"uhrt und ergaben eine
Elektronenaffinit"at von 1\,589\,618(5)~m$^{-1}$. Mit dem
zustandsempfindlichen Nachweis wurde die Li(2p)
Photoneutralisationsschwelle untersucht und die Elektronenaffinit"at zu
498\,490(17)~m$^{-1}$ bestimmt.

Doppelt angeregte Zust"ande von Li$^{-}$ und He$^{-}$ sind studiert
worden. Es wurde Resonanzstruktur im Li$^{-}$
Photoneutralisationsquerschnitt nahe der Li(3p) Schwelle untersucht. Die
Struktur wird durch einen autoneutralisierenden doppelt angeregten
$^{1}$P$^{\text{u}}$ Zustand von Li$^{-}$, der im Bezug auf den Li(3p)
Zustand gebunden ist, hervorgerufen. Gem"a"s einer begleitenden
Berechnung ist diese Struktur analog zum symmetrischen intra-Schalen
Zustand $(_{3}\{0\}_{3}^{+})\,^{1}\text{P}^{\text{u}}$ des
H$^{-}$. H"oherliegende asymmetrisch inter-Schalen Zust"ande
konvergieren zur Li(3p) Schwelle.

In He$^{-}$ sind drei doppelt angeregte Zust"ande, 1s3s4s\,$^{4}$S,
1s3p$^{2}$\,$^{4}$P und 1s3p4p\,$^{4}$P, untersucht worden.  Die
Feschbachresonanz verbunden mit dem 1s3s4s\,$^{4}$S Zustand ist im
He(1s2s\,$^{3}$S) + e$^{-}$($\epsilon$s) und
He(1s2p\,$^{3}$P$^{\text{o}}$) + e$^{-}$($\epsilon$p) partiellen
Photoneutralisationsquerschnitt beobachtet worden und die
1s3p$^{2}$\,$^{4}$P und 1s3p4p\,$^{4}$P Resonanzen wurden im
He(1s2p\,$^{3}$P$^{\text{o}}$) + e$^{-}$($\epsilon$p) partiellen
Photoneutralisationsquerschnitt beobachtet.

Die Energie aller Resonanzen [1s3s4s\,$^{4}$S: $E_{0}=
(2\,386\,803.1\pm4.2)~\text{m}^{-1}$, $\Gamma = 160(16)~\text{m}^{-1}$;
1s3p$^{2}$\,$^{4}$P: $E_{0}= 2.478\,2(55)\times 10^{6}~\text{m}^{-1}$,
$\Gamma = 40(3)\times 10^{3}~\text{m}^{-1}$; 1s3p4p\,$^{4}$P: $E_{0}=
2\,633\,297(40)$ $\text{m}^{-1}$, $\Gamma = 492(35)~\text{m}^{-1}$]
stimmt mit einer j"ungst erschienen ab initio Rechnung von Xi und Froese
Fischer [Phys. Rev. A, \textbf{53}(5), 3169, (1996)] bis zu einer
Genauigkeit von $10^{-4}$ "uberein. Die Breiten sind sowohl im
Experiment als auch in der Theorie ungenauer: Die Breite der
1s3s4s\,$^{4}$S Resonanz stimmt mit der Berechnung "uberein, aber f"ur
die 1s3p$^{2}$\,$^{4}$P Resonanz gibt es eine leichte Abweichung und die
1s3p4p\,$^{4}$P Resonanz ist nur halb so breit wie vorhergesagt.  }
\clearpage

%
\minisec{Riassunto}
{\small
\selectlanguage{italian}
Gli ioni negativi sono dei fragili sistemi quantistici con energie
legame di un ordine di grandezza inferiore a quelle degli atomi.  A
causa di queste basse energie di legame la correlazione tra gli
elettroni di valenza gioca un ruolo molto importante.  Per stati
doppiamente eccitati, dove due elettroni sono simultaneamente promossi,
la correlazione diviene dominante e quindi non pu\`o pi\`u essere
consderata una piccola perturbazione del modello a particelle
indipendenti.

La fotoneutralizzazione \`e un processo in cui uno ione negativo assorbe
un fotone ed emette un elettrone dello stato pi\`u esterno.  Lo studio
di questo processo \`e stato qui compiuto utilizzando una geometria
collineare in cui il raggio laser ed il fascio ionico sono accoppiati
per una lunghezza di 0.5 m.  Ci\`o migliora simultaneamente risoluzione
e sensibilit\`a.  Combinando misure con il raggio laser e il fascio
ionico co- e contro-propaganti lo spostamento Doppler pu\`o essere
eliminato \emph{a tutti} gli ordini.  La strumentazione che produce il
fascio ionico \`e provvista di una sorgente a \lq sputtering\rq\ e una
sorgente di plasma, che insieme permettono la produzione di ioni
negativi per quasi tutti gli elementi.  La radiazione laser usata
nell'indagine \`e generata da due laser a coloranti pompati da un laser
ad eccimeri.

\`E stato sviluppato un nuovo schema di rivelazione. 
L'atomo risultante dal fotoneutralizzazione \`e ionizzato
stato-selettivamente (spettroscopia a risonanza di ionizzazione).
Questo nuovo schema conserva i vantaggi di una geometria collineare e
allo stesso tempo offre la possibilit\`a di misurare sezioni d'urto
parziali efficaci di fotoneutralizzazione.

Le indagini sulla soglia di fotoneutralizzazione del Te$^{-}$ con
rivelazione delle particelle neutre hanno prodotto una misura
dell'affinit\`a elettronica di 1\,589\,618(5)~m$^{-1}$.  La
spettroscopia a risonanza di ionizzazione \`e stata utilizzata per lo
studio della soglia di fotoneutralizzazione del Li(2p) e si \`e
determinato che l'affinit\`a elettronica del litio
\`e 498\,490(17)~m$^{-1}$.

Sono stati studiati gli stati doppiamente eccitati di Li$^{-}$ e
He$^{-}$ e la struttura risonante della sezione d'urto efficace del
Li$^{-}$ vicino alla soglia del Li(3p).  La struttura deriva
dall'autoneutralizzazione degli stati doppiamente eccitati
$^{1}$P$^{\text{o}}$ del Li$^{-}$ che sono legati rispetto allo stato
del Li(3p).  Secondo alcuni calcoli questa struttura \`e analoga a
quella dello stato intraguscio
$(_{3}\{0\}_{3}^{+})$\,$^{1}$P$^{\text{o}}$ dell'H$^{-}$.  Gli stati
interguscio superiori ecciatati asimmetricamente, sono stati osservati e
la loro energia converge a quella del Li(3p).

Per l'He$^{-}$ abbiamo studiato i tre stati doppiamente eccitati
1s3s4s\,$^{4}$S, 1s3p$^{2}$\,$^{4}$P e 1s3p4p\-\,$^{4}$P.  Le risonanze
di Feschbach associate con lo stato 1s3s4s\,$^{4}$S sono state osservate
in entrambe le sezioni d'urto parziali efficaci degli stati
He(1s2s\,$^{3}$S) + e$^{-}$($\epsilon$s) e
He(1s2p\,$^{3}$P$^{\text{o}}$) + e$^{-}$($\epsilon$p), mentre le
risonanze 1s3p$^{2}$\,$^{4}$P e 1s3p4p\,$^{4}$P sono state osservate
nella sezione d'urto parziale efficace He(1s2p\,$^{3}$P$^{\text{o}}$) +
e$^{-}$($\epsilon$p).

Le posizioni di tutte le risonanze [1s3s4s\,$^{4}$S: $E_{0}=
(2\,386\,803.1\pm4.2)~\text{m}^{-1}$, $\Gamma = 160(16)~\text{m}^{-1}$;
1s3p$^{2}$\,$^{4}$P: $E_{0}= 2.478\,2(55)\times 10^{6}~\text{m}^{-1}$,
$\Gamma = 40(3)\times 10^{3}~\text{m}^{-1}$; 1s3p4p\,$^{4}$P: $E_{0}=
2\,633\,297(40)~\text{m}^{-1}$, $\Gamma = 492(35)~\text{m}^{-1}$] sono
in accordo con recenti calcoli ab initio di Xi e Froese Fischer
[Phys. Rev. A, \textbf{53}(5), 3169, (1996)] con un'accuratezza fino a
$10^{-4}$.  Le larghezze di riga, comunque, sono meno accurate sia
nell'esperimento che nella teoria: la larghezza di riga della risonanza
1s3s4s\,$^{4}$S \`e in accordo con i calcoli, ma per la risonanza
1s3p$^{2}$\,$^{4}$P c'\`e una leggera discrepanza e la risonanza
1s3p4p\,$^{4}$P \`e larga solo la met\`a di quanto previsto.  }
\clearpage

\selectlanguage{english}
%
\tableofcontents
\cleardoublepage
\addcontentsline{toc}{chapter}{List of figures}
\listoffigures
\cleardoublepage
\addcontentsline{toc}{chapter}{List of tables}
\listoftables
\cleardoublepage
\addcontentsline{toc}{chapter}{Preface}
%
\chapter*{Preface} 
\label{ch1}
This work was done in the atomic physics group of Professor Ingvar
Lindgren at the department of Atomic Physics, G{\"o}teborg University
and Chalmers University of Technology. The presence of both theorists
and experimentalists in the same group contributes to a stimulating
atmosphere of scientific commitment.

The theorists of our group have been mainly engaged in accurate
calculations for heavy atomic systems where relativistic, nuclear and
QED effects are important. On the experimental side research has been
conducted in the fields of hyperfine structure, combustion, trace
element analysis, optical tweezers, and the negative ions.

During my first year I participated in building up a new negative ion
beam apparatus. Meanwhile I also learned Swedish. To begin with the
experiments were mainly led by others but later I gradually ran the
apparatus more and more on my own.  My first real enterprise, an
investigation of calcium, was a failure, at least as far as publishable
results are concerned. Thereafter things worked out better and we
started our very successful investigations of the negative lithium
ion. In early 1995 we had a breakthrough with the application of a
state selective detection scheme based on resonance ionisation.  This
finding triggered a series of resonance structure investigations that
has so far not found an end. Also the most important experiments for
this thesis use the state selective detection scheme.

Above all I wish to cordially thank my supervisor Dag Hanstorp for his
guidance and encouragement far beyond his \lq work assignments\rq . His
commitment to supervision is perhaps best illustrated by the weekly
meeting held with each doctoral student.  I am deeply indebted to him
and wish to express my gratitude for his continued support. This
doctorate would not have been possible had Professor Ingvar Lindgren not
accepted me as his student, and therefore I wish to convey my thanks to
him and hope to have lived up to his expectations.

Many of people have contributed to render the work presented here
possible.  Among them I acknowledge James R. Petersons and Professor
David J. Pegg for many discussions and ideas concerning the studies of
resonance structure in the photodetachment cross section of calcium,
lithium and helium. My colleagues Ulric Ljungblad and Gunnar Haeffler
are acknowledged for countless stimulating discussions.

This thesis is based on the following publications:
\begin{itemize}
\item
U.~{Berzinsh}, G.~{Haeffler}, D.~{Hanstorp}, A.~{Klinkm{\"u}ller},
  E.~{Lindroth}, U.~{Ljungblad}, and D.~J. {Pegg}.
\newblock Resonance structure in the {L}i$^{-}$ photodetachment cross 
section.
\newblock {\em Phys. Rev. Lett.}, 74(24):4795--4798, June 1995.
\newblock e-print: physics/9703015.
\item
Gunnar {Haeffler}, Dag {Hanstorp}, Igor {Kiyan}, Andreas~E. 
{Klinkm{\"u}ller}, Ulric {Ljungblad} and David J.~{Pegg}.
\newblock Electron affinity of {L}i: {A} state-selective measurement.
\newblock {\em Phys. Rev.~A}, 53(6):4127--4131, June 1996.
\newblock e-print: physics/9703013.
\item
Gunnar {Haeffler}, Andreas~E. {Klinkm{\"u}ller}, Jonas {Rangell}, Uldis
  {Berzinsh}, and Dag {Hanstorp}.
\newblock The electron affinity of tellurium.
\newblock {\em Z. Phys. D}, 38:211--214, October 1996.
\newblock e-print: physics/9703012.
\item
Andreas~E. {Klinkm{\"u}ller}, Gunnar {Haeffler}, Dag {Hanstorp},
  Igor~Yu.  {Kiyan}, Uldis {Berzinsh}, Christopher {Ingram},
  David~J. {Pegg}, and James~R. {Peterson}.
\newblock Photodetachment study of the $1s3s4s\,^{4}\!S$ resonance in 
He$^{-}$.
\newblock {\em Phys. Rev.~A}, accepted for publication, 1997.
\newblock e-print: physics/9703011.
\item
D.~{Hanstorp}, G.~{Haeffler}, A.~E. {Klinkm{\"u}ller}, U.~{Ljungblad},
  U.~{Berzinsh}, I.~Yu. {Kiyan}, and D.~J. {Pegg}.
  \newblock Two electron dynamics in photodetachment.
  \newblock {\em Adv. Quant. Chem.}, 29, 1997.
  \newblock e-print: physics/9706013.
\end{itemize}
The results have also been presented at several international
conferences
\cite{Lju-94,Ber-95-3,Kli-96,Hae-96-4,Hae-96-5,Kli-97-2,Kli-97-3}.

All occurrences of author names in this thesis shall be understood as
\lq author et al.\rq\ if applicable. The reader, tempted to cry out
``foul'', is referred to the bibliography to find that there, contrary
to common practice in physics, the full list of authors and the title
are presented.

After having suffered from the stupidity of unit conversion from CGS and
other systems for many years I decided to use the SI
\cite{UIP-78,Tay-95} units throughout this thesis. To convert eV to
m$^{-1}$ we used the recommended factor 806\,554.10(24)~(m$^{-1}$/eV)
\cite{Coh-88}. For other eventual conversions we used the
factors from the same source \cite{Coh-88}. Theoretical results are
sometimes quoted in element specific atomic units.

%
\chapter{Introduction}
\label{ch2}
Negative ion research is a rapidly developing field of atomic
physics. Good introductions to the subject of negative ions are given in
the books of Massey \cite{Mas-38} and Smirnov
\cite{Smi-76}. The reviews by Buckman \cite{Buc-94}, Esaulov
\cite{Esa-86} and Andersen \cite{And-91} give an introduction to more
recent research. 

Negative ions are more sensitive to correlation effects than
isoelectronic atoms or positive ions, since for this member of a
sequence the core field is weakest and therefore the masking of the more
subtle inter-electronic interaction is reduced. This makes negative ions
a critical testing ground for theory since for highly correlated valence
electrons the independent particle model ceases to be a valid
approximation.

In doubly excited states, where an extra electron moves in the field of
an excited atom, the strength of the interaction between the two
electrons becomes comparable to the weakened interaction of each
electron with the core. Under these conditions it is possible to study
the dynamics of electron-electron correlation.

Most of the methods currently used to investigate negative ions are
based on the interaction with narrow bandwidth laser light
\cite{And-91}. Other powerful methods involve the scattering of
electrons on a neutral atom target. Weakly bound negative ions can also
be studied with field detachment in electric or magnetic fields.  In
electron scattering experiments resonances due to transient negative ion
states and thresholds appear as prominent structures in the scattering
cross section. This method allows one to cover large energy regions but
suffers from a relatively limited resolution of roughly 10~meV.

In field detachment experiments \cite{Opa-74,Nad-92} typically a fast
ion beam is exposed to an electric field. Negative ions with binding
energies of some tens of meV can be studied with this method. A
resolution of a few meV is realistic.

Photodetachment \cite{Dra-96-2} is another powerful technique to study
negative ions. Both products, the neutral atom and the photo-electron,
of the photodetachment can be detected. This leads to two principal
methods: Photo-electron spectroscopy \cite{Dra-96-4}, where the outgoing
electron is registered, and neutral particle detection, where the number
of produced atoms is counted.

In laser photo-electron spectroscopy a photon detaches an electron from
the negative ion and the energy of the escaping electron is
monitored. This permits one to determine the state in which the residual
atom is left, thus making it a channel specific method. If, in addition,
angular resolved experiments are performed, then the state of the
outgoing electron can be, apart from the spin orientation, fully
registered. The most severe limitation of the method is the resolution
that typically never exceeds 10~meV. Photo-electron spectroscopy is
usually performed with the laser beam perpendicularly intersecting the
ion beam.

Laser photodetachment with neutral particle detection gives a very high
resolution, especially if applied in a collinear geometry. Thus far,
most very precise electron affinity determinations have been made with
this method \cite{Hot-85,Blo-95}. Several such measurements are
presented in this thesis.

Recently this versatile method has been made channel specific by
selectively detaching the residual atom using resonance ionisation
\cite{Pet-95,Hae-96-1}. This
combination of collinear interaction and selective detection is rapidly
becoming a standard tool for negative ion investigations, since it
offers high sensitivity, resolution and selectivity. The method has been
developed by us at the Department of Atomic Physics in G{\"o}teborg
\cite{Hae-96-1} and, 
independently, in the research group of T. Andersen at Aarhus University
in Denmark \cite{Pet-95}. Several experiments presented in this thesis
have been performed using this new method.

Other accelerator based methods have also been developed
\cite{Har-90}. The group of T. Andersen in Aarhus, for example, has 
performed several experiments in a storage ring for negative ions. This
offers the possibility of Doppler tuning. By monitoring the stored
current they are also able to measure the lifetime of metastable
negative ions
\cite{And-93}.

Experiments on negative ions can also be performed on ions stored in
traps \cite{Lar-88}. A group at the University of Virginia has done
experiments on photodetachment of negative ions in the presence of
static magnetic fields \cite{Lar-85}. They found interesting periodic
modulations in the photodetachment cross section near the threshold.

With high power lasers it is also possible to induce multi-photon
processes in negative ions \cite{Cra-87,Blo-93}. For instance, excess
photon absorption in the photodetachment of Rb$^{-}$ has been observed
\cite{Sta-94}. Raman like coupling schemes \cite{Kri-93} have been
successfully applied to negative ions and the fine structure splitting
of Te$^{-}$ has been determined with this technique \cite{Tho-96-2}.

%
\chapter{Atomic negative ions}
\label{ch3}
In this chapter we briefly introduce some results from the theory of
atoms. We begin with a Hamiltonian, which contains, apart from the
Coulomb interaction, only the spin-orbit coupling. This can be thought
of as resulting from an heuristic derivation or as an approximation of
the Dirac equation.

An atomic negative ion is formed when an electron is attached to an
atom. The attachment process is discussed in section \ref{ch33}. With
the exception of nitrogen, the noble gases and the mercury group,
essentially all elements form negative ions. Negative ions are genuine
atomic systems, and consequently the description is very similar to
atoms.

The actual calculations of binding energies, fine structure or
excitation energies of negative ions, however, turns out to be a
formidable task. For instance the alkali earth metals negative ions are
bound with only some tens of meV, which is only a few times $10^{-7}$ of
the total electronic energy. Thus a very high precision has to be
attained on the total energy to obtain a good value for the electron
affinity.

The Hamiltonian for an atomic multi-electron singly charged negative ion
with spin-orbit coupling is \cite{Wei-78}
\begin{align}
\mathcal{H} &= \mathcal{H}_{0}^{\prime} + \mathcal{H}_{1}^{\prime} +
\mathcal{H}_{2}\label{eqch31a}\quad,\\
\intertext{where}
\mathcal{H}_{0}^{\prime} &= \sum_{i=1}^{Z+1}\left(
- \frac{\hbar^{2}}{2m}\nabla^{2}_{i} -
Z\frac{q^{2}}{4\pi\epsilon_{0}r_{i}}\right)\label{eqch31b}\quad,\\
\mathcal{H}_{1}^{\prime} &=
\sum_{i<j=1}^{Z+1}\frac{q^{2}}{4\pi\epsilon_{0}r_{ij}}\label{eqch31c}
\quad,\\
\mathcal{H}_{2} &= \sum_{i=1}^{Z+1} -\frac{\hbar^{2}}{2m^{2}c^{2}}
\frac{1}{r}
\frac{\partial V(r_{i})}{\partial r_{i}}\boldsymbol{\mathrm{l}}_{i}
\cdot
\boldsymbol{\mathrm{s}}_{i} \label{eqch31d}\quad.
\end{align}
The Hamiltonian $\mathcal{H}_{0}^{\prime}$ is the kinetic plus the
potential energy of an electron in the potential of a nucleus with a
charge $Z$, the Hamiltonian $\mathcal{H}_{1}^{\prime}$ is the mutual
electrostatic repulsion among the electrons, and finally Hamiltonian
$\mathcal{H}_{2}$ is the spin-orbit coupling in the central field
approximation.

In the central field approximation the spherical symmetric part of the
Coulomb repulsion \eqref{eqch31c} $\mathcal{H}_{1}^{\prime}$ is added to
$\mathcal{H}_{0}^{\prime}$ to obtain a one-electron operator $U(r)$,
\begin{align}
\sum_{i=1}^{Z+1} U(r_{i}) 
&=
- \sum_{i=1}^{Z+1}
Z\frac{q^2}{4\pi\epsilon_{0}r_{i}} + 
\left\langle
\sum_{i<j=1}^{Z+1}\frac{q^{2}}{4\pi\epsilon_{0}r_{ij}}
\right\rangle 
\quad ,\label{eqch34a}\\
\intertext{where the $\langle\;\rangle$ denotes the spherical average.
The new Hamiltonians $\mathcal{H}_{0}$ and $\mathcal{H}_{1}$ are:}
\mathcal{H}_{0} &= \sum_{i=1}^{Z+1}\left(- \frac{\hbar^2}{2m}\nabla^{2}
+ U(r_{i})
\right)\label{eqch34b} = \sum_{i=1}^{Z+1} h_{0i}\quad ,\\
\mathcal{H}_{1} &=
\sum_{i<j=1}^{Z+1}\frac{q^2}{4\pi\epsilon_{0}r_{ij}} -
\left\langle
\sum_{i<j=1}^{Z+1}\frac{q^2}{4\pi\epsilon_{0}r_{ij}} \right\rangle
\label{eqch34c}\quad .
\end{align}
Now the total Hamiltonian can be written as
\begin{equation}\label{eqch322}
\mathcal{H} = \mathcal{H}_{0} + \mathcal{H}_{1} + \mathcal{H}_{2} 
\quad . 
\end{equation}
The two limiting cases with 
\begin{equation}
\mathcal{H}_{0}\gg\mathcal{H}_{1}\gg\mathcal{H}_{2}\; ,\quad 
\text{or$\;$}
\quad\mathcal{H}_{0}\gg\mathcal{H}_{2}\gg\mathcal{H}_{1}
\quad ,\label{eqch324}
\end{equation}
correspond to LS-coupling and jj-coupling. In all other cases different
coupling schemes have to be used \cite{Cow-81-2}.

The central field equation is
\begin{equation}\label{eqch35}
h_{0i}\psi_{i} = \epsilon_{0i}\psi_{i}\quad ,
\end{equation}
where $\epsilon_{0i}$ is the energy eigenvalue of the i$^{\text{th}}$
electron.  For the i$^{\text{th}}$ electron the one-orbitals can be
expressed as the product of a radial function $R_{nl}$, an angular
function $Y_{lm}$, and a spin function $\xi$:
\begin{equation}\label{eqch36}
\psi_{i}(r_{i},\Omega_{i})=R_{n_{i}l_{i}}(r_{i})
Y_{l_{i}m_{i}}(\Omega_{i})\xi_{m_{s_{i}}}(s_{i})\quad .
\end{equation}
The eigenfunctions $\Psi$ of the multi-electron equation
\begin{equation}\label{eqch357}
\mathcal{H}_{0}\Psi = E_{0}\Psi \quad ,
\end{equation}
are anti-symmetrised products of central field orbitals, or, for
brevity, orbitals.  The energy eigenvalue of a multi-electron state
$E_{0}$ is the sum of one-electron energies $\epsilon_{0i}$ for the
respective configuration.  The angular function $Y_{lm}$ and the spin
function $\xi$ are identical to the one-electron case. To determine the
radial function $R_{nl}$ the Hartree-Fock method can be used.  With
these one-electron orbitals one can construct N-electron wavefunctions
$\Psi$ complying with the Pauli exclusion principle. This
anti-symmetrised wavefunction $\Psi$ is usually written as a Slater
determinant,
\begin{equation}\label{eqch37}
\Psi=\frac{1}{\sqrt{N!}}
\begin{vmatrix}
\psi_{1}(\lambda_{1}) & \psi_{2}(\lambda_{1}) & \dots &
\psi_{N}(\lambda_{1}) \\
\psi_{1}(\lambda_{2}) & \psi_{2}(\lambda_{2}) & \dots &
\psi_{N}(\lambda_{2}) \\
\dots & \dots & \dots & \dots \\
\psi_{1}(\lambda_{N}) & \dots & \dots & \psi_{N}(\lambda_{N})
\end{vmatrix}\quad ,
\end{equation}
where $\lambda_{i}$ is the $i^{\text{th}}$ set of spin and spatial
coordinates, $\psi_{j}(\lambda_{i})$ a spin orbital, and, $j$ a set of
quantum numbers. The parity of such a wavefunction $\Psi$ is given by
$(-1)^{l_{1}}(-1^{l_{2}})\dots (-1)^{l_{N}}$. The distribution of the
electrons with respect to $n$ and $l$ is called configuration, and
electrons with the same $n$ and $l$ are called equivalent. A
multi-configurational wave function is a superposition of several Slater
determinants,
\begin{equation}
\Psi = \sum_{i} c_{i} \Psi_{i} \quad .
\end{equation}

The starting points of most calculation are either, for low nuclear
charge where $\mathcal{H}_{1}$ dominates over $\mathcal{H}_{2}$
\eqref{eqch324},
\begin{align}
\mathcal{H}_{\text{a}} &= \mathcal{H}_{0} +
\mathcal{H}_{1}\label{eqch32a}\quad,\\ 
\intertext{or, for high nuclear charge where $\mathcal{H}_{2}$ dominates
over $\mathcal{H}_{1}$ \eqref{eqch324},}
\mathcal{H}_{\text{b}} &= \mathcal{H}_{0} +
\mathcal{H}_{2}
\label{eqch32b}\quad.
\end{align}
In both cases the second term of the sum, $\mathcal{H}_{1}$ and
$\mathcal{H}_{2}$ respectively, is treated as a perturbation to
$\mathcal{H}_{0}$. 

The Hamilton operator \eqref{eqch322} is rotation invariant, the
eigenfunctions of $\mathcal{H}$ consequently form a basis of the
irreducible representations of the rotation group. The spherical
harmonics $Y_{lm}(\Omega )$ in \eqref{eqch36} constitute such a
basis. Furthermore, the Hamiltonian is invariant under the exchange of
electrons (exchange of both space and spin coordinates), because
electrons are indistinguishable. The eigenfunctions accordingly form a
basis to the representations of the permutation group. In compliance
with the Pauli principle only the antisymmetric representations, written
for example as Slater determinants \eqref{eqch37}, occur for fermions
like electrons. The inversion symmetry is reflected by the definite
parity of the eigenfunctions of $\mathcal{H}$.

The operator $\mathcal{H}_{\text{a}}$ conserves the total orbital
angular momentum $\mathbf{L}$ and the total spin $\mathbf{S}$ in
addition to the total angular momentum $\mathbf{J}$, as is expressed by
the following commutators:
\begin{alignat}{3}
\big[ \mathcal{H}_{\text{a}},\boldsymbol{\mathrm{L}} \big] &= 0 \qquad &
& \mbox{if}\qquad&  
\boldsymbol{\mathrm{L}} &=
\sum_{i}\boldsymbol{\mathrm{l}}_{i}\label{eqch33a}\quad,\\
\big[ \mathcal{H}_{\text{a}},\boldsymbol{\mathrm{S}} \big] &= 0 \qquad&
& \mbox{if}\qquad&
\boldsymbol{\mathrm{S}} &=
\sum_{i}\boldsymbol{\mathrm{s}}_{i}\label{eqch33b}\quad,\\
\big[ \mathcal{H}_{\text{a}},\boldsymbol{\mathrm{J}} \big] &= 0 \qquad&
& \mbox{if}\qquad&
\boldsymbol{\mathrm{J}} &=
\boldsymbol{\mathrm{L}} + \boldsymbol{\mathrm{S}}\label{eqch33c}
\quad .\\
\intertext{The operator $\mathcal{H}_{\text{b}}$ conserves the total
angular momentum $\mathbf{J}$,}
\big[ \mathcal{H}_{\text{b}},\boldsymbol{\mathrm{J}} \big] &= 0 \qquad&
& \mbox{if}\qquad&
\boldsymbol{\mathrm{J}} &= \sum_{i}
\boldsymbol{\mathrm{j}}_{i}\label{eqch33d}\quad,\\ 
&&&&\boldsymbol{\mathrm{j}}_{i} &= \boldsymbol{\mathrm{l}}_{i} +
\boldsymbol{\mathrm{s}}_{i}\quad .\label{eqch33e}
\end{alignat}

The eigenfunctions of $\mathcal{H}_{\text{a}}$ can also be chosen to be
simultaneous eigenfunctions of $L^{2}, L_{z}, S^{2}$, and $S_{z}$, that
are written as $|LSM_{L}M_{S}\rangle$ plus an additional label that
describes for instance the radial part of the wavefunction. It is
possible to construct alternative simultaneous eigenfunctions of
$\mathcal{H}_{\text{a}}$ and $L^{2}, S^{2}, J^{2}$, and $J_{z}$. Such
functions can be written as $|LSJM_{J}\rangle$. The eigenfunctions of
$\mathcal{H}_{\text{b}}$ can only be chosen to be simultaneous
eigenfunctions of $J^{2}$ and $J_{z}$, written as $|JM_{J}\rangle$.

As long as only $\mathcal{H}_{0}$ is considered, the energy depends only
on the sum $n+l$, hence all levels within a configuration are
degenerate. In LS-coupling, used in conjunction with
$\mathcal{H}_{\text{a}}$ \eqref{eqch32a}, the non spherical part of the
Coulomb interaction $\mathcal{H}_{1}$ \eqref{eqch34c} splits a
configuration into terms whose energies depend on $L$ and $S$. The
spin-orbit interaction $\mathcal{H}_{2}$ \eqref{eqch31d} then splits
these terms further into levels whose energies depend of $J$. In
jj-coupling, used in conjunction with $\mathcal{H}_{\text{b}}$
\eqref{eqch32b}, a configuration is first split by the spin-orbit
interaction $\mathcal{H}_{2}$ \eqref{eqch31d} into terms with every
possible combination of the individual $j$ values and then by the
non-spherical part of the Coulomb interaction $\mathcal{H}_{1}$
\eqref{eqch34c} into levels. For intermediate nuclear charge values
other coupling schemes apply \cite{Cow-81-2}, but the resulting levels
can always be \emph{labelled} in LS-coupling nomenclature. 

\minisec{On correlation}

The troublesome part of the Hamiltonian $\mathcal{H}$ \eqref{eqch322} is
the mutual repulsion $\mathcal{H}_{1}$ \eqref{eqch34c} among the
electrons which induces correlated motion. Yet there is a clear
distinction between core electrons and valence electrons. Following
Amusia \cite{Amu-90-1} the interaction energy $E_{\text{int}}$ is
\begin{equation}\label{eqch39}
E_{\text{int}} = \frac{q^{2}}{4\pi\epsilon_{0}\overline{r}}
N_{\text{Sh}}\quad ,
\end{equation}
where $N_{\text{Sh}}$ is the number of electrons in a shell and
$\overline{r}$ is its mean radius. The mean kinetic energy
$E_{\text{kin}}$ is
\begin{equation}\label{eqch310}
E_{\text{kin}}=\frac{P^2}{2m}\quad ,
\end{equation}
where $P$ is the mean momentum and $m$ the electron mass. If the ratio,
\begin{align}\label{eqch311}
\eta &= \frac{E_{\text{int}}}{E_{\text{kin}}} \\
&= \sqrt[3]{N_{\text{Sh}}}\frac{\overline{r}}{a_{0}}\quad ,
\end{align}
is much less than one, the self consistent field dominates. For outer
electrons, especially in negative ions, the mean radius $\overline{r}$
can be much larger than the Bohr radius $a_{0}$, and thus correlation
effects become important or even dominating. 

The correlation energy is commonly defined as effects beyond the
independent particle model,
\begin{equation}\label{eqch312}
E_{\text{c}}=E_{\text{exp}} - E_{\text{HF}} - E_{\text{rel}}\quad ,
\end{equation}
where $E_{\text{exp}}$ is the experimental energy value, $E_{\text{HF}}$
is the Hartree-Fock energy, $E_{\text{rel}}$ is the relativistic
contribution and $E_{\text{c}}$ is called the correlation energy.  This
correlation energy increases roughly linearly with the nuclear charge
$Z$ and is very nearly equal for the iso-electronic positive ion, atom
and negative ion \cite{Alo-96}. The differences and similarities of
atoms and negative ions are further illuminated in the following
section.

\section{Structure of atomic negative ions}
\label{ch31}
Atoms in general have many bound discrete states. Most of the knowledge
about atomic systems is based on investigations of the energy position
of these bound states. The energy positions are mostly derived from line
spectra.  Negative ions do not have discrete spectra, owing to lack of
bound states, but only a continuum above the photodetachment
threshold. This continuum however contains resonance structure due to
excited autodetaching states of the negative ion. The short lifetime of
these autodetaching states leads to rather broad resonances, limiting
the precision and accuracy with which resonance energies can be
determined.

\begin{figure}
\begin{minipage}{\textwidth}
\parbox[b]{0.29\textwidth}{
\epsfig{file=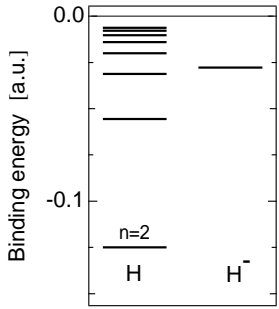, width=0.27\textwidth}
} \hfill
\parbox[b]{0.7\textwidth}{
\protect\caption[Binding energies of H and H$^{-}$]{\label{bi010}\sloppy
Binding energies of H and H$^{-}$: Energy levels for H and H$^{-}$ with
respect to their ionisation/detachment limits. The ground state binding
energies of H, 0.5 a.u., and H$^{-}$, approx.~0.025 a.u., differ by
roughly one order of magnitude, reflecting the different strength of the
binding potential (see figure \ref{bi020}). Note also the absence of a
Rydberg series for H$^{-}$ stemming from the lack of a long range
asymptotic potential.}  }
\end{minipage}
\end{figure}

Negative ions are fragile quantum systems with some striking properties
distinguishing them from other atomic systems such as positive ions and
neutral atoms. The higher binding energy of valence electrons in atoms
of around 10~eV (figure \ref{bi010}), compared to about 1~eV for
negative ions, stems from the strength of the Coulomb interaction. The
strength of an interaction has no definite relation to its range: there
are short ranged very strong interactions, the strong interaction
between hadrons, and long ranged very weak interactions, such as
gravitation. The binding energy is mainly determined by the strength of
the potential, or, loosely speaking, its depth at the origin. From the
lower binding energy of H$^{-}$ we can conclude that the binding
potential for the extra electron is relatively shallow.

\begin{figure}
\begin{minipage}{\textwidth}
\parbox[b]{0.54\textwidth}{
\epsfig{file=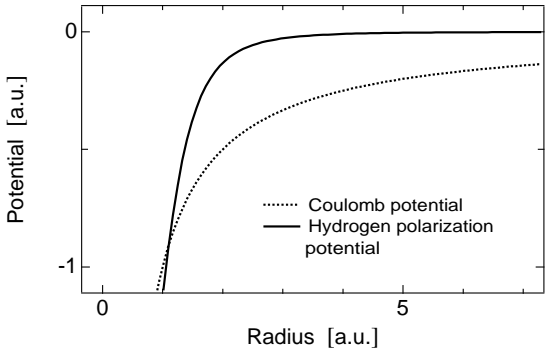, width=0.52\textwidth}
} \hfill
\parbox[b]{0.45\textwidth}{
\protect\caption[Asymptotic potentials of H and H$^{-}$]{\label{bi020}
\sloppy \small
Asymptotic potentials of H and H$^{-}$: The asymptotic Coulomb
potential, $V(r)=-1/(r+r_{0})\quad r_{0}=18\times 10^{-6}$ (proton
radius), is compared to the polarisation potential \eqref{eqch313} of H,
$V_{\text{pol}}(r)=-4.5/2(r^4+1)$, with no linear Stark effect.}  }
\end{minipage}
\end{figure}

Another characteristic of atoms and positive ions are the Rydberg series
converging on the detachment limit. These series exist as a consequence
of the long range nature of the Coulomb potential which, according to
Levinsons theorem, supports an infinite number of asymptotic levels. In
contrast, the asymptotic potential acting on an electron escaping from a
negative ion, figure \ref{bi020}, is only the polarisation potential
$V_{\text{pol}}$,
\begin{equation}\label{eqch313}
V_{\text{pol}}(r)=\frac{-\alpha_{\text{D}}}{2(r^{4}+r_{0}^{4})}\quad
\text{with}\quad r\gg r_{0}\quad ,
\end{equation}
where $\alpha_{\text{D}}$ is the dipole polarisability of the atom and
$r_{0}$ its diameter. This is a short ranged potential that does not
support Rydberg series.

In atoms and positive ions the lion's share of the ground state binding
energy is described by the independent particle model, and the remaining
discrepancy to the correct experimental value is due to electron
correlation \eqref{eqch312}. Even atoms have bound doubly excited states
with very high correlation, which are poorly described by the
Hartree-Fock model. In such states the independent particle model ceases
to be a valid approximation.

For negative ions, sometimes, not even the stability of the ground state
can be predicted from an independent particle point of view
\cite{Sal-96}. The electron correlation effects in negative ions are no
longer small corrections, but decisively contribute to the binding
energy. Roughly speaking we could view the situation for negative ion
binding energies as a shift of orders compared to atoms or positive
ions: what is a first order correction for an atom is zeroth order for a
negative ion. Hence negative ions are somewhat like a magnifying glass
for correlation effects among the outermost electrons. Since these
systems are moreover relatively easily accessible to ion-beam-laser
experiments, they have attracted a great deal of interest.

\section{Excited states}
\label{ch32}
Most of the bound excited states in negative ions are fine structure
levels \cite{Hot-85}. The ones where the configuration is different from
the ground state we call electronically excited.  In this thesis we will
call \emph{all} states above the ground state excited states.  Excited
states of negative ions can be singly or doubly excited and situated
below or above the detachment limit.  Only in rare cases, for example
platinum and iridium, even electronically excited states are bound
\cite{Hot-85}. Therefore practically all electronically excited states
of negative ions lie above the detachment limit and are consequently
autodetaching.

\minisec{Autodetachment}
Since autodetachment is such an important mechanism for the decay of
excited states of negative ions I hereby introduce the autodetachment
selection rules and outline their physical origin. The treatment closely
follows the one given by Brage \cite{Bra-91}.

Autodetachment can be crudely described as a transition from a discrete
state to an adjacent continuum \cite{Bra-91} (figure \ref{bi120}).  This
process is induced by a Breit-Pauli Hamiltonian
$\mathcal{H}_{\text{BP}}$
\cite{Dra-96-3,Bet-57}. The autodetaching state $\psi(E_{0})$  can, in a
first approximation, be written as a sum of a discrete state wave
function $\psi_{0}$ and a continuum wave function $\psi_{\text{cont}}$,
representing the open channel,
\begin{equation}\label{eqch314}
\psi(E_{0})=\psi_{0} + \psi_{\text{cont}}\quad ,
\end{equation}
where $E_{0}$ is the energy of the state $\psi_{0}$. If we assume
Fermi's Golden Rule to be valid, the autodetachment rate $\xi$ is
\begin{equation}\label{eqch315}
\xi = \left|\langle\psi_{\text{cont}}|\mathcal{H}_{\text{BP}} -
E_{0}|\psi_{0}\rangle\right|^{2}\quad .
\end{equation}
To calculate this matrix element is the subject of theoretical
investigations.

The Breit-Pauli Hamiltonian $\mathcal{H}_{\text{BP}}$ consists of
different parts,
\begin{equation}\label{eqch318}
\mathcal{H}_{\text{BP}} = \mathcal{H}_{\text{NR}} +
\mathcal{H}_{\text{RS}} + \mathcal{H}_{J}\quad ,
\end{equation}
where $\mathcal{H}_{\text{NR}}$ and $\mathcal{H}_{\text{RS}}$ are the
$J$ independent parts, consisting of non-relativistic and relativistic
shift operators, respectively. The operator $\mathcal{H}_{J}$ consists
of the $J$ dependent spin-orbit, spin-spin, and spin-other-orbit
operators.

Here we want to compile the relevant selection rules for autodetachment.
In all of the subsequent processes parity is conserved. The decay of an
autodetaching state can be induced by the Coulomb repulsion or the
relativistic terms in the Hamiltonian. The Coulomb interaction conserves
orbital angular momentum $L$, spin $S$, and the total angular momentum
$J$, and consequently the selection rules for Coulomb autodetachment are
$\Delta L=\Delta S =\Delta J =0$ (table \ref{tach31}).

A typical decay time for an autodetaching state is 10~fs. Autodetachment
induced by relativistic terms in the Hamiltonian \eqref{eqch318} are in
the order of a factor of $\alpha^{4}$ slower than Coulomb
autodetachment, but the selection rules are less restrictive (table
\ref{tach31}). These selection rules hold strictly for LS-coupling
only. For other couplings the selection rules on parity $\pi$ and total
angular momentum $J$ still hold, while the others are relaxed.

\begin{table}
\begin{center}
\caption[Autodetachment selection rules]{\label{tach31}\sloppy
Autodetachment selection rules: Conventionally if one speaks of
autodetachment only the Coulomb autodetachment is meant. In all cases
the parity is conserved. The labels SO, SOO mean spin-orbit and
spin-other-orbit induced, respectively, and SS stands for spin-spin
induced. These names refer to the type of interaction in the Breit-Pauli
Hamiltonian \eqref{eqch318} that induces the respective transition
\cite{Bra-91}.}\medskip
\begin{tabular}{cccc}
\hline\hline
Quantity & Coulomb & SO, SOO & SS \\ \hline
$\Delta L$ & 0 & 0, $\pm 1$ & 0, $\pm 1$, $\pm 2$ \\
$\Delta S$ & 0 & 0, $\pm 1$ & 0, $\pm 1$, $\pm 2$ \\
$\Delta J$ & 0 & 0 & 0\\
\hline\hline
\end{tabular}
\end{center}
\end{table}

If a state can decay via Coulomb autodetachment this mechanism will
completely dominate, owing to the much higher detachment rate than the
relativistically induced decays.  A given state that cannot decay via
Coulomb autodetachment may have small admixtures, caused by
configuration mixing, that couple to the continuum through Coulomb
autodetachment. The decay rate induced by these small admixtures is
likely to dominate the total decay rate. This mechanism is called
induced Coulomb autodetachment.

\minisec{Doubly excited states}
Among the excited states of negative ions, just as for atoms, we find
singly and doubly excited states. The expansion of the exact
wavefunction, as obtained from perturbation theory,
multi-configurational Dirac-Fock calculations (MCDF) or configuration
interaction (CI) calculations, in configurations is usually dominated by
one or a few terms. For doubly excited states the analogous expansion
contains many configurations with about equal weight. Hence, the state
can no longer be meaningfully labelled with one of these
configurations. It would be desirable to assign quantum numbers to
doubly excited states that reflect the symmetry of a two-electron state.

This goal has been pursued since the discovery of doubly excited states
in He \cite{Coo-63,Mad-63} and new approximate quantum numbers have been
derived. The issue of doubly excited states has not yet been treated in
standard textbooks, but there are numerous reviews and articles
\cite{Reh-78-2,Rau-90,Ber-89-2,Ros-91,Win-94,Lin-93-2,Lin-87-2} that
give a good introduction to the various aspects of doubly excited
states. Here we present only one of several possible descriptions.  In
this $K,T$ scheme 
\cite{SiH-75,SiH-76,HeS-75,Lin-76,KeH-78,HeK-80,HKP-80,Lin-82-2,%
Lin-82-1,FeB-86} doubly excited states are labelled,
\begin{equation}\label{eqch316}
_{n}(K,T)_{N}^{A}\,^{2S+1}L_{J}^{\pi}\quad ,
\end{equation}
with $n,N,S,L,\pi ,J$ having their usual meaning. The quantum numbers
$K,T$ and $A$ are specific two electron quantum numbers. They can take
the following values:
\begin{align}
T &= 0, 1, \dots , \min (L, N-1)\label{eqch317a}\\
K &= N-T-1, N-T-3, \dots , -(N-T-1) \label{eqch317b}\\
A &=
\begin{cases} (-1)^{S+T+\pi} & K>L-N\\
                   0 & K\le L-N\quad .\\
\end{cases}
\label{eqch317c}
\end{align}
The quantum numbers $K$ and $T$ are obtained \cite{Wul-73,Wul-83} from
the decomposition of $SO(4)\otimes SO(4)$ into $SO(4)$ representation
and label the so called
\lq doubly excited symmetry basis\rq . 

\begin{figure}
\begin{minipage}{\textwidth}
\parbox[b]{0.4\textwidth}{
\epsfig{file=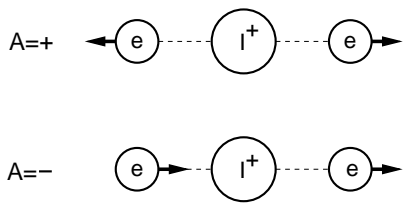, width=0.38\textwidth}
} \hfill
\parbox[b]{0.59\textwidth}{
\protect\caption[Significance of $A$ for doubly excited
states]{\label{bi040}\sloppy\small Significance of A for doubly excited
states: If the quantum number $A$, introduced in section \ref{ch32}, has
the value + the two electrons vibrate anti symmetrically as depicted in
the part (\textsf{a}) and for $A=-$ they stretch symmetrically as
indicated in part (\textsf{b}) of the figure. States with A=0 are singly
excited.}  }
\end{minipage}
\end{figure}

These labels are approximate in as much as the symmetry group of two
electrons with electrostatic repulsion is only almost $SO(4)\otimes
SO(4)$. Physically $K$ and $T$ have the following interpretation
\cite{Hei-93,Lin-93-2}: $K$ is proportional to $-\langle
\cos\theta_{12} \rangle$, where $\theta_{12}$ is the angle between the
two electrons with respect to the nucleus, and $T$ measures to what
extend the two electron orbits are not coplanar. Both $K$ and $T$ thus
describe angular correlation.

The quantum number $A$ was introduced \cite{Lin-85} to characterise radial
correlation. A value $A=0$ corresponds to singly excited states, $A=+$
can be envisioned as an anti-symmetric stretching vibration of the
electrons with respect to the ion core, see figure
\ref{bi040}, and $A=-$ as a symmetric stretch. Based on these quantum
numbers, approximate selection rules, the propensity rules, for
radiative and non-radiative transitions can also be derived
\cite{RoB-90}.
%
\section{Formation}
\label{ch33}
To study negative ions requires ways to produce them in sufficient
numbers.  Even though atomic negative ions in general have a low binding
energy they can be formed in a great variety of processes
\cite{Mas-38}. Here we present only a selection that is related to the
experiments performed in this thesis. There are different classes of
formation processes: Radiative processes, dissociative processes, three
body collisions and electron capture from bound states. Among the
radiative processes, radiative attachment,
\begin{equation}\label{eqch319}
\text{A} + \text{e}^{-} \mapsto \text{A}^{-} + \hbar\omega\quad ,
\end{equation}
is the time inverse of photodetachment \eqref{eqch401}. The probability
for this process, however, is low since it is approximately the ratio of
the transit time of a thermal electron to pass an atom, about 1~fs, to
the radiation time for a photon, about 10~ns, making it ineffective in
the production of negative ion beams of any appreciable intensity.  This
inefficiency can also be understood as a consequence of angular momentum
conservation, since the electron has to hit the atom with an impact
parameter in a very narrow range.  Another radiative negative ion
formation process is the polar photodissociation,
\begin{equation}\label{eqch320}
\text{A}_{2} + \hbar\omega \mapsto \text{A}^{+} + \text{A}^{-}\quad ,
\end{equation}
which has been observed for instance for iodine \cite{Mas-38}. 

The low probability of \eqref{eqch319} increases a lot if a third body
participates in the collision. Then the angular momentum balance can be
fulfilled in a manifold of different ways. Negative ion creation via
three body collisions leads therefore to much higher cross sections than
two body processes.  Three body collisions can involve electrons or more
massive particles as \lq third\rq\ body. The latter case usually leads
to higher cross sections for negative ion creation. This type of
negative ion formation can be fairly efficient in dense gaseous media
\cite{Smi-76}.

In a collision of two neutral atoms the more electro-negative one can
capture an electron from the other,
\begin{equation}
\text{A} + \text{B} \mapsto \text{A}^{-} + \text{B}^{+}\quad
\label{eqch321a}.
\end{equation}
In negative ion beam experiments often a positive ion beam is charge
exchanged in a vapour cell. The charge exchange is a two step process
where first the positive ion $A^{+}$ is neutralised
\begin{align}\label{eqch321b}
\text{A}^{+} + \text{B} &\mapsto \text{A} + \text{B}^{+}\quad ,\\
\intertext{and then captures another electron,}
\text{A} + \text{B} &\mapsto \text{A}^{-} + \text{B}^{+}\quad
.\label{eqch321c}
\end{align}

Charge exchange is favoured if the electron affinity of A is about equal
to the ionisation potential of B. Therefore charge exchange is most
efficient for a positive ion beam in an alkali metal vapour
\cite{Tyk-78}.  The conversion efficiency is between 0.1~\% and 10~\%.

Since charge exchange is a scattering process the negative ion beam will
have a slightly higher divergence than the initial positive ion beam.
In our apparatus, however, this effect is negligible.

%
\chapter{Photodetachment of negative ions}
\label{ch4}
Photodetachment of negative ions is the process where a photon
$\hbar\omega$ detaches an electron e$^{-}$ from a negative ion A$^{-}$:
\begin{equation}\label{eqch401}
\text{A}^{-}+\hbar\omega\mapsto\text{A}+\text{e}^{-}\quad.
\end{equation}
The photodetachment of negative ions exhibits a characteristic threshold
behaviour. By virtue of energy conservation a minimum energy of the
photon $\hbar\omega$, equal to the electron affinity, is required to
detach an electron from the negative ion. More details of this process
are discussed in section \ref{ch41}. Conservation of parity and total
angular momentum lead to the electric dipole selection rules (E1
transitions), that severely restrict the number of accessible final
states.

Photodetachment induced by an electromagnetic field can be treated as a
small time-dependent perturbation and can therefore be calculated with
Fermi's Golden Rule. The pulses delivered by our laser system (section
\ref{ch52}) are weak in this sense and therefore essentially only induce
one-photon transitions. The photodetachment cross section is analogous
to the transition probability. Under the mentioned restrictions the
photodetachment cross section $\sigma (\omega)$ (in a.u.) for an
$N$-electron ion is given by \cite{Amu-90-3}
\begin{equation}\label{eqch402}
\sigma (\omega) = \frac{4\pi^2\alpha}{\omega}\int
\left|M_{0\nu}\right|^{2} \delta(E-E_{0}-\omega)d\nu\quad,
\end{equation}
with the transition matrix element
\begin{equation}\label{eqch403}
M_{0\nu} = \sum_{q=1}^{N}\int\Psi_{0}^{\ast}(x_{1}\dots
x_{n})e^{-i\boldsymbol{k}\cdot\boldsymbol{r}}(\mathbf{e}\cdot
\mathbf{P}_{q})\Psi_{\nu}(x_{1}\dots x_{N})dx_{1}\dots x_{N}\quad,
\end{equation}
where $E-E_{0}$ is the energy difference between the initial and final
state, $\omega$ the energy of the incident light, and $\nu$ labels the
state of the outgoing electron. The vectors $\mathbf{e}$ and
$\mathbf{P}$ are the polarisation of the light and the momentum of the
electron respectively.  The selection rules for an E1 electric dipole
transition ($e^{-i\boldsymbol{k}\cdot\boldsymbol{r}}\approx 1$ in
\eqref{eqch403})
\cite{Wei-78} are:
\begin{align}\label{eqch44}
\Delta E &= \hbar\omega\quad, \\
\Delta J&=0,\pm 1 \quad\text{and}\quad J_{f}+J_{i} \ge 1 \quad,\\
\Delta M &= 0,\pm 1 \quad,\\
\pi_{f} &= -\pi_{i}\quad,
\end{align}
where the indices \textit{f} and \textit{i} stand for final and initial,
respectively. 

This photodetachment cross section is a total cross section in as much
as no further distinction is made regarding the final atom state on the
right side of \eqref{eqch401}. But it is possible to calculate partial
cross sections for photodetachment. This means that a specific state of
the residual atom, a certain orbital angular momentum of the outgoing
electron \emph{and} the coupling of these two angular momenta have to be
specified.

In the following sections we first introduce the Wigner threshold law 
(section \ref{ch41}), the various types of resonances (section
\ref{ch42}), and finally the asymptotic behaviour of the photodetachment
cross section (section \ref{ch43}). To give a rough idea about the
structure of the photodetachment cross section, we present an
analytically solvable model negative ion with realistic features, namely
the H$^{-}$ ion with the extra electron bound in a zero range
$\delta$-potential \cite{Dem-88-3}.

%
\begin{figure}
\begin{minipage}{\textwidth}
\parbox[b]{0.55\textwidth}{
\epsfig{file=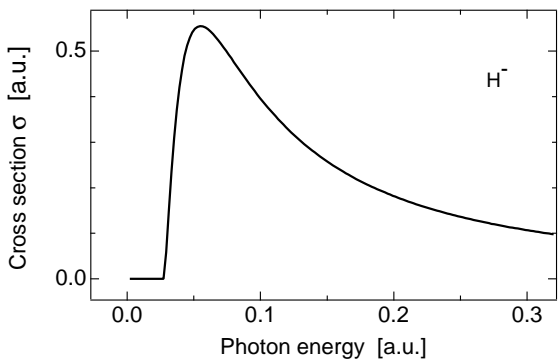, width=0.53\textwidth}
} \hfill
\parbox[b]{0.44\textwidth}{
\protect\caption[Analytic model of H$^{-}$ photodetachment cross
section]{\label{bi050}\sloppy Analytic model of H$^{-}$ photodetachment
cross section: The photodetachment cross section of an analytical model
of H$^{-}$ given by
\eqref{eqch45}. Above the threshold the cross section rises
steeply to a maximum at $E=2E_{\text{D}}$ and asymptotically falls as
$\omega^{-3/2}$ (see section \ref{ch43}). This cross section agrees
qualitatively with experimental findings.}  }
\end{minipage}
\end{figure}

To calculate the photodetachment cross section $\sigma (\omega)$ of this
model system three simplifying assumptions are necessary. First, the
bound outermost electron is assumed to occupy a hydrogen-like orbital;
second, the detached electron is approximated as a plane wave, and
finally the binding potential is modelled by an attractive zero range
potential $V=-A\delta (r)$. The solution for this model is (in a.u.)
\cite{Amu-90-4}:
\begin{equation}\label{eqch45}
\sigma (\omega )=\frac{16\pi}{3c}\sqrt{E_{\text{D}}} \frac{\sqrt{(
\omega - E_{\text{D} })^{3} } }{\omega^{3} }\quad ,
\end{equation}
with $E_{\text{D}}$ being the detachment limit.

\begin{figure}\centering
\subfigure[Photodetachment cross section of
Li$^{-}$.]{\epsfig{file=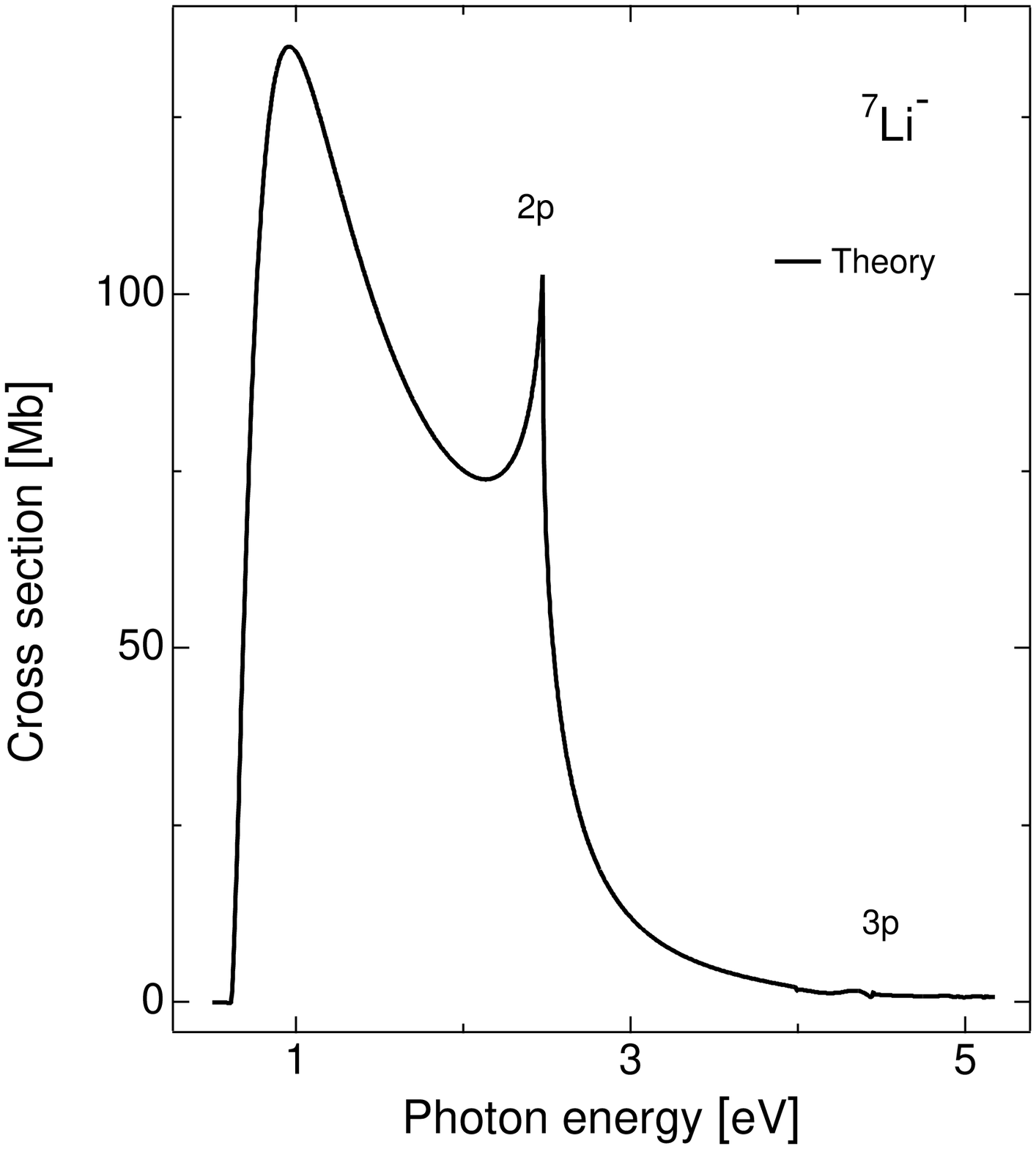, width=0.44\textwidth}}\hfill
\subfigure[Magnification at the $n$=3 thresholds]{
\epsfig{file=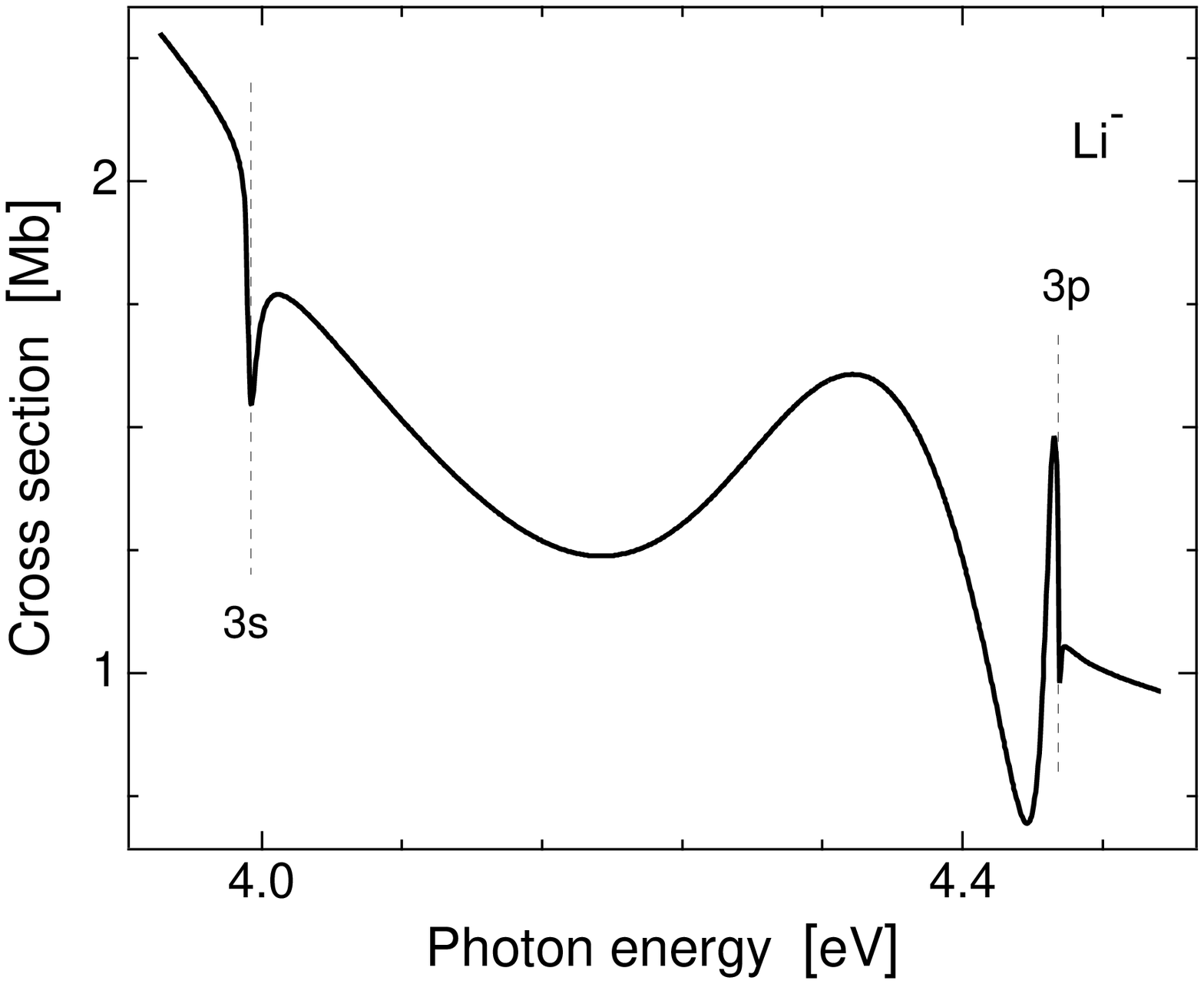, width=0.49\textwidth}}
\protect\caption[Calculated total photodetachment cross section of
Li$^{-}$]{\label{bi060}\sloppy Calculated photodetachment cross section
of Li$^{-}$ \cite{Lin-94-2}: Part (a) shows how this cross section
qualitatively resembles the analytic model H$^{-}$ cross section
\eqref{eqch45} of figure \ref{bi050}, but with additional features at
threshold of new detachment channels, resembling the cusp at the Li(2p)
threshold. Part (b) shows structures at higher thresholds. They are
further discussed in section \ref{ch631}.}
\end{figure}

This cross section, shown in figure \ref{bi050}, has most of the
characteristics of photodetachment cross sections: A steep rise of the
cross section above the threshold that reaches a maximum at $\omega =
2E_{\text{D}}$ followed by a slow decrease of the cross section towards
higher photon energies. Qualitatively this is the behaviour of the
photodetachment cross section of negative ions. In most real negative
ions there will be additional structure present. As can be seen in the
Li$^{-}$ photodetachment cross section in figure \ref{bi060}(a),
calculated by Lindroth \cite{Lin-94-2}, this model gives a good general
picture of the photodetachment cross section.
%
\section{Threshold behaviour}
\label{ch41}
The Wigner threshold law \cite{Wig-48} is a general structure in
inelastic scattering cross sections. According to this law the energy
dependence of the photodetachment cross section $\sigma (E)$ near the
threshold is given by 
\begin{equation}\label{eqch46}
\sigma (E) \sim \begin{cases}
	\left( \sqrt{E-E_{j}}\,\right)^{2l+1}  &  E\ge
	E_{j}\quad , \\
	0 & E<E_{j}\quad ,
	\end{cases}
\end{equation}
where $l$ is the lowest allowed angular momentum in the continuum
channel $j$ with a threshold energy $E_{j}$. This is valid provided the
binding potential $V(r)$ is short ranged so that
\begin{equation}\label{eqch47}
\lim_{r \rightarrow\infty} r^{2}V(r) = 0\quad .
\end{equation}
No resonance should be located at the threshold, because this leads to a
modified threshold behaviour \cite{Bae-86}.

\begin{figure}\centering
\begin{minipage}{\textwidth}
\parbox[b]{0.47\textwidth}{
\epsfig{file=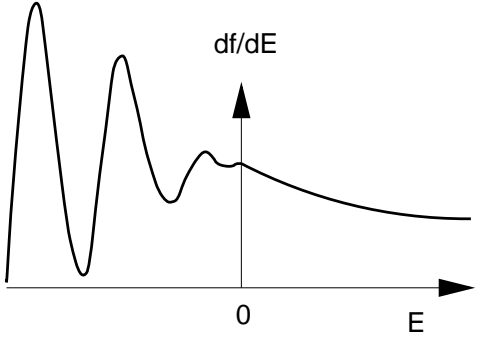, width=0.45\textwidth}
} \hfill
\parbox[b]{0.52\textwidth}{
\protect\caption[Schematic oscillator strength across a
threshold]{\label{bi075}\sloppy 
Schematic oscillator strength across a threshold: To the left of zero
the oscillator strength is depicted as measured with a finite
resolution. To the right of zero the oscillator strength density is
shown.}  }
\end{minipage}
\end{figure}

The photodetachment process, in contrast to photo-ionisation, starts
with zero cross section at the threshold. In the case of atoms or
positive ions the photo-ionisation cross section is non-zero already at
the threshold, stemming from continuity with the Rydberg series
converging to the ionisation limit, schematically depicted in figure
\ref{bi075}. The absence of Rydberg series in negative ions, by the same
token, leads to zero cross section at the threshold.

\begin{figure}
\begin{minipage}{\textwidth}
\parbox[b]{0.48\textwidth}{
\epsfig{file=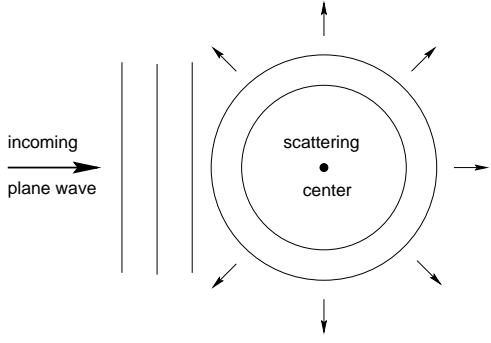, width=0.46\textwidth}
} \hfill
\parbox[b]{0.51\textwidth}{
\protect\caption[Scattering geometry]{\label{bi080}\sloppy Scattering 
geometry: An incoming plane wave from the left is scattered from a
scatterer. The scattered part of the wave is an angularly modulated
spherical wave \eqref{eqch49}. For clarity the outgoing plane wave is
omitted.} }
\end{minipage}
\end{figure}

We will now sketch the origin of this threshold law as derived from
inelastic scattering \cite{Fri-90,Joa-75-1}. Inelastic scattering
induces a transition from the initial state of the target particle, here
a negative ion, to a different final state. The normalised asymptotic
wave function is
\begin{equation}\label{eqch49}
\psi(\boldsymbol{r}) = \delta_{ji} e^{ik_{i}z} +
\frac{e^{i\boldsymbol{k}_{j}\cdot\boldsymbol{r}}}{r} f_{j,i}(\Omega
)\;,\quad r\rightarrow\infty\quad .
\end{equation}
Hence, here
\begin{equation}\label{eqch410}
k_{j} = \sqrt{\frac{2\mu (E-E_{j})}{\hbar^{2}}}\quad 
\end{equation}
is the asymptotic wavenumber of the outgoing electron in the open
channel $j$ and $\mu$ is the reduced mass of the electron. The first
term in
\eqref{eqch49} is the plane wave incoming along the z-axis 
(figure \ref{bi080}) that represents the incoming particle current, and
the second term is the outgoing angularly modulated spherical wave.
Most of the plane wave passes the scattering centre forming the so
called forward scattered part. The angular modulation $f_{j,i}(\Omega )$
of the spherical wave contains most of the information concerning
processes at the scattering centre. Most prominently, the outgoing
particle current through a specific solid angle is given by
$f_{j,i}(\Omega)$,
\begin{equation}\label{eqch413}
\frac{d\sigma_{i,j}}{d\Omega} = \frac{k_{j}}{k_{i}}
|f_{j,i}(\Omega)|^{2}\quad .
\end{equation}

The scattering amplitude $f_{j,i}(\Omega)$ is essentially the matrix
element for the transition between the final state \eqref{eqch49} and
the incoming plane wave,
\begin{equation}\label{eqch411}
f_{j,i}(\Omega) = -\frac{\mu}{2\pi\hbar^{2}}\sum_{n}\int
e^{-i\boldsymbol{k}_{j}\cdot\boldsymbol{r}^{\prime}} V_{j,n}
\psi_{n}(\boldsymbol{r})d\boldsymbol{r}^{\prime}\quad .
\end{equation}
Here the potential operator $V_{j,n}$ determines the $\Omega$ dependence
of $f_{j,i}$. To further clarify this point we expand the plane wave and
the final state in spherical harmonics. Inserting this in \eqref{eqch411}
and using the orthogonality of the spherical harmonics we obtain
\begin{multline}\label{eqch412}
f_{j,i}(\Omega)
= -\frac{\mu}{2\pi\hbar^{2}} \sum_{n}\int \Big[ 4\pi \sum_{l=0}^{\infty}
\sum_{m=-l}^{L}
(-i)^{l}\text{j}_{l}(\boldsymbol{k}_{j}\cdot\boldsymbol{r}^{\prime})
(-1)^{-m} Y_{l,m}(\Omega_{k_{j}}) \Big]\\
\times V_{j,n}(r)\alpha_{n,l^{\prime},m^{\prime}} \delta_{ll^{ \prime}}
\delta_{-mm^{\prime}}dr\quad .
\end{multline}
Herein j$_{l}$ is a Bessel function, $Y_{l,m}$ a spherical harmonic, and
$\alpha_{n,l^{\prime},m^{\prime}}$ are the expansion coefficients of the
final state.  In the vicinity of the threshold, $k$ is small and the
Bessel function j$_{l}(\boldsymbol{k}_{j}\cdot\boldsymbol{r}^{\prime})$
can be approximated as $k^{l}$ \cite{Arf-85} . Close to the threshold
all other terms in \eqref{eqch412} can be considered as constant.  To
obtain the integrated cross section for inelastic scattering we have to
square $f_{j,i}(\Omega )$ and multiply by $k_{j}/k_{i}$, as can be seen
from
\eqref{eqch413}. This leads to the Wigner law for cross sections
$\sigma_{ij}(E)$ at the threshold of inelastic scattering channels
\eqref{eqch46}. 

\begin{figure}\centering
\epsfig{file=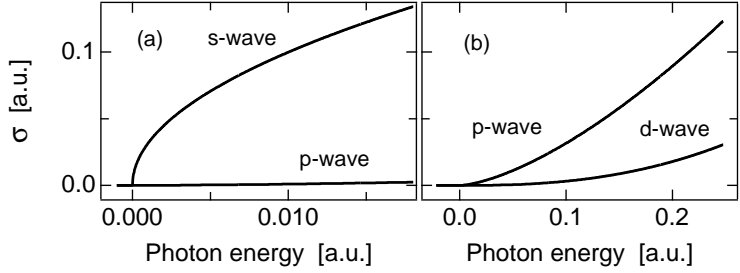, width=0.7\textwidth}
\protect\caption[Wigner threshold law]{\label{bi090}\sloppy
Wigner threshold law \eqref{eqch46}: In the left graph (\textsf{a}) the
photodetachment cross sections for a s-wave and p-wave detachment are
compared. The p-wave cross section increases relatively slowly because
it starts with zero slope at the threshold. In the right part
(\textsf{b}) the p-wave and d-wave cross sections are compared.}
\end{figure}

Close to photodetachment thresholds usually only the cases $l=0,1$, also
called s-wave and p-wave, are important. The s-wave has an onset with an
infinite slope (figure \ref{bi090}) which facilitates a precise
determination of the threshold position. All higher orbital angular
momenta lead to thresholds with zero slope at the threshold.

The sharp onset of the s-wave detachment cross section is a remarkable
feature, and one can question how sharp it can become. In this case
there is no limit due to the finite lifetime of excited states, such as
for the width of spectral lines.  Photodetachment involves a transition
between two \emph{stable} states. Both the initial and the final state
have an infinite lifetime, and consequently no width can be associated
with it. Only imperfect isolation from the environment, such as stray
electric and magnetic fields, cause a \lq broadening\rq\ of the
threshold.

To fit our data we used a function $\sigma_{\text{W}}(E)$ derived from
the Wigner law \eqref{eqch46} that includes a constant background $a$,
an amplitude $b$, the photon energy $E$, and a threshold energy $E_{0}$:
\begin{equation}\label{eqch416}
\sigma_{\text{W}}(E)= a + b \left[ \sqrt{ \frac{|E-E_{0}|}{2}+
\frac{(E-E_{0})}{2}}\thinspace\right]^{2l+1}
\quad .
\end{equation}
The two terms under the square root assure that for $E\le E_{0}$ the
radix is zero. For $E>E_{0}$ the sum of these terms gives the correct
value as in \eqref{eqch46}.
\begin{figure}\centering
\subfigure[Modified s-wave threshold]{\epsfig{file=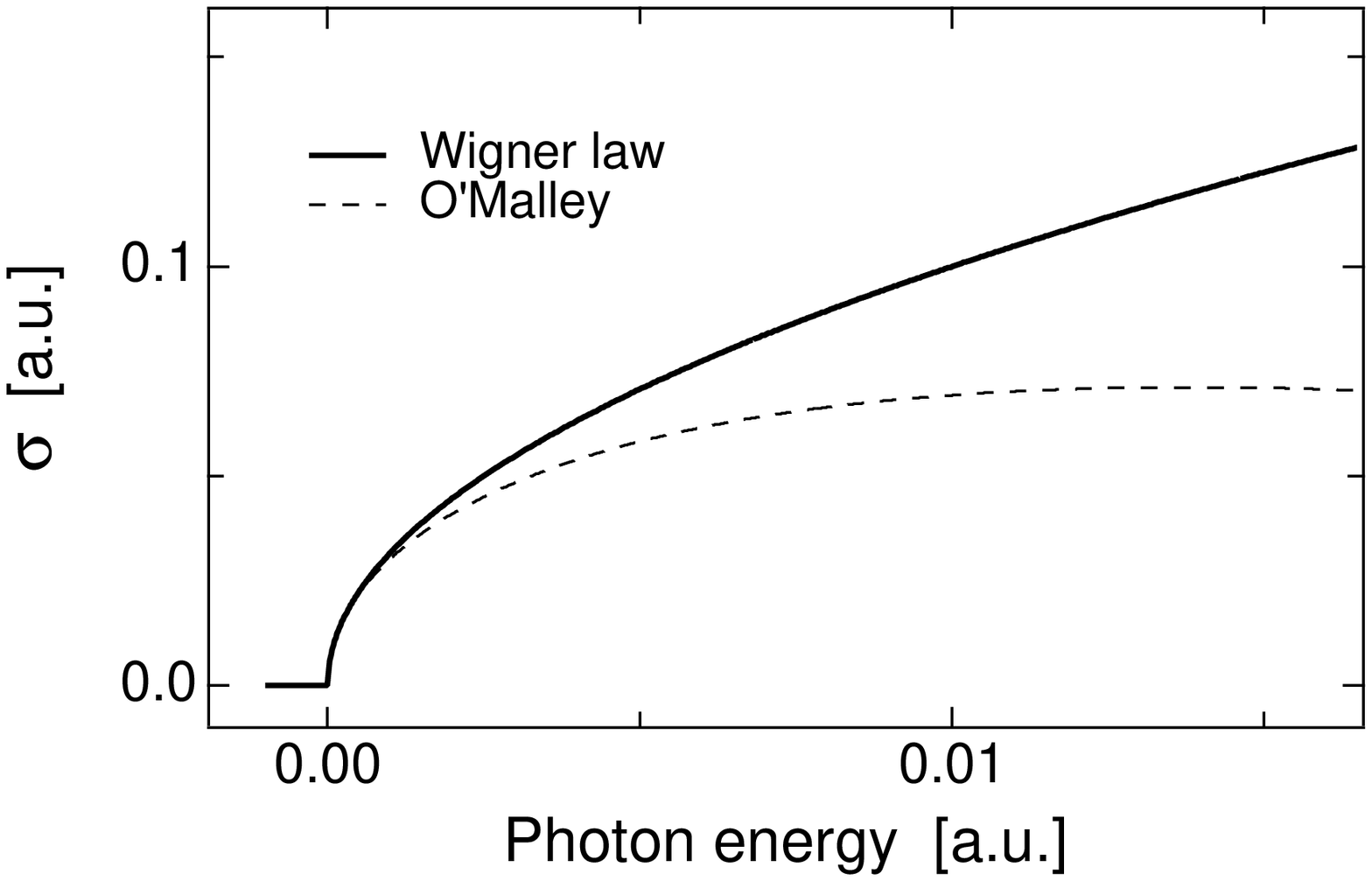,
width=0.48\textwidth}}\hfill
\subfigure[Modified p-wave threshold]{\epsfig{file=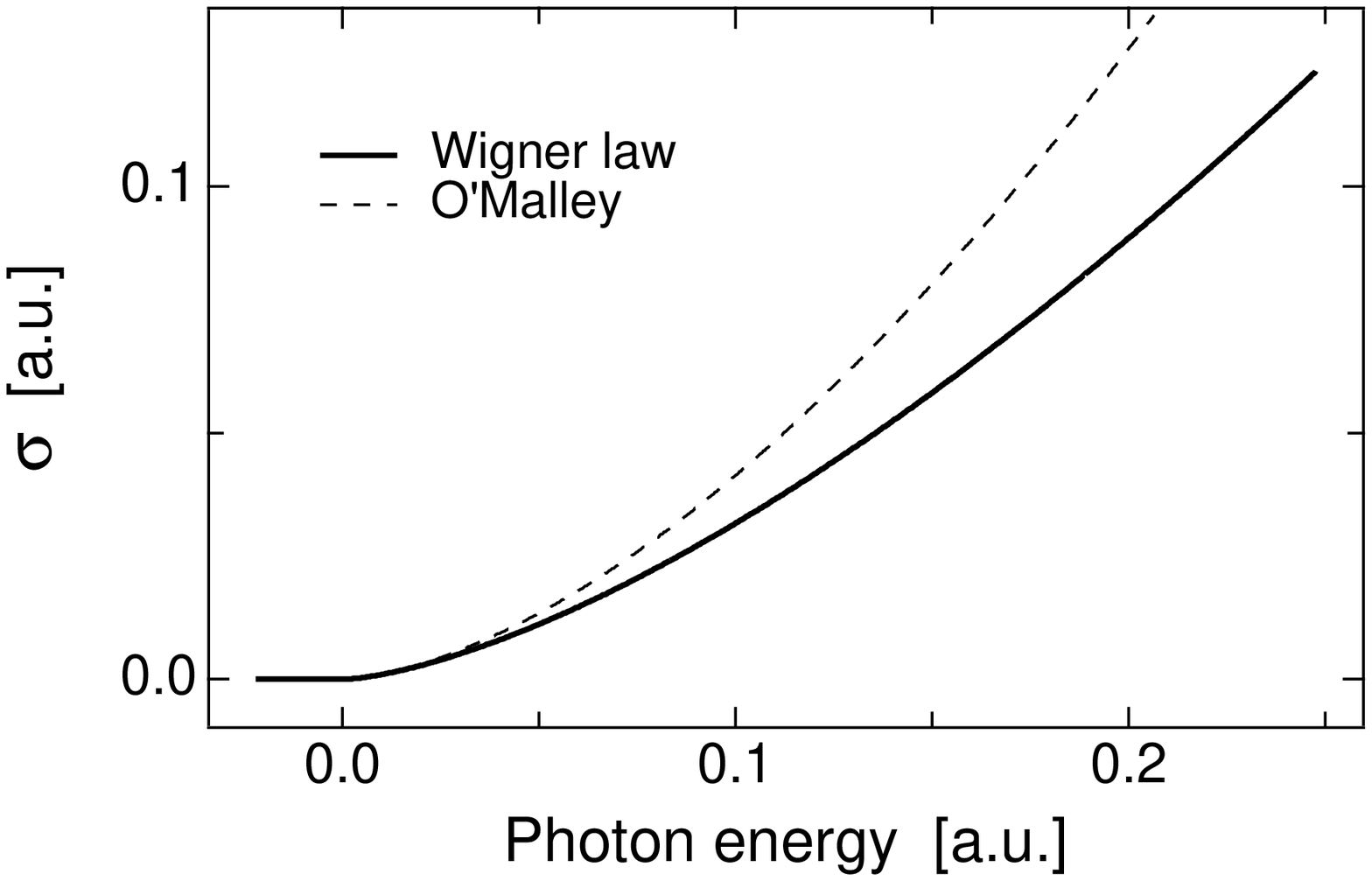,
width=0.48\textwidth}}
\protect\caption[O'Malleys modification of the Wigner
law]{\label{bi100}\sloppy O'Malleys \cite{OMa-65} modification of the
Wigner law: O'Malley has calculated a correction
\eqref{eqch418} to the
Wigner law which accounts for the interaction between the residual atom
and the outgoing electron through the polarisation potential
\eqref{eqch313} of the atom. The correction is proportional to the
dipole polarisability $\alpha_{\text{D}}$ of the residual atom, and is
here plotted for $\alpha_{\text{D}} =10$.}
\end{figure}

A simple and general law such as the Wigner law is likely to have
limited range of validity above the threshold. One restriction is due to
the interaction of the outgoing electron with the residual atom through
the polarisation potential \eqref{eqch313}. The modification of the
Wigner law, depicted in figure \ref{bi100}, caused by this interaction
has been calculated by O'Malley \cite{OMa-65,Hot-73,Hot-73-2} to be
\begin{align}\label{eqch418}
\sigma_{\text{OM}}(E) &= k^{2l+1}\Big[ 1 - \frac{4\alpha_{\text{D}}
k^{2}\ln k}{(2l+3)(2l+1)(2l-1)} + O(k^{2}) \Big]\quad ,\\
\intertext{with}
k &= \sqrt{2E}\quad .
\end{align}

Farley \cite{Far-89} has proposed an analytic model (called zero core
contribution model) to calculate the photodetachment cross section near
the threshold, given the core radius and the electron affinity.  With
these two numbers as input he succeeds in correctly calculating the
photodetachment cross section higher above the threshold than it is
possible with the O'Malley \cite{OMa-65} form \eqref{eqch418}. The
proposed model contains fairly easily accessible parameters and could
therefore become an option for the description of measured
photodetachment cross section in the vicinity of thresholds.

\section{Resonance structure}
\label{ch42}
Apart from the overall smooth variation of the photodetachment cross
section there occur sharper structures. Some of them are resonances due
to autodetaching states of the negative ion, others are Wigner cusps
located at the threshold of s-wave detachment channels. Among the
resonances it is possible to distinguish Feshbach resonances, due to
negative ion states below the atomic parent state, and shape resonances,
due to negative ion states bound by a centrifugal barrier above the
atomic parent state.

In the simplest analysis the resonance states are assumed to be isolated
i.e. well separated in energy from the threshold region and other
resonance states.  In some cases resonances close to thresholds can be
described by modified threshold laws \cite{Bae-86,Pet-85,Wal-94-4}.  The
simple case of one state being embedded in one continuum leads to the
well known Beutler-Fano \cite{Fan-61} profile in the total
photodetachment cross section.

Here we will first discuss Feshbach resonances, then briefly a
situation with two levels embedded in one continuum and thereafter
discuss shape resonances and finally introduce Wigner cusps.

\begin{figure}
\begin{minipage}{\textwidth}
\parbox[b]{0.24\textwidth}{
\epsfig{file=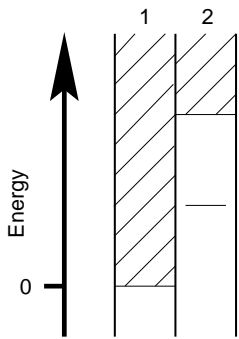, width=0.22\textwidth}
} \hfill
\parbox[b]{0.75\textwidth}{
\protect\caption[Two channel level scheme]{\label{bi120}\sloppy
\small Two
channel level scheme: The discrete level in the closed channel 2 would
be stable without the interaction with the open channel 1. Due to this
interaction the discrete level may be shifted and broadened. This leads
to a continuum structure called Beutler-Fano profile (figures
\ref{bi130} and \eqref{eqch419a}).}
}
\end{minipage}
\end{figure}

\begin{figure}\centering
\subfigure[Asymmetric Fano
profiles]{\epsfig{file=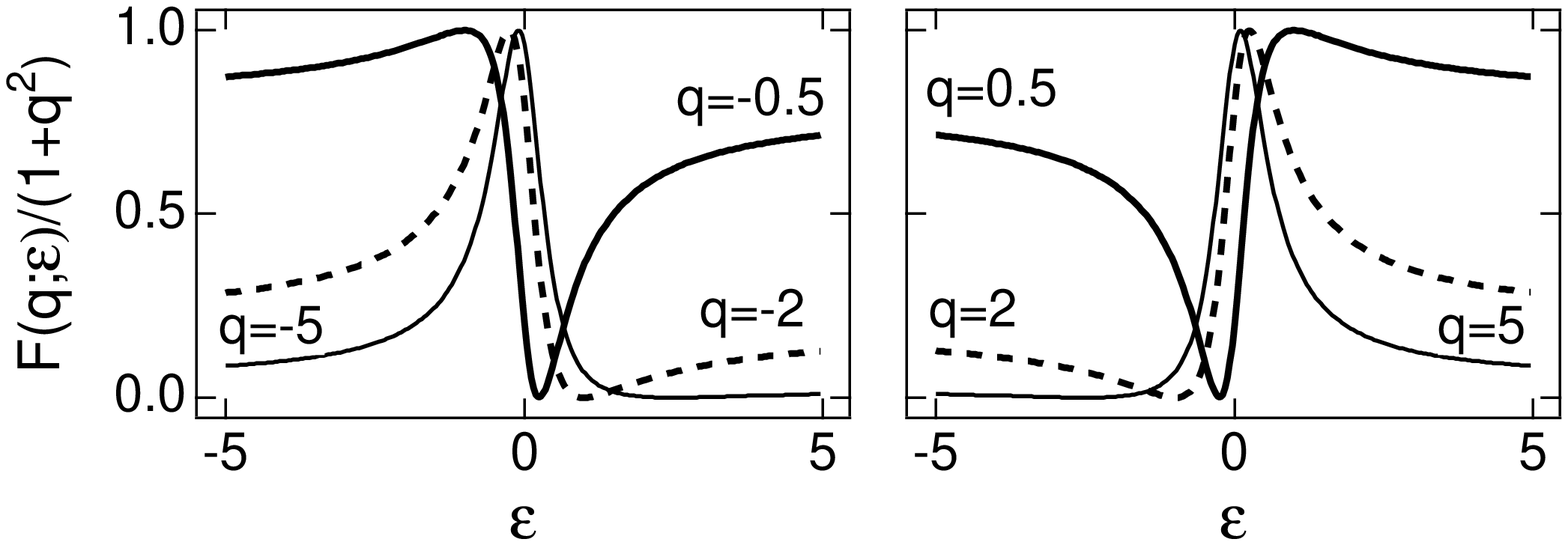, width=0.57\textwidth}}\hfill
\subfigure[Symmetric Fano
profiles]{\epsfig{file=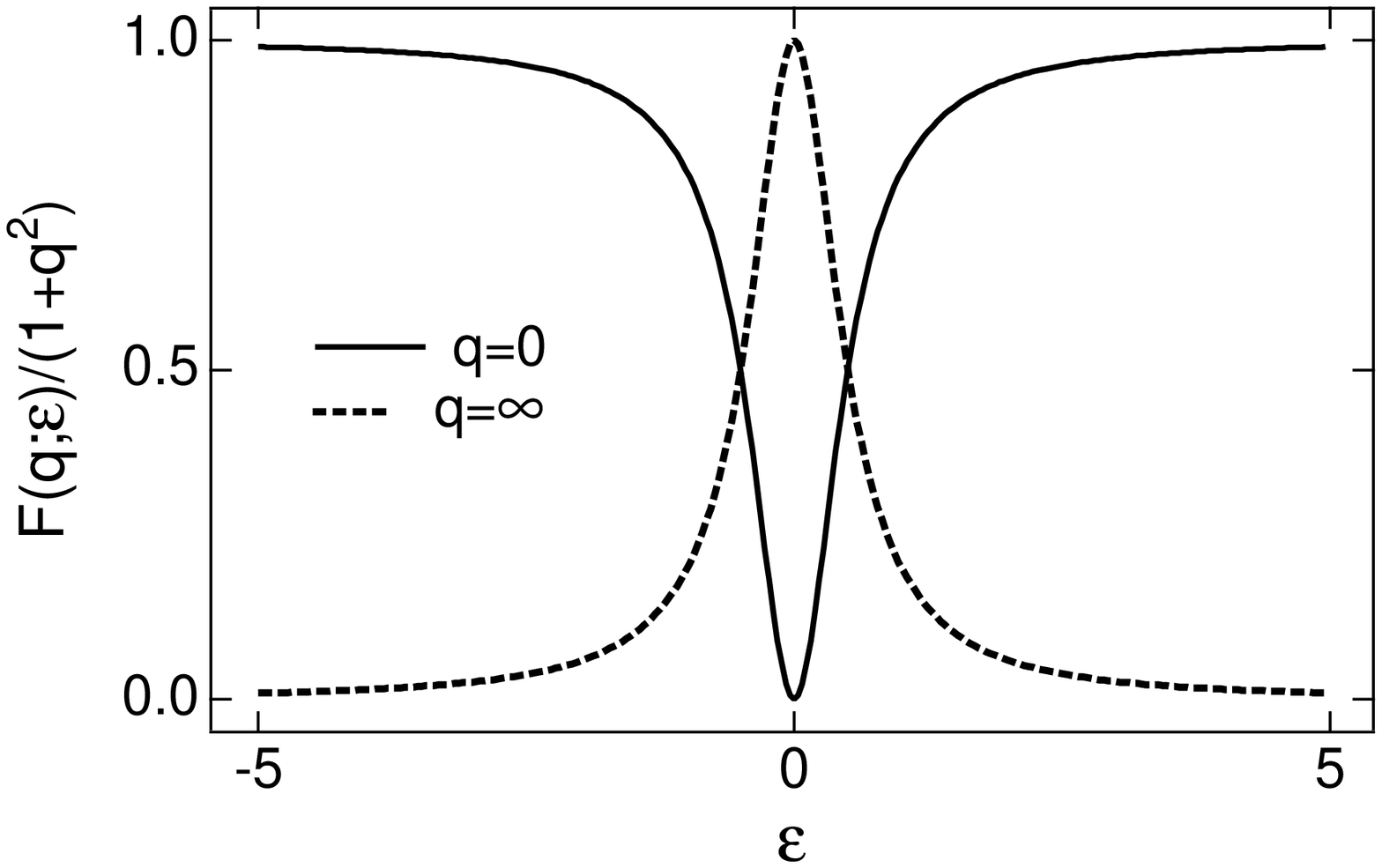, width=0.41\textwidth}}
\protect\caption[Beutler-Fano profile]{\label{bi130}\sloppy
Beutler-Fano profile: Normalised Beutler-Fano profiles
\eqref{eqch420} for various $q$ values. The curves for $q$ and $-q$
are mirror images of each other, as shown in part (a) of the figure, and
the profile becomes a symmetric Lorentzian line in the limit of $q=0$
and $q=\infty$, as can be seen in part (b) of the figure.}
\end{figure}

\minisec{The Beutler-Fano profile}
A Feshbach resonance is due to an isolated state of a negative ion
embedded in one continuum as schematically shown in figure
\ref{bi120}. The isolated state in channel \textsf{2} would be stable
without the continuum channel \textsf{1} to which it can decay by
autodetachment. This model was first investigated by Fano
\cite{Fan-61}. The structure in the photodetachment cross section owing
to this kind of interaction is shown in figure \ref{bi130} and called
a Beutler-Fano profile.

In the vicinity of an autodetaching state of the negative ion the total
photodetachment cross section is given by
\begin{align}
\sigma_{\text{BF}}(E) &= a + b F(q;\epsilon )\label{eqch419a}\quad \\
\intertext{with}
F(q;\epsilon ) &= \frac{(q+\epsilon )^{2}}{1+
\epsilon^{2}}\label{eqch419b}\quad \\
\intertext{where}
\epsilon &= \frac{E-E_{0}}{\Gamma /2}\quad .\label{eqch419c}
\end{align}
$E_{0}$ is the resonance energy, $\Gamma$ the width, $q$ the shape
parameter, $E$ the photon energy, $a$ a background and $b$ the amplitude
of the resonance structure.

This formula \eqref{eqch419a} only strictly applies for an isolated
state embedded in one continuum. An isolated state embedded in several
continua can be described by \eqref{eqch419a} if all but one continuum
contribute only to a non-resonant background, accounted for by the
parameter $a$ in \eqref{eqch419a}. The Beutler-Fano model is suited only
to describe total cross sections.

The Beutler-Fano profile is also used with a different parametrisation
involving a correlation parameter $\rho^{2}$ with a value between 0 and
1,
\begin{equation}\label{eqch420}
\sigma_{\text{BF}}(E) = \sigma_{0} \big[ (1-\rho^{2}) + \rho^{2}F(q;\epsilon
)\big]\quad ,
\end{equation}
with $\sigma_{0}$ being the background cross section. The Fano
parameters $\rho$ and $q$ can be expressed in terms of transition matrix
elements as, for instance, in \cite{Fer-87}.

Another, parametrisation has been suggested by Shore \cite{Sho-68},
\begin{equation}\label{eqch421}
\sigma_{\text{Sh}}(E) = \sigma_{0} + \frac{A\epsilon + B}{1+ 
\epsilon^{2}} \quad .
\end{equation}
The parameters $A$ and $B$ are called the Shore parameters and
$\sigma_{0}$ is a constant accounting for a non-resonant background.

The unsatisfactory situation of only having a description for resonance
structure in the total cross section was resolved by Starace
\cite{Sta-77}. He addressed a model with one level coupled to two
continua. The cross section $\sigma_{\text{SF}}$ then is,
\begin{multline}\label{eqch422}
\sigma_{\text{SF}}(E) = \frac{\sigma_{0}}{1+\epsilon^{2}}\bigg\{
\epsilon^{2}
+ 2\big[ q\Re (\alpha_{\mu}) - \Im (\alpha_{\mu})\big]\epsilon + \\
\big[ 1 - 2q\Im (\alpha_{\mu}) - 
2\Re (\alpha_{\mu}) + (q^{2} +
1)|\alpha_{\mu}|^{2} \big]\bigg\}\quad ,
\end{multline}
in which $\sigma_{0}, \epsilon$ and $q$ retain their meaning as in the
Beutler-Fano model \eqref{eqch419b} and the new complex parameter
$\alpha_{\mu}$ describes the branching into the different channels.

The appearance of all these parametrisations suggests a generalisation,
given for example in \cite{Fer-87}, to a Lorentz line times a quadratic
polynomial in $\epsilon$,
\begin{equation}\label{eqch423}
\sigma_{\text{P}}(E) = \sigma_{0}\frac{C_{1} + C_{2}\epsilon +
\epsilon^{2}}{1+\epsilon^{2}}\quad .
\end{equation}
The aforementioned parametrisations are all special cases of this one
with
\begin{xalignat}{2}
C_{1} &= 1 + \frac{B}{\sigma_{0}}\; , & B &= \sigma_{0} (C_{1} -1 )
\quad ,\label{eqch427} \\
C_{2} &= \frac{A}{\sigma_{0}}\; , & A &= \sigma_{0} C_{2}\quad ,
\label{eqch428}\\
\intertext{for the Shore form \eqref{eqch421}, and}
C_{1} &=
\begin{gathered}[t]
1-2q\Im (\alpha_{\mu}) - 2\Re (\alpha_{\mu})\\ \quad{}+
(q^{2}+1)|\alpha_{\mu}|^{2}\quad ,
\end{gathered}
& & \label{eqch429}\\
C_{2} &= 2q\Re (\alpha_{\mu}) - 2 \Im (\alpha_{\mu})\quad ,& &
\label{eqch430}\\
\intertext{for the Starace-Fano form \eqref{eqch422}, and}
C_{1} &= \rho^{2}q^{2} + 1 - \rho^{2}\; , & q_{1,2} &= 
\frac{C_{1} -1}{C_{2}}\pm\sqrt{ \frac{(1-C_{1})^{2}}{C_{2}^{2}} - 1
}\quad ,\label{eqch431} \\
C_{2} &= 2\rho^{2}q \; , & \rho^{2} &= \frac{C_{2}}{2q}\quad ,
\label{eqch432} 
\end{xalignat}
for the Beutler-Fano from \eqref{eqch420}. Hence, the Beutler-Fano and
the Shore parametrisation are \emph{not} equivalent and the latter form
has been used in this work since it applies to resonance structure in
partial as well as total cross sections. From a fit of the generalised
parametrisation \eqref{eqch423} the parameters of the Beutler-Fano
\eqref{eqch420} and Shore form \eqref{eqch421} can be obtained from  
\eqref{eqch431}\eqref{eqch432} and \eqref{eqch427}\eqref{eqch428}
respectively.

A resonance that properly has to be described with the theory of Starace
\cite{Sta-77} can also be fitted with the generalised form
\eqref{eqch423},  as
\eqref{eqch429} and \eqref{eqch430} reveal, but from the obtained
parameters $\sigma_{0}$, $C_{1}$, $C_{2}$, $E_{0}$, and $\Gamma$, it is
not possible to extract $\alpha_{\mu}$, thus these two forms are not
equivalent. They refer to different physical systems, but nonetheless
the width $\Gamma$ and the energy $E_{0}$ are properties of the
underlying negative ion state, and consequently \emph{not} depending on
the parametrisation.

\begin{figure}\centering
\subfigure[Three channels]{\epsfig{file=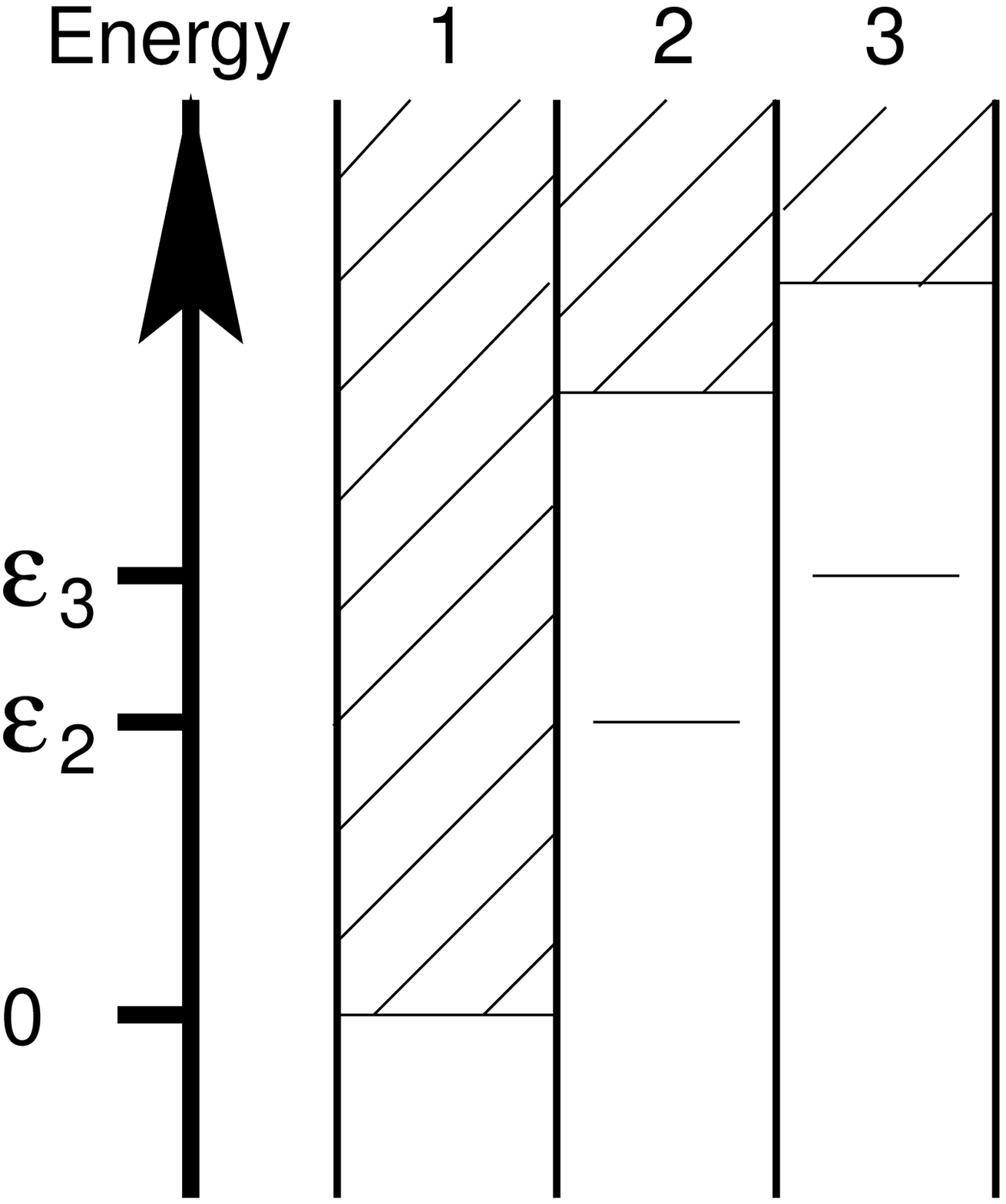,
width=0.35\textwidth}}\hfill
\subfigure[Photodetachment cross section]{\epsfig{file=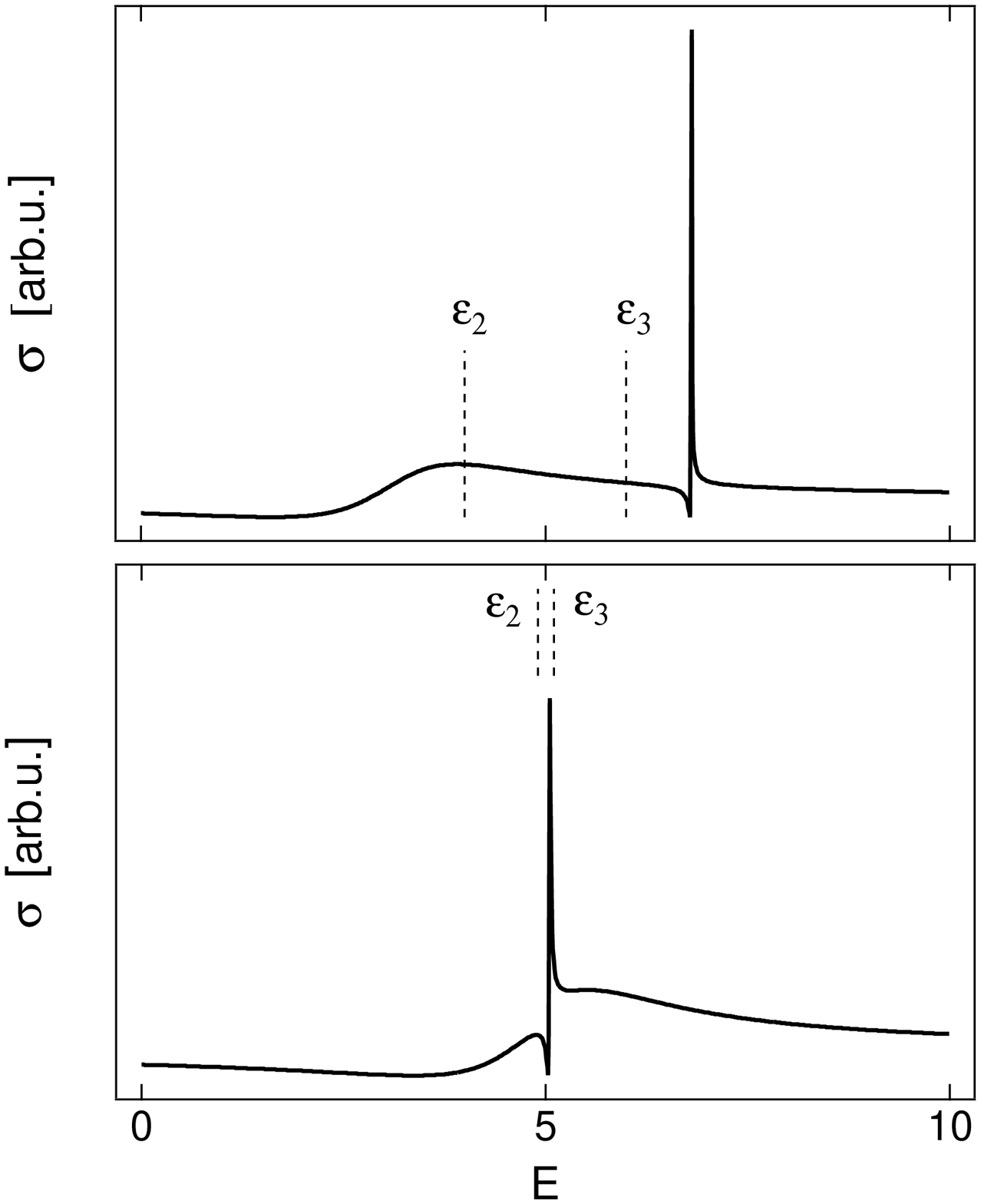,
width=0.48\textwidth}}
\protect\caption[Three channel photodetachment
model]{\label{bi150}\sloppy Three channel photodetachment model: The two
closed channels 2,3 interact with the one open channel 1, as depicted in
part (a). This leads to similar effects as for two channels but with a
richer structure. In the upper part of (b) the two levels are well
separated ($\epsilon_{2}$=4,$\epsilon_{3}$=6) and give rise to a broad
and very narrow resonance. In the lower part of (b) the two levels are
close together ($\epsilon_{2}$=4.9, $\epsilon_{3}$=5.1), resulting in a
narrow Beutler-Fano profile cutting into a broader one.}
\end{figure}

More complicated systems with more than one level embedded in one
continuum lead to a much richer and complex continuum structure. To
illustrate this we present here a situation with two closed channels
(number \textsf{2,3} in figure \ref{bi150}(a)) interacting with an open
one, number
\textsf{1}.  

This model is treated in \cite{Fri-90-1}, and here we only present the
results, in graphical form.  The graphs in figure \ref{bi150} should be
understood as follows: the positions of the discrete states in part (b)
are \emph{nominal} positions disregarding the interaction with the
continuum. These nominal positions are also indicated by the vertical
dashed lines in part (b). The coupling of these levels to the continuum
leads to new eigenstates and eigenenergies of the extended system. The
cross section of this coupled system is given by the solid line in the
figures of part (b).  Two different situations are shown: one where the
two levels are well separated, and a second below where the two levels
are close together. The first situation leads to a broad and narrow
resonance structure, both of them well away from their nominal position
without the coupling to the continuum. The two resonances share their
width, or in other words, the sum of their widths is equal to the sum of
the individual level width, but here one resonance carries almost all
width while the other is very narrow. The second situation gives rise to
one narrow Beutler-Fano profile that cuts into another broader one.

%

\begin{figure}\centering
\epsfig{file=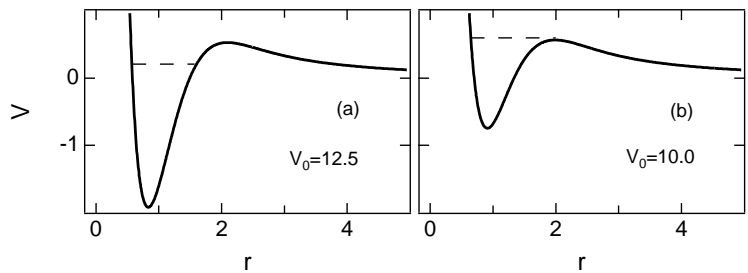, width=0.7\textwidth}
\protect\caption[Shape resonance model]{\label{bi170}\sloppy Shape
resonance model: In a binding potential with a repulsive centrifugal
barrier \eqref{eqch425} states with positive energy can exist, but they
decay by tunnelling through the barrier. The width $\Gamma$ of a shape
resonance increases as the energy $E$ of the state. In the left part
(\textsf{a}) the energy $E=0.21$ and $\Gamma=0.06$, whereas in the right
part (\textsf{b}) $E=0.6$ and $\Gamma\approx 0.5$ \cite{Fri-90-2}.}
\end{figure}

\minisec{Shape resonances}
This type of resonances can occur in attractive potentials with
a centrifugal  barrier. The attractive asymptotic polarisation potential
$V_{\text{pol}}$ \eqref{eqch313} with a repulsive centrifugal
term has a maximum $V_{\text{max}}$ \cite{Gre-95-2},
\begin{equation}\label{eqch424}
V_{\text{max}} = \frac{(l+1/2)^{4}}{8\alpha_{\text{D}} }\quad ,
\end{equation}
in atomic units, $\alpha_{\text{D}}$ denoting the atomic dipole
polarisability.  Any shape resonance must be situated below
$V_{\text{max}}$. The width of shape resonances is usually about equal
to the excess energy above the parent state. This is confirmed by a
model calculation \cite{Fri-90-2} for a potential $V_{\text{Shape}}$ (in
a.u.),
\begin{equation}\label{eqch425}
V_{\text{Shape}} = - V_{0} e^{-r^{2}} + \frac{l(l+1)}{2r^{2}}\quad .
\end{equation}
Width and energy of a shape resonance in this potential for two
different values of $V_{0}$ are shown in figure \ref{bi170}. 

%

\begin{figure}
\begin{minipage}{\textwidth}
\parbox[b]{0.42\textwidth}{
\epsfig{file=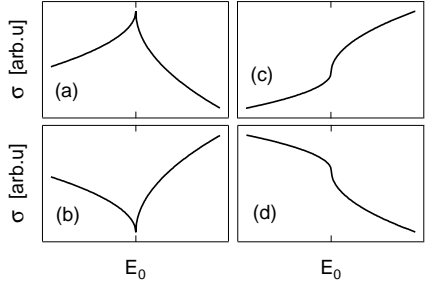, width=0.4\textwidth}
} \hfill
\parbox[b]{0.57\textwidth}{
\protect\caption[Wigner cusp]{\label{bi180}\sloppy
Wigner cusp: The four possible shapes of cusps \eqref{eqch426} 
that occur at the threshold of a new $\epsilon $s continuum. The
upward one (\textsf{a}) has been observed in \cite{Bae-85} at the Li(2p)
threshold in figure
\ref{bi060}(a), and a downward one (\textsf{b}) at the
Li(3p) threshold shown in figure \ref{bi060}(b).}  }
\end{minipage}
\end{figure}

\minisec{Wigner cusps}
A Wigner cusp is not a resonance but a discontinuity in the scattering
cross section at the threshold of an s-wave continuum. They can appear
in four different variants, all shown in figure \ref{bi180}, with the
common feature of an infinite slope at the threshold energy $E_{0}$. The
cross section $\sigma_{\text{cu}}$ near a Wigner cusp can be
parametrised as
\begin{equation}\label{eqch426}
\sigma_{\text{cu}} = 1\pm G\sqrt{ \frac{E_{0}-E}{2}
+ \frac{|E_{0}-E|}{2} }\pm H\sqrt{\frac{E-E_{0}}{2} +
\frac{|E-E_{0}|}{2} }\quad .
\end{equation}
$E$ is the photon energy, $E_{0}$ the threshold energy and $G$ and $H$
are constants describing the amplitude on either side of the cusp. The
cusps in figure \ref{bi180} have $G=0.57$ and $H=1$.
 
%

\section{Asymptotic behaviour}
\label{ch43}
The asymptotic decrease of the photodetachment cross section is due to
dwindling overlap of the initial and final state wave function. The
higher the excess energy of the detached electron is, the more rapid its
wave function oscillates. Since the initial state is fixed, the overlap
is given by the product of this function with the ever more rapidly
oscillating function of the outgoing electron.

\begin{figure}
\begin{minipage}{\textwidth}
\parbox[b]{0.39\textwidth}{
\epsfig{file=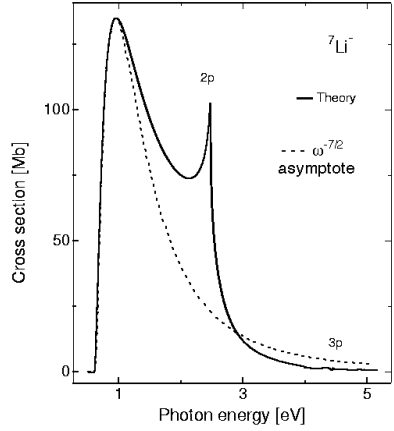, width=0.37\textwidth}
} \hfill
\parbox[b]{0.605\textwidth}{
\protect\caption[Asymptotic photodetachment cross section]{\label{bi190}
\sloppy Asymptotic photodetachment cross section: The Li$^{-}$
photodetachment cross section, calculated by Lindroth \cite{Lin-94-2} is
compared to a $\omega^{-7/2}$ asymptote. } }
\end{minipage}
\end{figure}

The model cross section for H$^{-}$ \eqref{eqch45} drops like
$\omega^{-3/2}$ for large $\omega$. For negative ions the asymptotic
fall off should theoretically be between $\omega^{-1/2}$ and
$\omega^{-7/2}$
\cite{Amu-90-5}. In figure \ref{bi190} we compare the calculated 
Li$^{-}$ photodetachment cross section \cite{Lin-94-2} with a modified
\eqref{eqch45} that matches the top and falls off as $\omega^{-7/2}$. In
the case of He$^{-}$ the decay exponent is below the predicted lower
limit of $-1/2$.

%
\chapter{Experimental apparatus}
\label{ch5}
\begin{figure}
\begin{minipage}{\textwidth}
\parbox[b]{0.41\textwidth}{
\epsfig{file=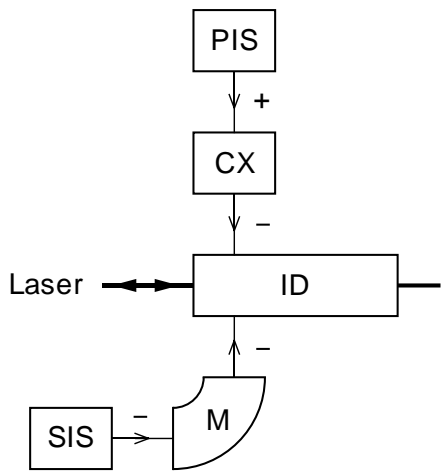, width=0.41\textwidth}
} \hfill
\parbox[b]{0.545\textwidth}{
\protect\caption[Ion beam machine]{\label{bi210}\sloppy
The ion beam machine: \textsf{PIS}, {plasma ion source}; \textsf{CX},
charge exchange chamber; \textsf{ID}, interaction-detection chamber;
\textsf{M}, sector magnet; \textsf{SIS}, sputter ion source. Velocity
filtered positive ions from the plasma ion source (figure \ref{bi240})
are charge exchanged in the charge exchange chamber (figure
\ref{bi260}). The negative ions then enter the interaction-detection
chamber (figure \ref{bi360} page
\pageref{bi360}). Negative ions from the sputter ion source (figure
\ref{bi230}) are mass selected by the sector magnet and then directed
into the interaction-detection chamber.} }
\end{minipage}
\end{figure}

The apparatus used in this work, shown schematically in figure
\ref{bi210}, consists of two ion sources with independent acceleration
stages and a common collinear interaction-detection chamber. Beam
handling and the two sources are described in section \ref{ch51}. The
laser system will be the subject of section \ref{ch52} and the detection
system is treated in section \ref{ch53}. In section \ref{ch61} we will
discuss the data acquisition and processing.

The two different ion sources allow us to produce almost any negative
ion. From both sources the negative ion beam is guided into the
interaction-detection chamber where the ion beam is either collinearly
merged with the laser beam or perpendicularly intersected by the
light. This flexibility makes this apparatus a versatile tool for
negative ion investigations.

In the collinear geometry laser and ion beams are merged over a 0.5~m
long path defined by two 3~mm apertures. This gives a large interaction
volume, leading to a high sensitivity.  The velocity compression
\cite{Kau-76} in a fast ion beam reduces the Doppler broadening and
thereby enhances the precision. By combining energy positions from
measurements with co- and counter-propagating laser and ion beams one
can eliminate the Doppler shift to \emph{all} orders.  In the collinear
geometry, just as in a crossed beam arrangement, it is possible to
let the ions interact with more than one laser beam, a possibility that
we exploited in the measurements involving state selective detection
through resonance ionisation presented in chapter \ref{ch622}.  

For laser photo-electron spectroscopy it is required to let the laser
and ion beams intersect perpendicularly.  This beam geometry is also
suitable when high light intensities are necessary to induce nonlinear
effects such as multi-photon absorption. In the experiments presented in
this thesis none of these options has been used.
\section{The negative ion beam apparatus}
\label{ch51}
Our negative ion beam machine has two legs (figure \ref{bi210}), one
starting with a sputter ion source (\textsf{SIS}) and the other one
beginning with a plasma ion source (\textsf{PIS}). From the sputter ion
source negative ions are extracted directly, whereas from the plasma
source positive ions are extracted and charge exchanged to form a
negative ion beam.

\begin{figure}
\begin{minipage}{\textwidth}
\parbox[b]{0.48\textwidth}{
\epsfig{file=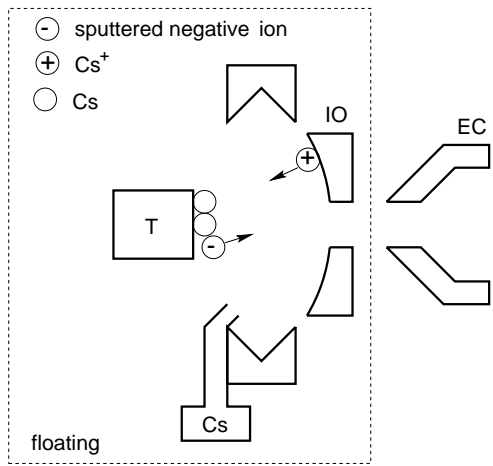, width=0.46\textwidth}
} \hfill
\parbox[b]{0.51\textwidth}{
\protect\caption[Sputter ion source]{\label{bi230}\sloppy\small
Sputter ion
source: \textsf{T}, sputter target; \textsf{Cs}, cesium reservoir;
\textsf{IO}, spherical or helical ioniser; \textsf{EC}, electrostatic
extractor. Cesium atoms from the reservoir are ionised on the ioniser
and accelerated towards the target on negative potential. On the target
Cs$^{+}$ ions are neutralised and form a layer on the target. When
sputtered atoms pass this layer they pick up an electron. In the field
between the target and the ioniser the negative ions are accelerated
towards the extractor on ground potential and leave the source
\cite{Ros-95}.}}
\end{minipage}
\end{figure}

In the sputter ion source negative ions are produced by accelerating
positive cesium ions from the ioniser, shown in figure
\ref{bi230}, towards the sputter target. The target consists of a
heat conducting material that contains the atoms of which we strive to
form negative ions. On the targets surface a thin layer of cesium is
formed. When target atoms are sputtered they have to penetrate this
cesium layer, and upon doing so, an electron can be attached. The
negative ion is accelerated towards the extractor and leaves the
source. This source can produce micro ampere currents
\cite{Mid-89} of negative ions from metals with high electron
affinities. Metals with lower electron affinities are usually less
prolific.

\begin{figure}\centering\hfill
\subfigure[Ion optics in the plasma ion beam
line]{\epsfig{file=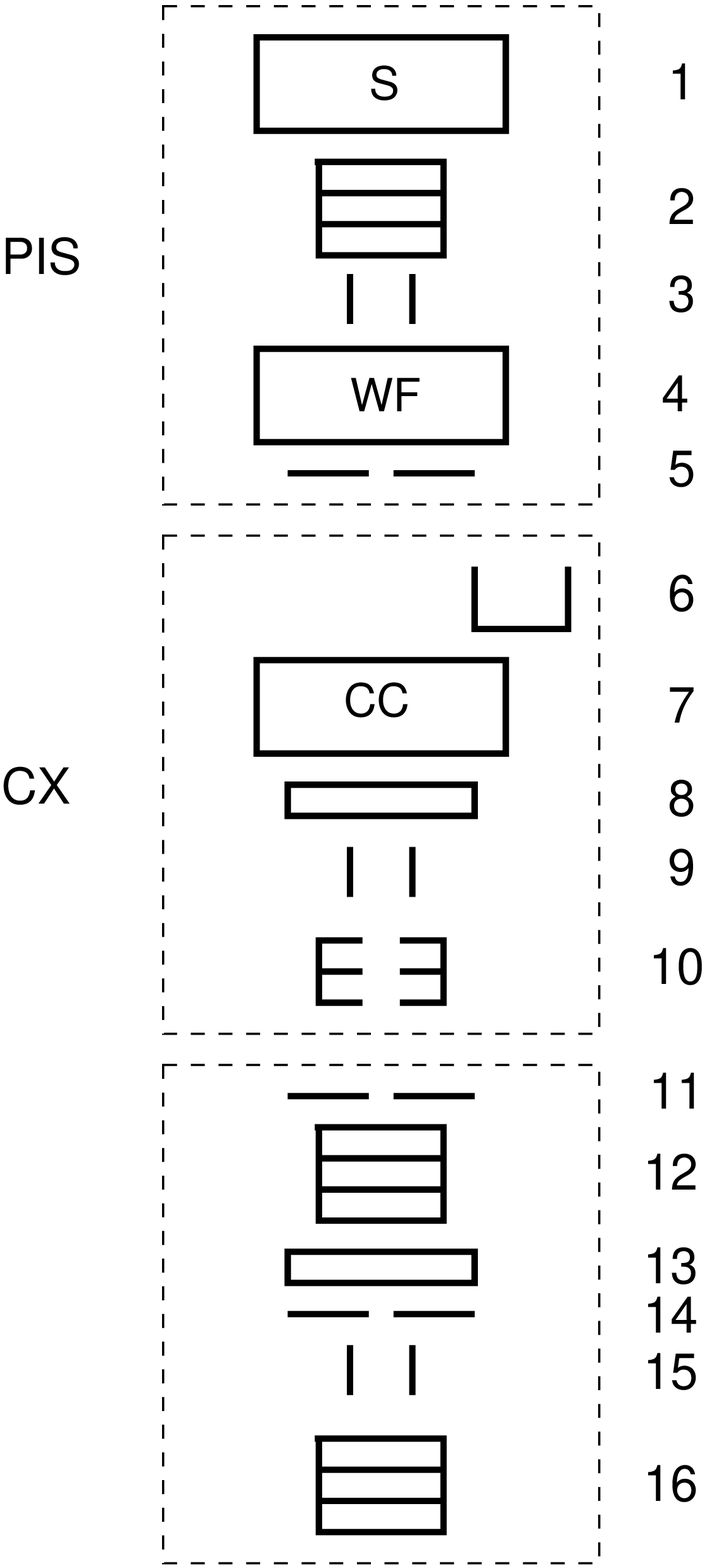, width=0.35\textwidth}}\hfill
\subfigure[Ion optics in the sputter ion beam
line]{\epsfig{file=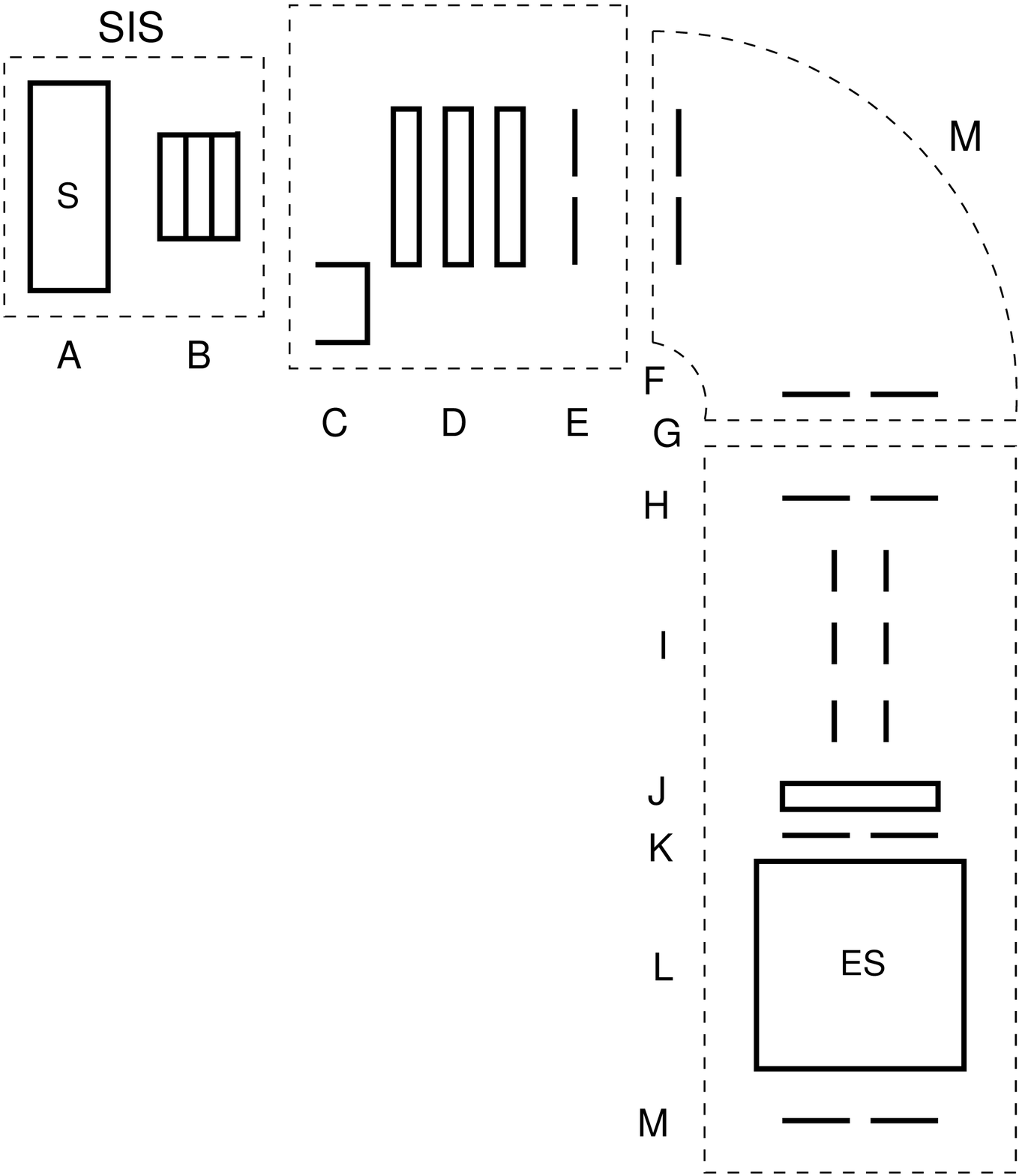, width=0.63\textwidth}\hfill}
\protect\caption[Ion optics in the beam
lines]{\label{bi270}\sloppy Ion optics in the beam lines: In
part (a) there are shown \textsf{1}, plasma ion source;
\textsf{2,12,16}, einzellense,
\textsf{3,9,15}, horizontal steering plate; \textsf{4}, Wien filter;
\textsf{5,11,14}, aperture; \textsf{6}, movable
Faraday cup; \text{7}, charge exchange cell; \textsf{8,13}, vertical
steering plate; \textsf{10}, triple aperture lense after
\cite{Har-76}. In part (b) there are shown \textsf{A}, sputter ion
source; \textsf{B}, einzellens; \textsf{C}, movable Faraday cup;
\textsf{D,I}, cylinder lens; \textsf{E,F,G,H,K,M}, aperture;
\textsf{J}, deflection plate; \textsf{L}, electron spectrometer. The
compression lens consists of three pairs of plates, the two outer pairs
are on ground potential and the middle pair is fed with one voltage per
plate to both compress and steer the beam. The electron spectrometer
\textsf{ES} will in the near future be moved to place indicated in the
figure.}
\end{figure}

The commercial sputter ion source in our laboratory delivers a beam of
about 1~mm diameter with a divergence of approximately 40~mrad. With
some ion optics (figure \ref{bi270}(b)) it is possible to form a
reasonably collimated beam in the interaction region.

\begin{figure}
\begin{minipage}{\textwidth}
\parbox[b]{0.45\textwidth}{
\epsfig{file=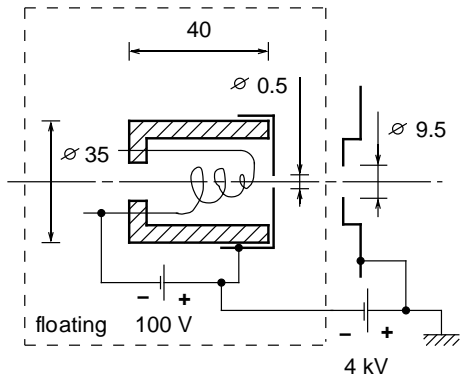, width=0.43\textwidth}
} \hfill
\parbox[b]{0.54\textwidth}{
\protect\caption[Plasma ion source]{\label{bi240}\sloppy\small
Plasma ion source: The body of the source is machined in BN$_{3}$
ceramics. From the discharge between the filament and the anode cape
positive ions are extracted through the hole in the anode. Through the
larger hole from the left a charge holder with a compound containing the
desired atoms can be inserted. The approximate dimensions are given in
mm.}  }
\end{minipage}
\end{figure}

\begin{figure}
\begin{minipage}{\textwidth}
\parbox[b]{0.59\textwidth}{
\epsfig{file=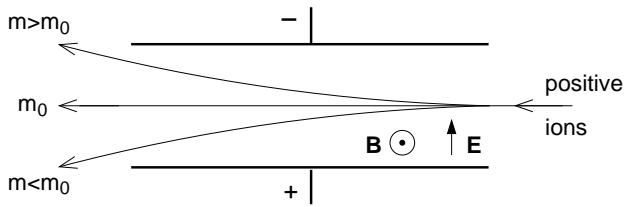, width=0.59\textwidth}
} \hfill
\parbox[b]{0.39\textwidth}{
\protect\caption[Wien filter]{\label{bi250}\sloppy\small
Wien filter: For singly charged positive ions
of mass $m_{0}$ the speed dependent Lorentz force is exactly
balanced by the electric force on the ion, all others are deflected out
of the beam.}}
\end{minipage}
\end{figure}

\begin{figure}
\begin{minipage}{\textwidth}
\parbox[b]{0.67\textwidth}{
\epsfig{file=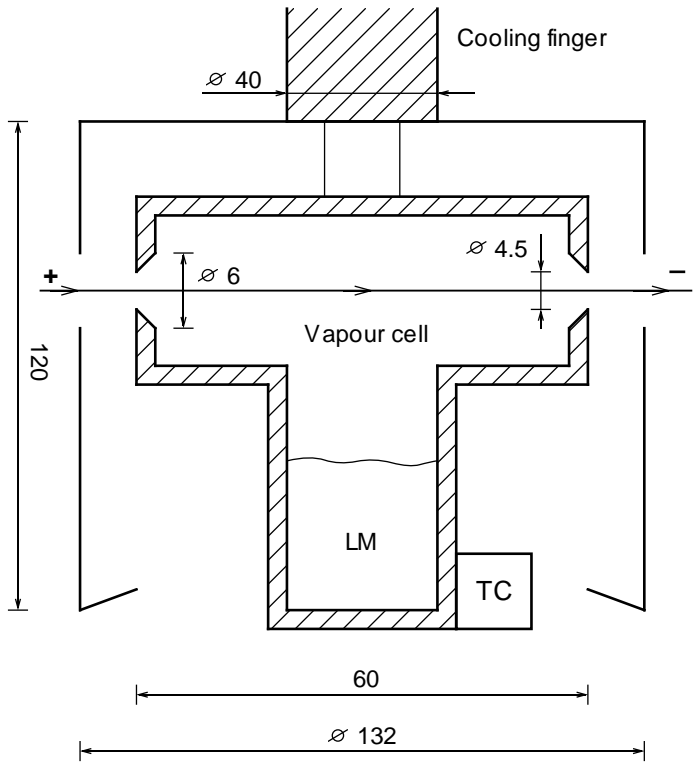, width=0.65\textwidth}
} \hfill
\parbox[b]{0.315\textwidth}{
\protect\caption[Charge exchange cell]{\label{bi260}\sloppy Charge
exchange cell: \textsf{LM}, liquid metal; \textsf{TC}, thermo
couple. The mass selected positive ions from the plasma ion source
(figure \ref{bi240}, \ref{bi210}) enter the cesium vapour from the left
and leave the cell, with two more electrons attached, as negative
ions. The cooled cylinder, mounted on a water cooled finger, around the
vapour cell serves to avoid excessive cesium contamination of the vacuum
chamber. The approximate dimensions are given in mm.}}
\end{minipage}
\end{figure}

The other leg of the ion beam machine starts with the plasma ion source
and a charge exchange chamber. From the plasma ion source, figure
\ref{bi240}, positive ions are extracted from the discharge between the
filament and the anode cap.  The positive ions are mass selected with a
Wien filter, figure
\ref{bi250}, and directed into a cesium vapour in the charge exchange
chamber, figure \ref{bi260}. 

To run the source with a gaseous medium, like helium, the desired gas is
admitted through the hole on the left side in figure
\ref{bi240}. Through the hole to the right in figure \ref{bi240}
positive ions are extracted from the discharge between the molybdenum
filament (cathode) and the anode plate. To create ions from condensed
matter we insert a charge holder, loaded with a compound containing the
desired atoms, from the left in figure \ref{bi240} and heat the filament
with a current of up to 18~A, to evaporate the compound. A buffer gas,
for example nitrogen, is admitted to create a steady discharge. In this
discharge most compounds are destructed and we see mostly atomic masses
and some very stable dimers, such as N$_{2}$.

The extraction, according to the manufacturer, ensues from an area of
about 25~$\mu$m diameter with a beam divergence of 250~mrad.  From this
point-like source a reasonably collimated beam can be formed with the
relatively simple ion optics shown in figure \ref{bi270}(a).

To perform well controlled experiments it is mandatory to create a mass
selected beam. The resolution should suffice to separate the desired
negative ion from the nearly always present hydrides, with a mass number
one higher than the ion. A resolution of 200 is therefore enough. This
can be accomplished with Wien filters (also called velocity filter) or
sector magnets.

In a Wien filter, figure \ref{bi250}, crossed electric and magnetic
fields act on the ion. For a parallel beam with uniform speed the
required resolution is achieved with a moderate electric field of
$\approx 15$~$\frac{\text{kV}}{\text{m}}$ and a magnetic field of
$\approx 0.1$~T. Currently our Wien filter has a resolution of about
50. For the lighter elements this is still good enough. Our sector
magnet has a mass resolution of roughly 200 and is therefore well
adapted to the heavier negative ions produced by the sputter ion source.

The charge exchange is achieved in a saturated cesium vapour heated to
typically 380~K. This is estimated to give a vapour pressure of a few
milli Pascal. To optimise the charge exchange we have varied the
reservoir temperature and found the efficiency had become constant above
a certain temperature. We interprete this as being due to saturation of
the charge exchange process. No investigations of the charge exchange
process have been undertaken here. Due to the scattering nature of the
charge exchange the beam acquires additional divergence, which we for
$^{7}$Li with an energy of 4.5~keV estimated to be less than 1~mrad.

\begin{figure}
\begin{minipage}{\textwidth}
\parbox[b]{0.52\textwidth}{
\epsfig{file=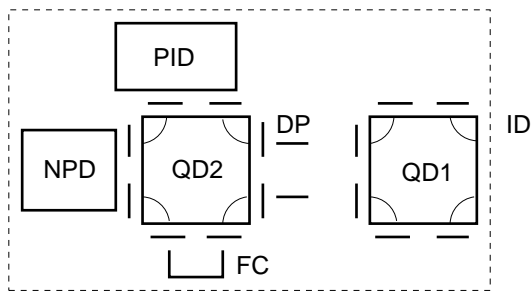, width=0.5\textwidth}
} \hfill
\parbox[b]{0.47\textwidth}{
\protect\caption[Ion optics in the interaction-detection
chamber]{\label{bi290}\sloppy\small Ion optics in the 
interaction-detection
chamber: \textsf{NPD}, neutral particle detector (figure \ref{bi360}
page \pageref{bi360}); \textsf{PID}; positive ion detector (figure
\ref{bi360}); \textsf{QD1,QD2}, electrostatic quadrupole deflectors;
\textsf{DP}, deflection plates. Around the quadrupole deflectors there 
are apertures.  Between the quadrupole deflectors the beam is shielded
stray from electric fields by a stainless steel tube.} }
\end{minipage}
\end{figure}

To deflect and focus the ion beams there are numerous deflection plates
and electrostatic lenses installed in the system, figure
\ref{bi270},\ref{bi290}. The horizontal and vertical deflection plates 
are used to steer the beam. Thanks to careful alignment of the beam
path, using a theodolite, the steering voltages are only a few Volt. The
electrostatic quadrupoles, figure \ref{bi290}, in the
interaction-detection chamber bend the ion beam by $90^{\circ}$. The
einzellense in the plasma ion source chamber and in the focusing section
are used to maximise the current through the system by appropriately
focusing the ion beam. From the sputter ion side a similar system is
used, with two differences: the compression lenses are necessary to
compensate for the focusing properties of the magnet, and these lenses
are also used to steer the beam since they are fed with two independent
voltages.

During the design the system has been simulated with a ray tracing
program called \textsf{SIMION} \cite{Hae-96-3}, to find the right
position and strength for the lenses.  The lens design follows
suggestions given in Hartings book \cite{Har-76} on electrostatic
lenses. We have chosen a design that minimises abberations under the
geometric restrictions of our system.  It has also been estimated which
portion of the initial phase space spanned by the ion beam can possibly
be transfered into the interaction region \cite{Hae-96-3}.
\section{The laser system}
\label{ch52}
\begin{figure}\centering
\subfigure[Commonly pumped dye lasers]{\epsfig{file=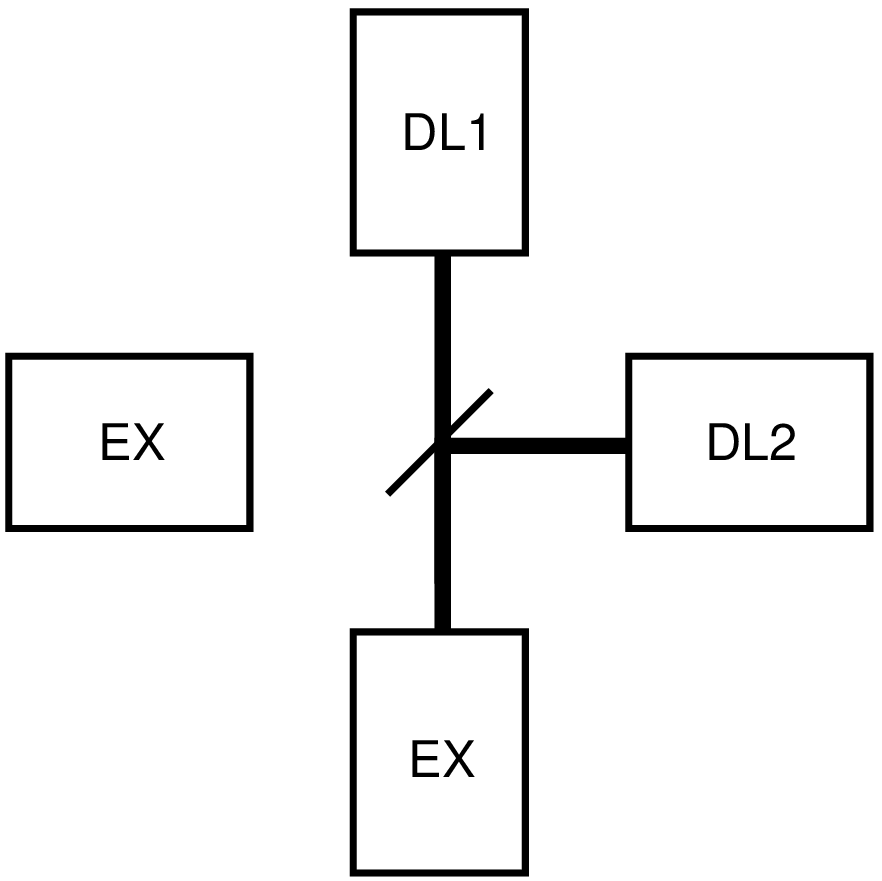,
width=0.385\textwidth}}\hfill
\subfigure[Separately pumped dye lasers]{\epsfig{file=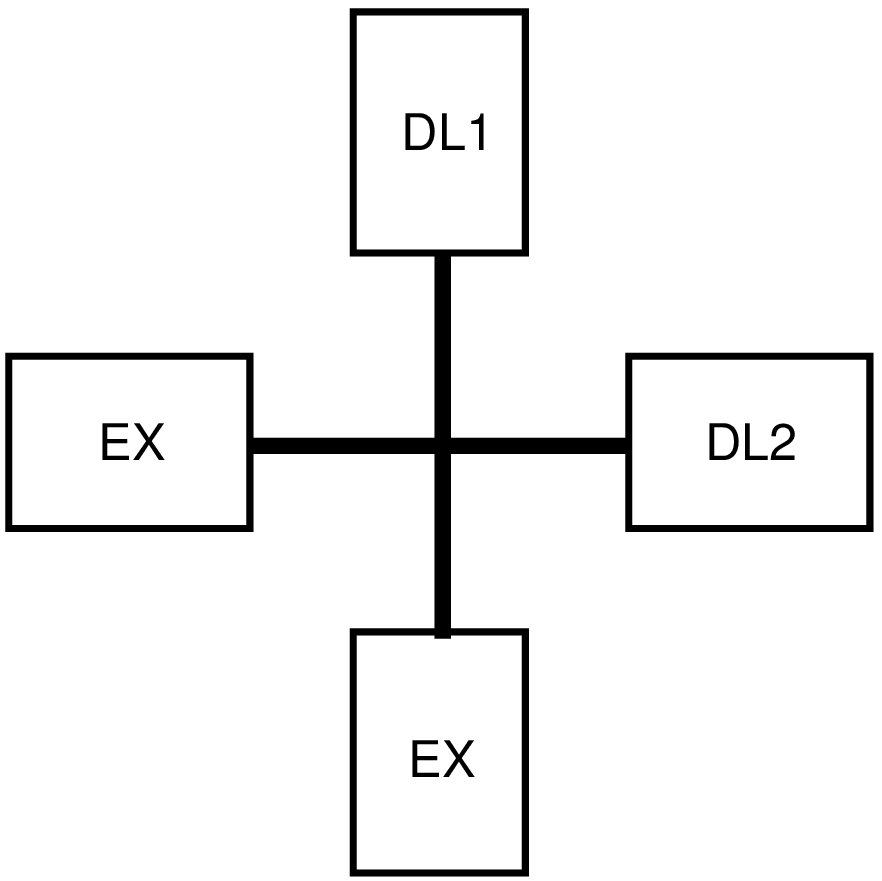,
width=0.385\textwidth}}
\caption[Laser arrangement]{\label{bi300}\sloppy Laser arrangement: The
excimer and dye lasers are placed such  that both excimer lasers can
pump either one of the dye lasers or both at once. The excimer lasers
are placed about 1.5~m from the dye lasers.}
\end{figure}

Our laser system consists of two excimer and two dye lasers. Normally
one excimer laser pumps one dye laser, as indicated in figure
\ref{bi300}(b). Each set of excimer plus dye laser can be used
individually for high resolution measurements.  It is also possible to
pump both dye lasers by a common excimer laser as shown in figure
\ref{bi300}(a).

The XeCl excimer lasers deliver pulses of 15~ns duration with an energy
of 100~mJ to 200~mJ and 308~nm wavelength. One of them has a repetition
rate of up to 200~Hz and the other one has a maximum repetition rate of
20~Hz. The dye laser pumped with these pulses emit pulses of the same
duration and repetition rate with an energy ranging from 1~mJ to 10~mJ,
for the fundamental, and 100~$\mu$J to 1~mJ per pulse for the doubled
light.
\begin{figure}
\begin{minipage}{\textwidth}
\parbox[b]{0.27\textwidth}{
\epsfig{file=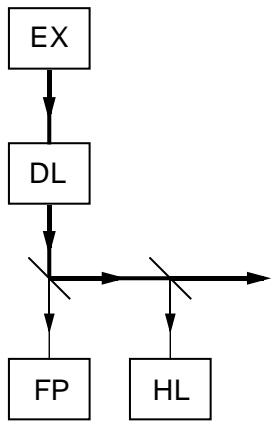, width=0.25\textwidth}
}\hfill
\parbox[b]{0.72\textwidth}{
\caption[Laser system overview]{\label{bi295}\sloppy Laser system
overview: \textsf{EX}, excimer laser; \textsf{DL}, dye laser;
\textsf{FP}, Fabry Perot etalon; \textsf{HL}, hollow cathode lamp. The
wavelengths of the excimer pumped dye lasers is determined by combining
an absolute calibration from transitions in the hollow cathode lamp with
the frequency markers provided by the fringes of the Fabry-Perot
etalon.}}
\end{minipage}
\end{figure}

The dye lasers generate tunable light from 800~nm to 330~nm, and 400~nm
to 200~nm with frequency doubling. The line width is 20~m$^{-1}$
(6~GHz). By inserting a Fabry Perot etalon in the oscillator beam path
of the dye laser the bandwidth can be narrowed to 4~m$^{-1}$ (1.2~GHz).

For experiments involving resonance ionisation 
the two laser pulses have to be synchronised. This is
accomplished by simply pumping both dye lasers by a common excimer
laser, as shown in figure \ref{bi300}(a).

The wavelength of the laser light is determined by combining an absolute
calibration from transitions in the hollow cathode lamp with the
frequency markers provided by the fringes of the Fabry Perot etalon.
With high contrast Fabry Perot fringes and good hollow cathode lamp
lines we can calibrate to a precision of approximately 1~m$^{-1}$.

The accuracy of a fitted energy position we judge by
the statistical scatter of the fitted parameters and by the estimated
error of the individual values.  Generally the scattering and the
individual error bars were, as expected, of comparable size.
\begin{figure}
\epsfig{file=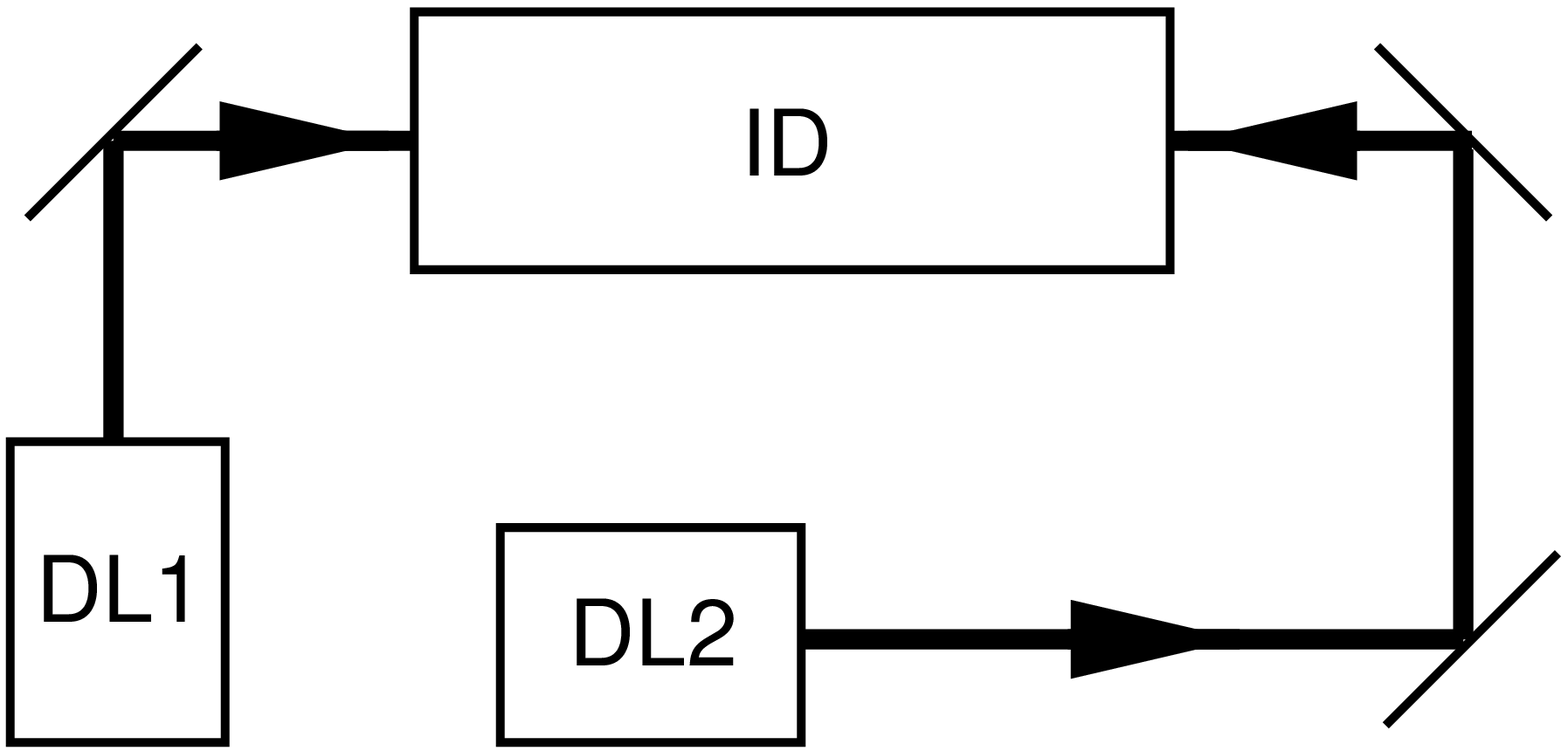, width=0.41\textwidth}\hfill
\epsfig{file=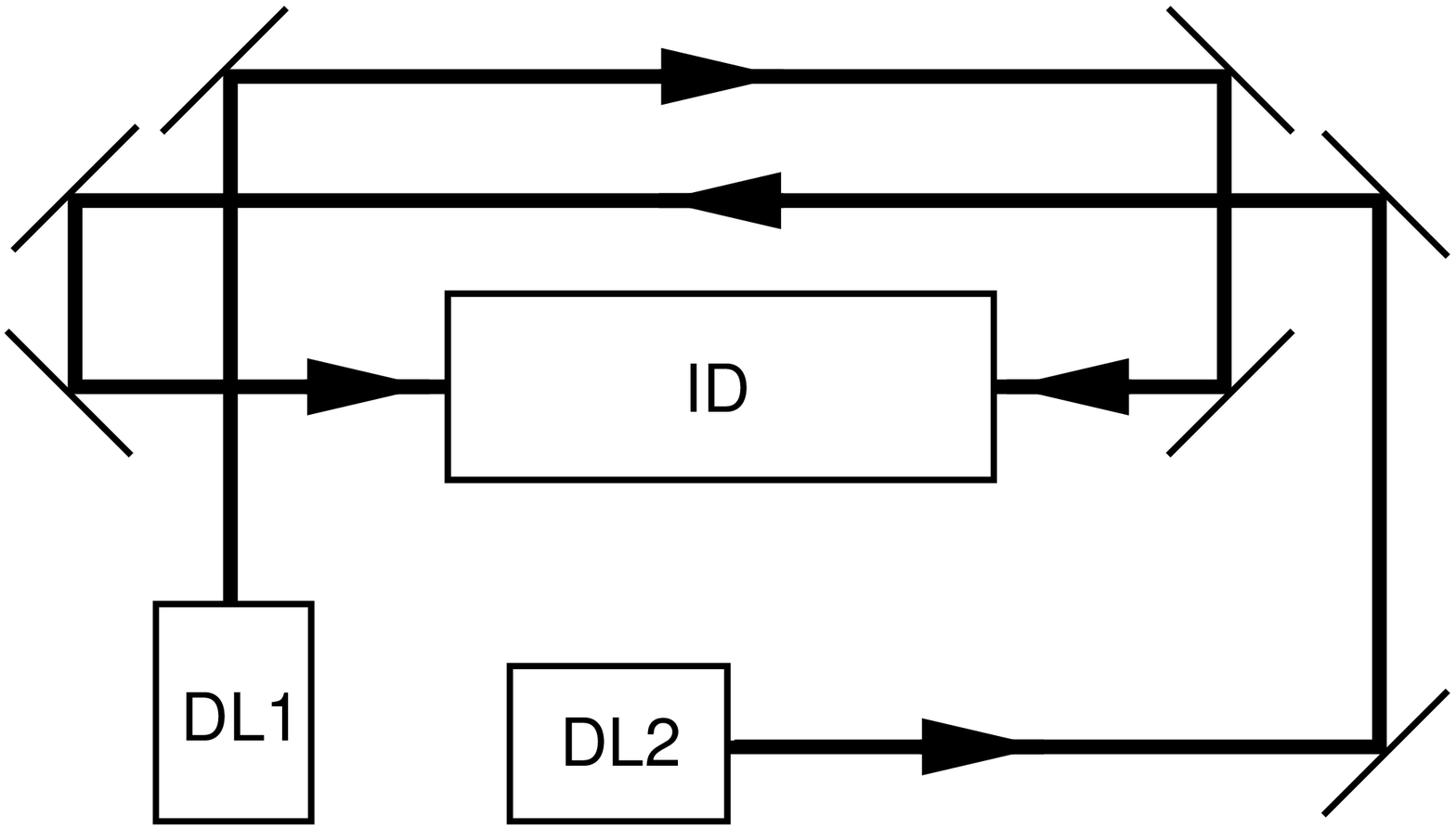, width=0.54\textwidth}
\protect\caption[Two colour laser beam path]{\label{bi340}\sloppy Two
colour laser beam path: In the interaction-detection chamber \textsf{ID}
the laser beams are always running opposite to each other. The left
part of the figure shows the photodetachment laser beam 1 (from
\textsf{DL1}) co-propagating with the the ion beam and the right part
counter-propagating. The beam reversal requires some rearranging of
optics and cannot be undertaken during a scan.}
\end{figure}
By combining two measurements with co and counter-propagating laser and
ion beams one can eliminate the Doppler shift to \emph{all} orders.  To
obtain these measurements the laser beam(s) have to be reversed.
The geometric mean of the two
measurements gives a Doppler free value for the measured energy $E_{0}$,
\begin{equation}\label{eqch51}
E_{0} = \sqrt{E_{0}^{\text{r}} E_{0}^{\text{b}}}\quad ,
\end{equation}
if $E_{0}^{\text{b,r}}$ is the blue- respectively red shifted value for
co- and counter-propagating laser and ion beams.  

This combining assumes the beam energy, or rather the speed of the ions,
to be constant. For a one-laser experiment where the beam reversal is
made during a scan this clearly is a valid assumption. For two-laser
experiments with reversed beams, shown in figure \ref{bi340}, beam
reversal typically is performed within the next days. Here the question
of stability is far less trivial. The ionisation step $\omega_{2}$
(figure \ref{bi450}) of the state selective excitation scheme is tuned
to a Rydberg state, and this transition is used to check our day to day
stability of the beam speed. In all cases we found the laser wavelength
to still be in resonance after a restart of the beam one or even a few
days later. We therefore believe that the above named combination of red
and blue shifted value also makes sense if the ion beam has to be
restarted to reverse the laser beam.

\section{The detection system}
\label{ch53}
\begin{figure}\centering
\epsfig{file=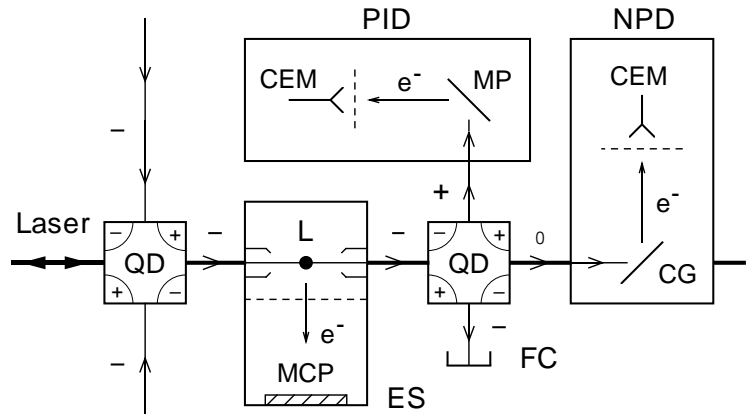, width=0.7\textwidth}
\protect\caption[Interaction-detection chamber]{\label{bi360}\sloppy
Interaction-detection chamber: \textsf{QD}, electric quadrupole
deflector; \textsf{PID}, positive ion detector; \textsf{CEM}, channel
electron multiplier; \textsf{L}, perpendicular laser beam (for electron
spectroscopy); \textsf{MCP}, micro channel plate; \textsf{ES},
time-of-flight electron spectrometer; \textsf{MP}, metal plate;
\textsf{FC}, Faraday cup; \textsf{NPD}, neutral particle detector;
\textsf{CG}, conducting glass plate. Laser and ion beams are coaxially 
merged over 0.5~m between the quadrupole deflector. For electron
spectroscopy \emph{only} the laser beam \textsf{L} perpendicularly
intersecting the ion beam is used.}
\end{figure}

The detection chamber, figure \ref{bi360}, renders four different
detection schemes possible: angular resolved photodetachment, laser
photo electron spectroscopy, photodetachment with neutral particle
detection and photodetachment resonance ionisation spectroscopy with
positive ion detection. In all cases particles are counted.  Electrons
are registered with micro channel plates \textsf{MCP} or channel
electron multipliers \textsf{CEM}. Neutral atoms or positive ions are
counted with a secondary emission detector (\textsf{PID,NPD})
\cite{Han-92-2} shown in figure \ref{bi360}.  In the secondary emission
detectors the electrons emitted upon impact of the atom or positive ion
on the plate are registered with a \textsf{CEM}.

\begin{figure}\centering
\epsfig{file=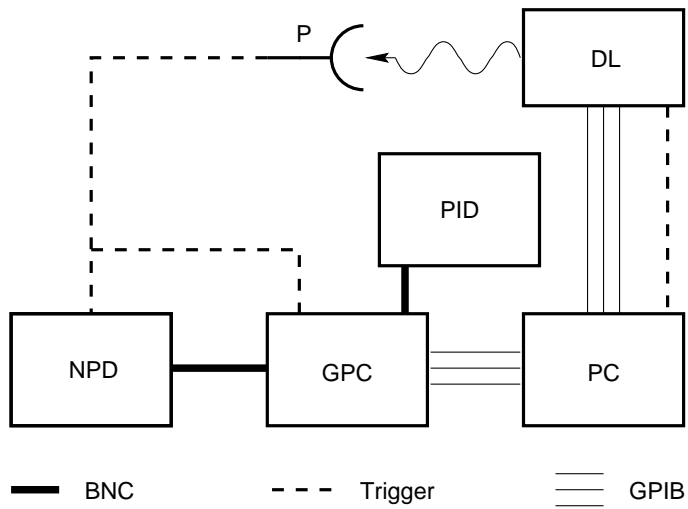, width=0.65\textwidth}
\protect\caption[Logic of the detection system]{\label{bi370}\sloppy
Logic of the detection system: \textsf{NPD}, neutral particle detector;
\textsf{P}, photodiode; \textsf{GPC}, gated photon counter;
\textsf{PID}, positive ion detector; \textsf{DL}, dye laser;
\textsf{PC}, data acquisition computer; \textsf{GPIB}, general purpose
interface bus. The apparatus operates under the control of the
\textsf{PC} that sends, via \textsf{GPIB}, a start signal to the dye
laser, which then delivers a preset number of pulses without a further
start signal. The trigger pulses from the dye laser to the \textsf{PC}
are counted to recognise the end of a burst. The neutral particle
detector and the gated photon counter are triggered from a photodiode
(figure
\ref{bi380}). The positive ion detector does not need triggering. After
a burst of pulses the data from the gated photon counter are transfered
via \textsf{GPIB} to the \textsf{PC}. Thereafter a new cycle can start.}
\end{figure}

\begin{figure}
\begin{minipage}{\textwidth}
\parbox[b]{0.5\textwidth}{
\epsfig{file=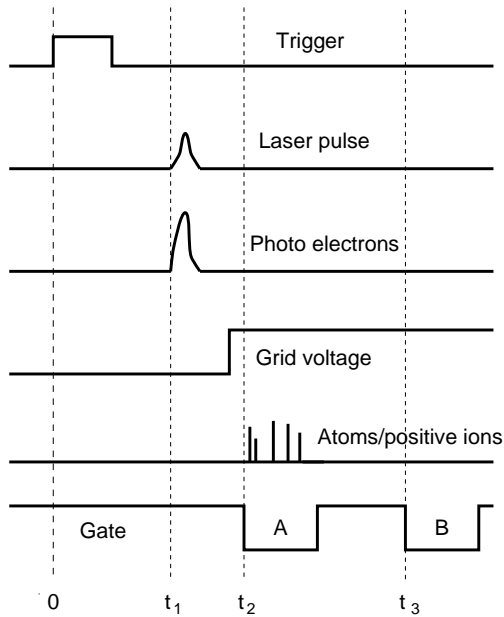, width=0.48\textwidth}
} \hfill
\parbox[b]{0.49\textwidth}{
\protect\caption[Time structure of the signal]{\label{bi380}\sloppy\small
Time structure of the signal: at the time $\text{t}_{1}\approx
1\mathrm{\mu}$s after the dye laser trigger, the excimer laser fires and
the light pulse almost instantaneously creates photo electrons on the
plate of the neutral particle detector (figure
\ref{bi360}). To prevent the photo electrons from reaching the
\textsf{CEM} (figure \ref{bi360}) the grid between the plate and the
\textsf{CEM} is held at negative potential with respect to the
plate. Shortly before t$_{2}\approx 2\mathrm{\mu}$s this grid is
grounded. At t$_{2}$ the gate \textsf{A} of the photon counter
starts. The delay $\text{t}_{2}-\text{t}_{1}$ is the time of flight for
the positive ions or neutral particles which arrive first at the
detector. The width of gate \textsf{A} and \textsf{B} is equal to the
time of flight ($\approx$5~$\mu$s) for the 0.5~m length of the
interaction region. In gate \textsf{B} we count the background with a
delay of $\text{t}_{3}\approx 40\mathrm{\mu}$s.}  }
\end{minipage}
\end{figure}

The pulses from any of the hitherto mentioned detectors are
pre-amplified and then counted by a gated photon counter, figure
\ref{bi370}. This gated detection based on the time structure of the
signal, figure \ref{bi380}, allows us to exclude all events not
originating from the spatial and temporal overlap of laser and ion beams
from being counted.

We empirically found the neutral particle and positive ion detectors
start to saturate at a count rate $cr$ per pulse of,
\begin{equation}
cr \approx \frac{1}{2}\sqrt{m}\quad ,
\end{equation}
with $m$ denoting the mass number of the negative ion in question. The
mass dependence is an indirect one, because it is really the mass
dependent beam speed for a given beam energy (3~kV to 4~kV) that lies
behind this dependency.

It has been estimated that every positive ion or neutral atom created in
the interaction region is detected, owing to the very high efficiency of
the secondary emission detectors \cite{Han-92-2}. This makes the
apparatus very sensitive and well adapted to measurements on small
photodetachment cross section.

The negative ion current is monitored with the Faraday cup \textsf{FC}
in the interaction-detection chamber. On the beam path there are several
apertures on which a current can be measured. This serves as a valuable
aid in adjusting the beam.

The electron spectrometer \textsf{ES} in figure \ref{bi360} is of time
of flight type. On the entrance the ion beam is decelerated and focused
to improve resolution and signal level. The resolution is about
10~meV. 

Angular resolved photodetachment studies can be used to distinguish
between different photodetachment channels. A carbon cylinder with
approximately radial holes can be inserted in the beam path in the
interaction-detection chamber to perform angular resolved
photodetachment studies \cite{Han-89-1}. Neither electron spectroscopic
nor angular resolved measurements are presented in this thesis.

\section{Data acquisition and processing}
\label{ch61}
All data are acquired with a \textsf{PC} and stored in 7-bit ASCII
files. The \textsf{PC} controls the whole experiment via \textsf{GPIB},
figure \ref{bi370}. A measurement cycle is initiated by a start message
to one of the dye lasers. The micro computer integrated in the dye laser
then generates a trigger pulse which is shown in the upper trace in
figure
\ref{bi380}. This pulse triggers the excimer laser(s) to emit a pulse at
t$_{1}$.

Since this time t$_{1}$ is subject to a substantial jitter we trigger
the rest of the equipment from a photodiode \textsf{P} picking up light
from the excimer laser. In this way we are able to trigger the light
related detection of interaction products with a jitter of only
$\approx$0.5~ns, thus reducing fluctuations in the detection. The
quality of these trigger pulses is also carefully checked with a fast
oscilloscope.

The neutral particle detector is triggered. To avoid overloading the
\textsf{CEM} with photo electrons, the grid between the plate, figure
\ref{bi360}, and the \textsf{CEM} is held at negative potential with
respect to the plate. This voltage has to be switched off a few hundred
nanoseconds before the first neutral particles from the interaction zone
impinge on the plate at the time t$_{2}$ in figure
\ref{bi380}. The photo electrons and neutral particles can be separated
because the photo electrons are created on the plate while the neutral
particles have to travel about 0.15~m from the interaction region to the
detector with a speed of about 100\,000~$\frac{\text{m}}{\text{s}}$.
The delay time $\text{t}_{2} - \text{t}_{1}$ is the time of flight for
the neutral particles from the interaction region between the quadrupole
deflectors \textsf{QD1,2} to the neutral particle detector
\textsf{NPD}. 

The photodetachment events are counted during the gate \textsf{A}, as
indicated in the lowermost trace of figure \ref{bi380}. The width of
this gate is equal to the time of flight, about 5~$\mu$s, for the 0.5~m
long interaction region. The photon counter \textsf{GPC} uses a
threshold height for the pulses to discriminate the events caused by
neutral particles from noise. To subtract an eventual background of
neutral particles due to collisional detachment we count the events in a
second gate \textsf{B} with a delay t$_{2}$ of about 40~$\mu$s.

The positive ion detector \textsf{PID} functions the same way as the
neutral particle detector \textsf{NPD} except that this detector is not
triggered. We found the effect of laser pulse related disturbances to be
so small that no effort for further suppression was made.

After a preset number of laser pulses the number of events counted by
the photon counter \textsf{GPC} is transfered via \textsf{GPIB} to the
\textsf{PC} where it is recorded in a file. After that a start
request is sent to the dye laser and a new cycle may start. 

The program has recently been refined \cite{Ran-96} and allows now a
different step length and pulse number in different parts of the scan.
This comes in useful when scans with relatively short data relevant
parts are made and in between only the Fabry Perot fringe count has to
be kept.

\begin{figure}\centering
\subfigure[Fit logbook page]{\fbox{\epsfig{file=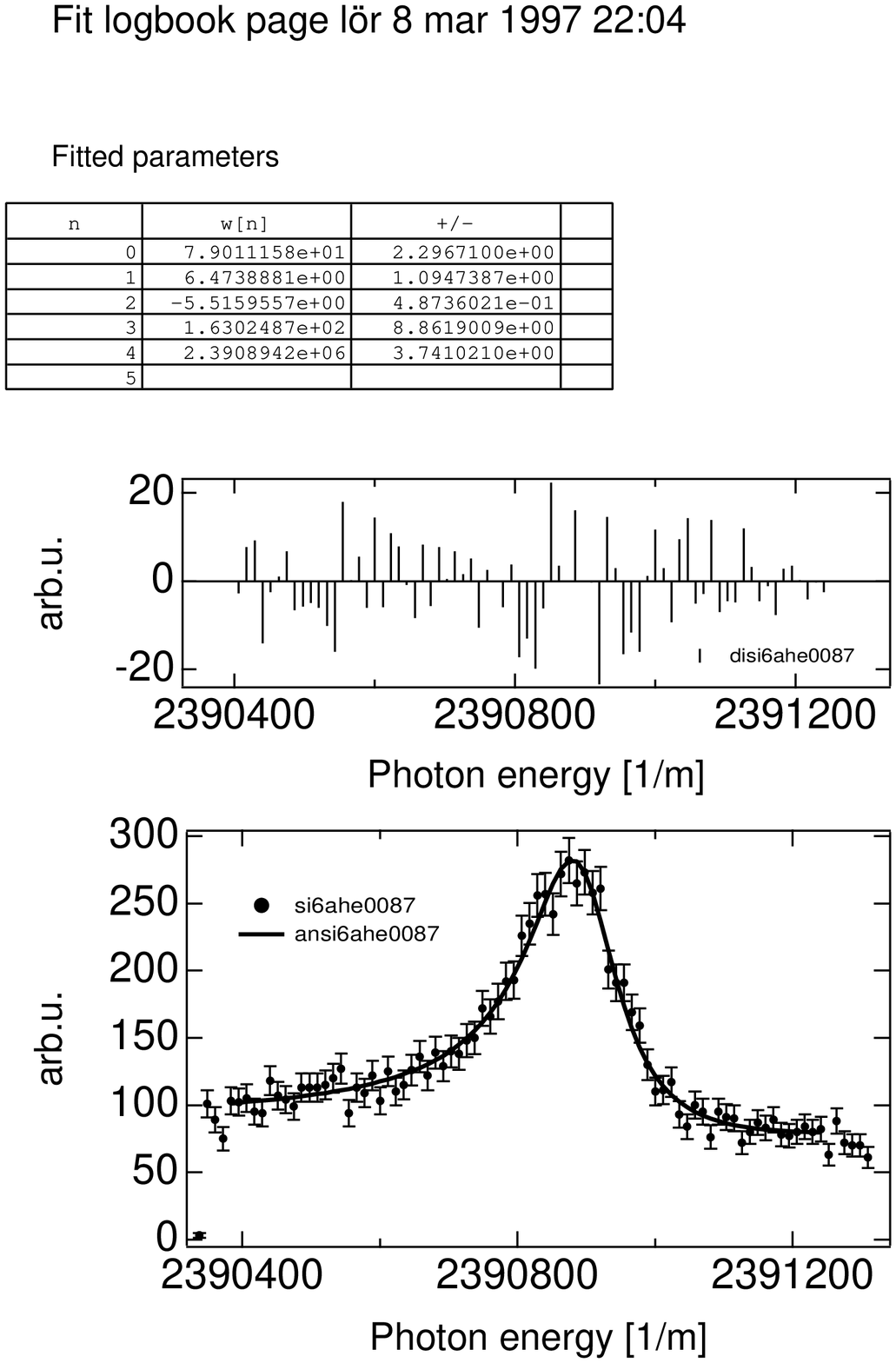,
width=0.42\textwidth}}}\hfill
\subfigure[Overview logbook page]{\fbox{\epsfig{file=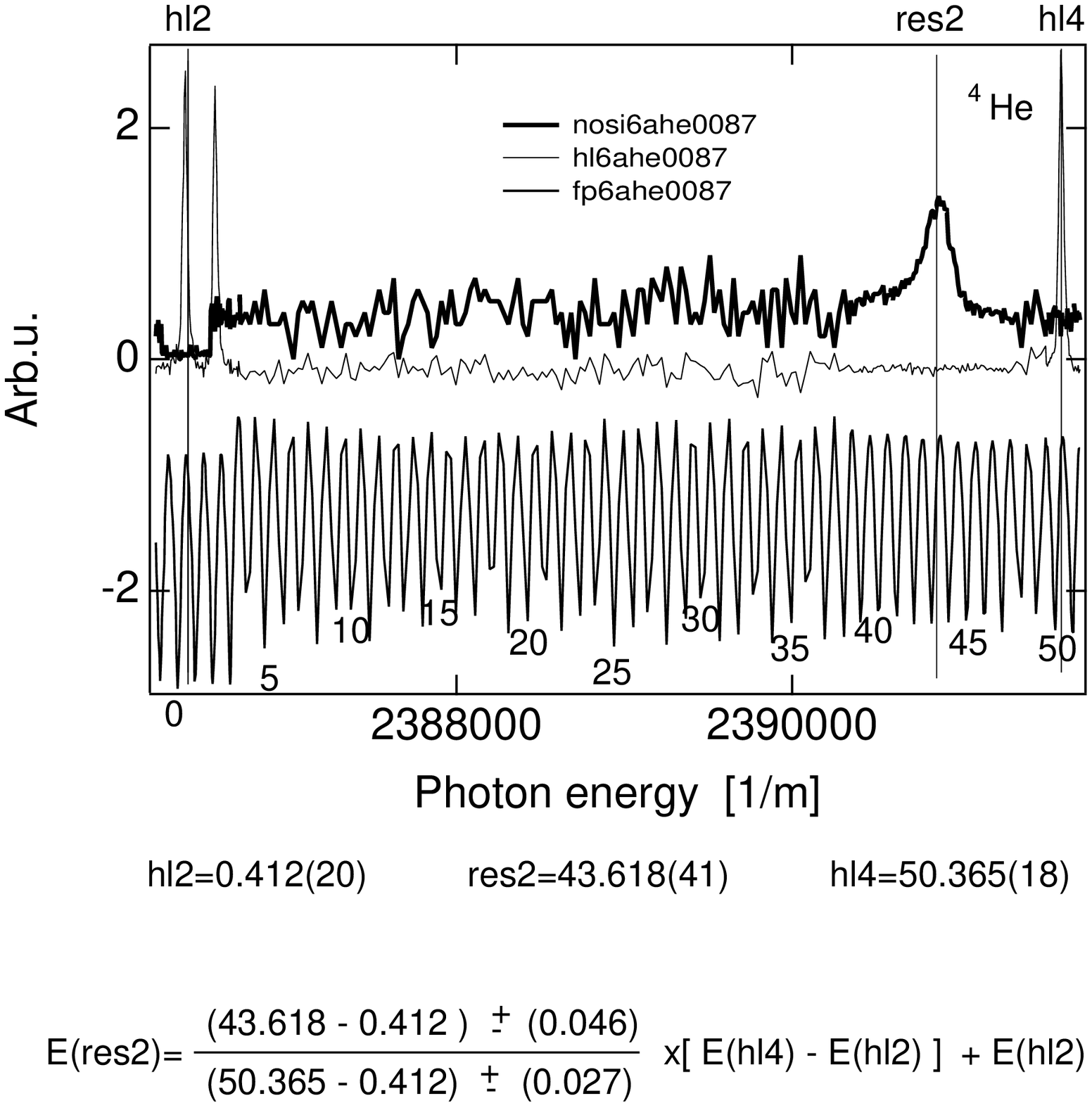,
width=0.48\textwidth}}}
\protect\caption[Logbook pages]{\label{bi390}\sloppy Logbook
pages: The logbook page (a) is archived for every fit containing
the parameter values, their uncertainty, the data with the best fit,
and, the fit residuals to roughly judge the quality of the fit. Part (b)
shows, from top to bottom, the signal per laser pulse, the hollow
cathode lamp signal and the Fabry Perot fringes.  This page is used to
screen the data and to collect the fitted parameters. The positions of
\textsf{hl2},
\textsf{res2} and \textsf{hl4} are given in Fabry Perot fringe
numbers, which together with the energy values for the calibration
lines, give the calibrated position of the Doppler shifted resonance.}
\end{figure}

From the \textsf{PC} the data files are transfered to a computer, here
\textsf{Linux/Unix} machines, and automatically preprocessed for further
evaluation with the program \textsc{Igor} 2.02 on a Macintosh
computer. The data files are also archived as
\texttt{gnuzip}-compressed \texttt{tar}-files. All files are graphed for
screening and archiving purposes, as shown in figure \ref{bi390}(b).

To fit the parameters of the respective functions to our data, we used
the Levenberg Marquardt method \cite{PFT-89}.  This method elegantly
interpolates between a more global search for the minimum of $\chi^{2}$
and a fine local search. The quality of a non-linear fit can depend on
the initial values for the fit routine. The graphical interface of
\textsc{Igor} allows one to quickly find adequate start values. We
found it seldom necessary to hold parameters constant during a fit.  In
the case of very noisy data we also used a robust fit algorithm that
reduces the relative significance of outlier data points described in
\cite{PFT-89}. When fitting to the photodetachment signal the data were
weighted with their inverse error, all other data were fitted with equal
weight for all points. When the data are weighted with their inverse
error then $\chi^{2}$ also has an absolute meaning and can be used to
judge the statistical quality of the data. Of all fits a logbook page is
printed and archived, as shown in figure \ref{bi390}(a)

There is no straight forward way of obtaining reliable errors for the
fitted parameters \cite{PFT-89}. The uncertainties given by the routine
are the square root of the diagonal elements of the correlation
matrix. These values we use as guideline in estimating the error
bars. Always a serious attempt is made to have sufficient statistics to
also judge the statistical scatter of the fitted parameters. Usually
both error estimates give comparable results.

%
%
\chapter{Threshold studies}
\label{ch62}
The most accurate methods to determine the electron affinity of
atoms is to measure the threshold energy for the photodetachment process
\cite{Hot-85}. Especially for s-wave threshold this gives
very accurate results owing to the very steep onset of the
photodetachment cross section. Here we present one classical
photodetachment threshold study on tellurium and an electron affinity
measurement on lithium with a novel state selective method.

Neutral particle detection is a well developed method and has been
demonstrated to be very sensitive and accurate
\cite{And-91,Han-92-1,Neu-85,Ber-95-4,Blo-89}.  In
spite of some limitations, we attained a result for the electron
affinity of tellurium, that ranks among the most accurate electron
affinities of the periodic system \cite{Blo-95}. To reach this level of
accuracy we paid special attention to a reliable absolute calibration,
which we ensured by using two completely independent sets of calibration
lines in two different hollow cathode lamps.  The measurement is
presented in section \ref{ch621}.

The measurement of the tellurium electron affinity is almost a textbook
example: an easily accessible s-wave threshold in an element of which
the sputter ion source efficiently produces negative ions. 

With the affinity measurement on lithium, presented in section
\ref{ch622}, we mainly intended to demonstrate a new state selective
detection scheme. The state selectivity is accomplished by resonance
ionisation of the residual atom of the photodetachment process. This new
method is introduced here and in the subsequent experiment we assume
familiarity with its principles.  The emphasis for this experiment was
to demonstrate the feasibility of the scheme and to explore its
potential. We found this technique to work very well and were able to
improve the accuracy of the lithium electron affinity by an order of
magnitude.

\section{Tellurium electron affinity}
\label{ch621}
\begin{figure}
\begin{minipage}{\textwidth}
\parbox[b]{0.29\textwidth}{
\epsfig{file=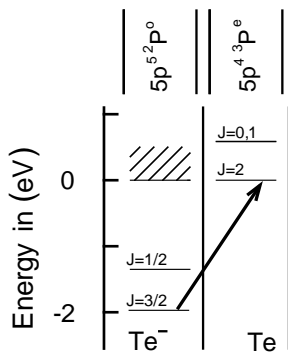, width=0.27\textwidth}
} \hfill
\parbox[b]{0.7\textwidth}{
\protect\caption[Tellurium excitation scheme]{\label{bi410}\sloppy
Tellurium excitation scheme: Selected states of Te/Te$^{-}$. The arrow
indicates the transition induced in this experiment. } }
\end{minipage}
\end{figure}

Previous to this measurement \cite{Hae-96-2} the electron affinity of
tellurium was known to have a poor precision compared to other elements
with similarly high electron affinities. We aimed to improve this value,
by determining the Doppler free photodetachment threshold position by
studying the process:
\begin{equation}\label{eqch61}
\text{Te}^{-}\text{(5p$^{5}$\,$^{2}$P$_{3/2}$)} + \hbar \omega
\mapsto
\text{Te}\text{(5p$^{4}$\,$^{3}$P$_{2}$)} +
\text{e}^{-}(\epsilon\text{s})\quad .
\end{equation}
Close to the threshold the s-wave detachment dominates over the also
allowed d-wave detachment (figure \ref{bi090}). As can be seen in figure
\ref{bi410}, there is additional contribution to the neutral atom
count rate due to the processes:
\begin{equation}\label{eqch62}
\text{Te}^{-}\text{(5p$^{5}$\,$^{2}$P$_{1/2}$)} + \hbar \omega
\mapsto  
\text{Te(5p$^{4}$\,$^{3}$P$_{0,1,2}$)} + \text{e}^{-}\quad .
\end{equation}
The contribution of these processes to the signal, however, varies
slowly with the photon energy since they are far above their thresholds.
Furthermore, these signals are relatively small since the sputter ion
source predominantly produces ground state negative ions.

The light pulses with 10~$\mu$J were generated by one of our excimer
pumped dye lasers operating with Rhodamine 6G. We calibrate with four
lines from neon and argon given in table \ref{NeArRefLine}.

\begin{figure}
\begin{center}
\epsfig{file=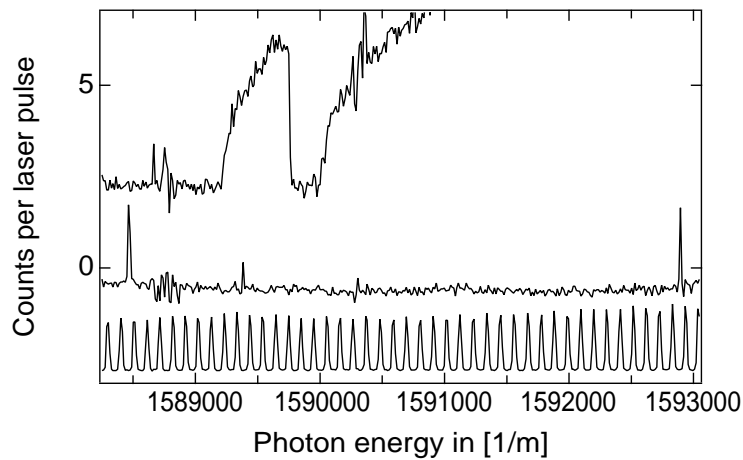, width=0.7\textwidth}
\end{center}
\protect\caption[Tellurium scan overview]{\label{bi420}\sloppy\small
Tellurium scan overview: The three curves show the photodetachment
signal, the reference lines (Ne) and the Fabry-Perot fringes. In the
photodetachment signal, the left threshold is obtained with
counter-propagating laser and ion beams and the right threshold with
co-propagating beams.  The background is partly due to photodetachment
of the Te$^{-}(5p^{5}\,^{2}$P$_{1/2}$) state. The signal saturates at
about 7 counts per laser pulse.  In the vicinity of the calibration
lines and the photodetachment thresholds the step-length was 1\,pm and
100 shots per point were taken.  Elsewhere the step-length was 5\,pm and
10 shots per point were taken. The vertical scale (Counts per laser
pulse) is only valid for the photodetachment signal.}
\end{figure}

For the data acquisition we used a newly written program \cite{Ran-96}.
It allows to divide a scan into several parts of which each has its
individual step length and pulse number per step. It is possible to
increase the data quality on the short stretches where the investigated
structure is located. In between only the Fabry-Perot fringe count has
to be kept, which is attainable with less laser pulses and more sparse
data points. Compared to a whole scan at full resolution we reduced the
scan time by about a factor of ten. An example of such a scan is given
in figure \ref{bi420}.
\begin{table}
\begin{center}
\caption[Tellurium calibration
lines]{\label{NeArRefLine} \sloppy Tellurium calibration lines:
Transition energy of the calibration lines used in this experiment.  We
calculated the neon transition energies from the tabulated level
energies.  The Ar transition energies are calculated from the levels
presented in Tab \textsc{v} of
\protect\cite{Min-73}. The levels are designated in Paschen
notation.}\medskip
\begin{tabular}{lcr@{.}l}
\hline\hline
Element & Transition & \multicolumn{2}{c}{Line  (m$^{-1}$)}\\ \hline
Ne \cite{Cha-94,Kau-72} & 2p$_{2}$ $\rightarrow$ 4d$_{6}$ &
1\,592\,921&6(4)\\
Ne \cite{Cha-94,Kau-72} & 2p$_{5}$ $\rightarrow$ 3s$_{2}$ & 
1\,588\,439&6(6)\\
Ar \cite{Min-73} & 2p$_{6}$ $\rightarrow$ 5d$_{4}$ & 
1\,592\,259&8(5)\\
Ar \cite{Min-73} & 2p$_{2}$ $\rightarrow$ 5s$^{\prime\prime}_{1}$ & 
1\,587\,650&8(5)\\
\hline\hline
\end{tabular}
\end{center}
\end{table}

\begin{figure}
\begin{minipage}{\textwidth}
\parbox[b]{0.54\textwidth}{
\epsfig{file=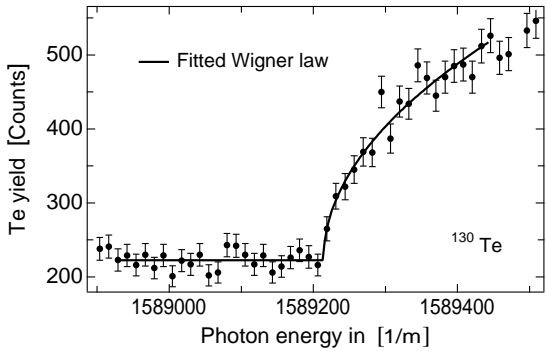, width=0.52\textwidth}
} \hfill
\parbox[b]{0.45\textwidth}{
\protect\caption[Te(5p$^{4}$)+$\epsilon$s
threshold]{\label{bi430}\sloppy\small 
Te(5p$^{4}$)+$\epsilon$s threshold: Measurement of the relative
photo\-de\-tachment cross section near the Te(5p$^{4}$)+$\epsilon$s
threshold with counter-propagating  laser and ion beams.  The solid 
line is a
fit of the Wigner law \eqref{eqch416}. 
The error bars represent the shot noise. Each data point was
acquired during 100 laser pulses.  The background is mainly due to
photodetachment of Te$^{-}$(5p$^{5}$\,$^{2}$P$_{1/2}$) 
\eqref{eqch62}.}  }
\end{minipage}
\end{figure}

A typical measurement is shown in figure \ref{bi430}.  To obtain the
photodetachment threshold energy $E_{0}$ we fitted \eqref{eqch416} to
the data and took the geometric mean of the fitted blue and red shifted
thresholds as described in \eqref{eqch51}. The average threshold energy
$E_{0}$ is
\begin{equation}\label{eqch63}
E_{0} = 1\,589\,618(5)\; \text{m}^{-1}\quad .
\end{equation}

There are two major contributions to the uncertainty of $E_{0}$. There
is a statistical error of 1~m$^{-1}$ corresponding to the spread of
fitted threshold values. Second, there is an uncertainty related to the
laser intensity profile. We have estimated this uncertainty by analysing
the Fabry-Perot fringes and the atomic reference lines to be less than
4~m$^{-1}$, which is one fifth of the laser frequency bandwidth.

%

%

\begin{figure}
\begin{minipage}{\textwidth}
\parbox[b]{0.58\textwidth}{
\epsfig{file=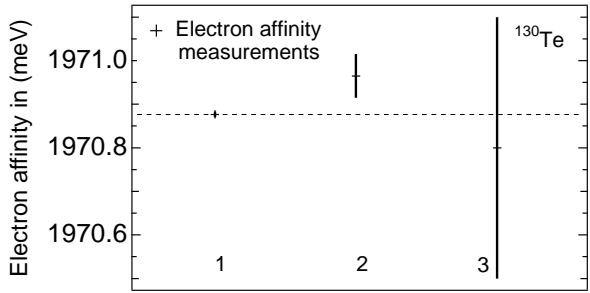, width=0.56\textwidth}
} \hfill
\parbox[b]{0.41\textwidth}{
\protect\caption[Tellurium electron affinity]{\label{bi440}\sloppy
\small
Tellurium electron affinity: Graphical comparison of the different
experimental tellurium electron affinity values with their respective
uncertainties. From left to right the values are as follows: \textsf{1},
this work; \textsf{2}, Th{\o}gersen \cite{Tho-96-2}; \textsf{3},
Slater \cite{Sla-77}. See also table \ref{tach61}.}
}
\end{minipage}
\end{figure}
%

%
\begin{table}
\begin{center}
\caption[Tellurium electron affinity]{\label{tach61}\slshape Tellurium
electron affinity: Comparison of the different measurements. The value
in m$^{-1}$ is only given if the publication contains the
value.}\medskip
\begin{tabular}{lr@{.}lr@{.}l}
\hline\hline
Author & \multicolumn{2}{c}{Affinity in m$^{-1}$} &
\multicolumn{2}{c}{Affinity in meV} \\ \hline
Slater (1977) \cite{Sla-77} & \multicolumn{2}{c}{} & 1970&8(3) \\
Th{\o}gersen (1996) \cite{Tho-96-2} & 1&589\,69(4)$\times 10^{6}$ &
1970&965(50)\\ 
This work (1996) \cite{Hae-96-2} & 1&589\,618(5)$\times 10^{6}$ &
1970&876(7) \\
\hline\hline
\end{tabular}
\end{center}
\end{table}

In figure \ref{bi440} we compare our electron affinity with other
measurements. The affinity from \cite{Tho-96-2} is not the main result
but a ``windfall profit'' from fine structure measurements on
Te$^{-}$. The numerical values are presented in table \ref{tach61}.

We have shown that the background contribution from photodetachment of
Te$^{-}$($^{2}$P$_{1/2}$) is not critical for the accuracy achievable
with our apparatus. An important part of this experiment was the use of
the computer program that can vary both the wavelength step and the
acquisition times for each wavelength. This has proven to be most
valuable to improve the statistics and simultaneously decrease the total
acquisition time, hence diminishing the probability of long term
drifts. This new value of the electron affinity of tellurium is 45 times
more accurate than the value of Slater
\cite{Sla-77} and fall within their error bars. The more recent
value of Th{\o}gersen
\cite{Tho-96-2} agrees with ours within two error bars.

\section{Lithium electron affinity}
\label{ch622}
Before the advent of photodetachment with state selective ionisation, it
was very difficult to determine electron affinities of negative ions
with p-wave photodetachment thresholds. As a consequence the
electron affinity of lithium was not very well determined.

Electron affinity measurements using neutral particle detection could in
general only hope to reach a high precision if the photodetachment
threshold is of s-wave character. The state selective detection now
allows to choose a higher lying atomic state that gives rise to an
s-wave continuum. The electron affinity is then obtained by subtracting
the atomic excitation energy from the measured threshold energy. We
demonstrated the feasibility of such experiments with an improved
electron affinity of lithium. This section provides the general outline
of how the scheme is implemented in practice.

The idea to combine laser photodetachment and resonance ionisation
\cite{Hur-79} was born out of the struggle to reduce the background 
in measurements on doubly excited states of lithium, extending the
results presented in section \ref{ch631} to higher thresholds. Such
measurements, exploiting the advantages of state selective detection,
have now been carried out in our group by Ljungblad
\cite{Lju-96}. Even though the method has grown beyond its adolescence
the background suppression is still one of its main features.

Resonance ionisation of the residual atoms, left by the photodetachment,
renders final state selective detection possible while retaining the
high sensitivity and superior precision and accuracy of the collinear
geometry.  Through the state selective ionisation it is possible to
exclusively monitor photodetachment events leading to the formation of
the residual atom in a specific excited state.

With this method it is possible to isolate the partial cross sections
associated with a specific final state of the residual atom.  A partial
photodetachment cross section is labelled by the state of the residual
atom, the orbital angular momentum of the outgoing electron and the
coupling between the angular momenta of the atom and the electron. Since
we detect the state of the residual atom A$^{\ast}$, it is often
convenient to speak of \emph{all} partial cross sections associated with
a specific atomic state as \lq A$^{\ast}$ partial cross sections\rq
(plural!), where A$^{\ast}$ is the atom in an excited state.  This
detection scheme enriches the experimental possibilities in two
important cases: firstly electron affinities of virtually any element
can be determined using the advantageous s-wave thresholds, since it is
almost always possible to find an accessible s-wave threshold higher up
in the continuum. Secondly resonance structure due to autodetaching
states can be studied in different partial cross sections.

One of the earlier measurements of the lithium electron affinity by
Feldman \cite{Fel-76} used infra-red output of a laser pumped optical
parametric oscillator to investigate the p-wave threshold. Another
paper \cite{Bae-85} treats the sharp cusp at the Li(2p) threshold
(figure \ref{bi060}) and attains also, as a side result, a value for the
electron affinity of lithium.

\begin{figure}
\begin{minipage}{\textwidth}
\parbox[b]{0.75\textwidth}{
\epsfig{file=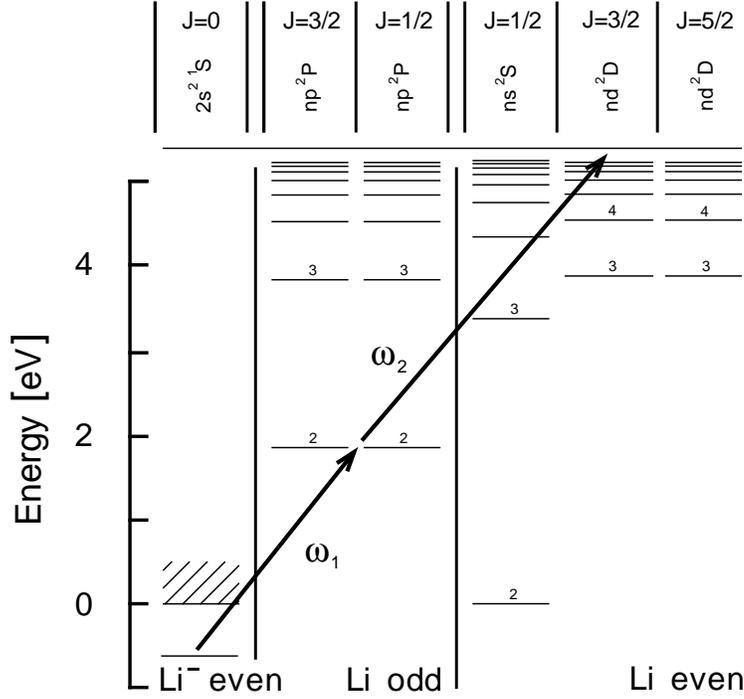, width=0.73\textwidth}
} \hfill
\parbox[b]{0.235\textwidth}{
\protect\caption[Li$^{-}$ excitation scheme \textsc{I}]{\label{bi450}
\sloppy
\small Li$^{-}$ excitation scheme \textsc{I}: Selected bound states of
Li/Li$^{-}$ grouped according to their parity and total angular
momentum.  The arrows indicate transitions induced in this experiment. }
}
\end{minipage}
\end{figure}

Here we use a novel excitation scheme depicted in figure \ref{bi450}.
This two-colour state selective photodetachment scheme is simple in
concept. One laser of frequency $\omega_{1}$ is used to photodetach
Li$^{-}$ ions producing an excited lithium atom and a free electron. A
second laser of frequency $\omega_{2}$ resonantly excites lithium atoms
left in the 2p state to a Rydberg state, which subsequently is field
ionised. Hence, the entire process can be represented by the following
steps:
\begin{align}\label{eqch64}
\text{Li}^{-}\text{(2s$^{2}$\,$^{1}$S$_{0}$)} + \hbar\omega_{1} &\mapsto
\text{Li(2p\,$^{2}$P$_{1/2,3/2}$)} + \text{e}^{-}\quad ,\\
\text{Li(2p\,$^{2}$P$_{1/2,3/2}$)} + \hbar\omega_{2} &\mapsto
\text{Li($n$s,$n$d)}\quad ,\label{eqch64b}\\
\text{Li($n$s,$n$d)} &\leadsto
\text{Li}^{+}\text{(1s$^{2}$\,$^{1}$S$_{0}$)} + \text{e}^{-}\quad
,\label{eqch64c} 
\end{align}
where $\leadsto $ denotes field ionisation and Li($n$s,$n$d) corresponds
to a highly excited Ryd\-berg atom state with $n\approx 23$. State
selectivity is accomplished in the resonant ionisation step since only
Li(2p) atoms can be ionised via the intermediate Rydberg state. In this
manner we were able to isolate a particluar photodetachment channel, in
this case the Li(2p) channel, and investigate the partial
photodetachment cross section by measurement the yield of Li$^{+}$ ions.

\begin{figure}
\begin{minipage}{\textwidth}
\parbox[b]{0.46\textwidth}{
\epsfig{file=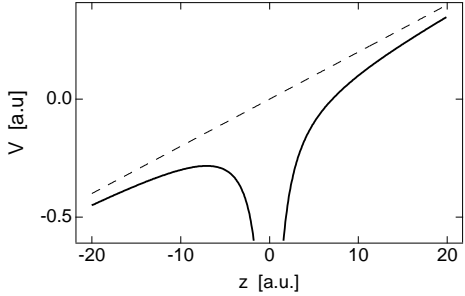, width=0.44\textwidth}
} \hfill
\parbox[b]{0.53\textwidth}{
\protect\caption[Static electric field ionisation]{\label{bi460}\sloppy
Static electric field ionisation: Potential energy of a hydrogen atom in
an uniform electric field with a field strength of 0.02 along the
z-axis. All states of the atom are unbound and can tunnel through the
barrier into the continuum. }
}
\end{minipage}
\end{figure}

Field ionisation of slow atoms in a static electric field, depicted in
figure \ref{bi460}, is due to tunneling predominantely along the field
axis.  For hydrogen this problem can be solved exactly \cite{Lan-74} and
for all other atomic systems semi-classical formulae have been derived
\cite{Fab-93,Smi-66,Dem-2-81,Dem-64}. For us  the most improtant result
is the critical field strength $\mathcal{E}_{\text{cr}}$ as a function
of the principal quantum number $n$ (in a.u.),
\begin{equation}\label{eqch65}
\mathcal{E}_{\text{cr}} = \frac{1}{16n^4}\quad ,
\end{equation}
above which essentially all atoms are field ionised. For the field
strength of 150~$\frac{\text{KV}}{\text{m}}$ at the entrance of the
second quadrupole \textsf{QD2} the corresponding $n$ is 22.

\begin{figure}\centering
\subfigure[Lithium Rydberg series]{\epsfig{file=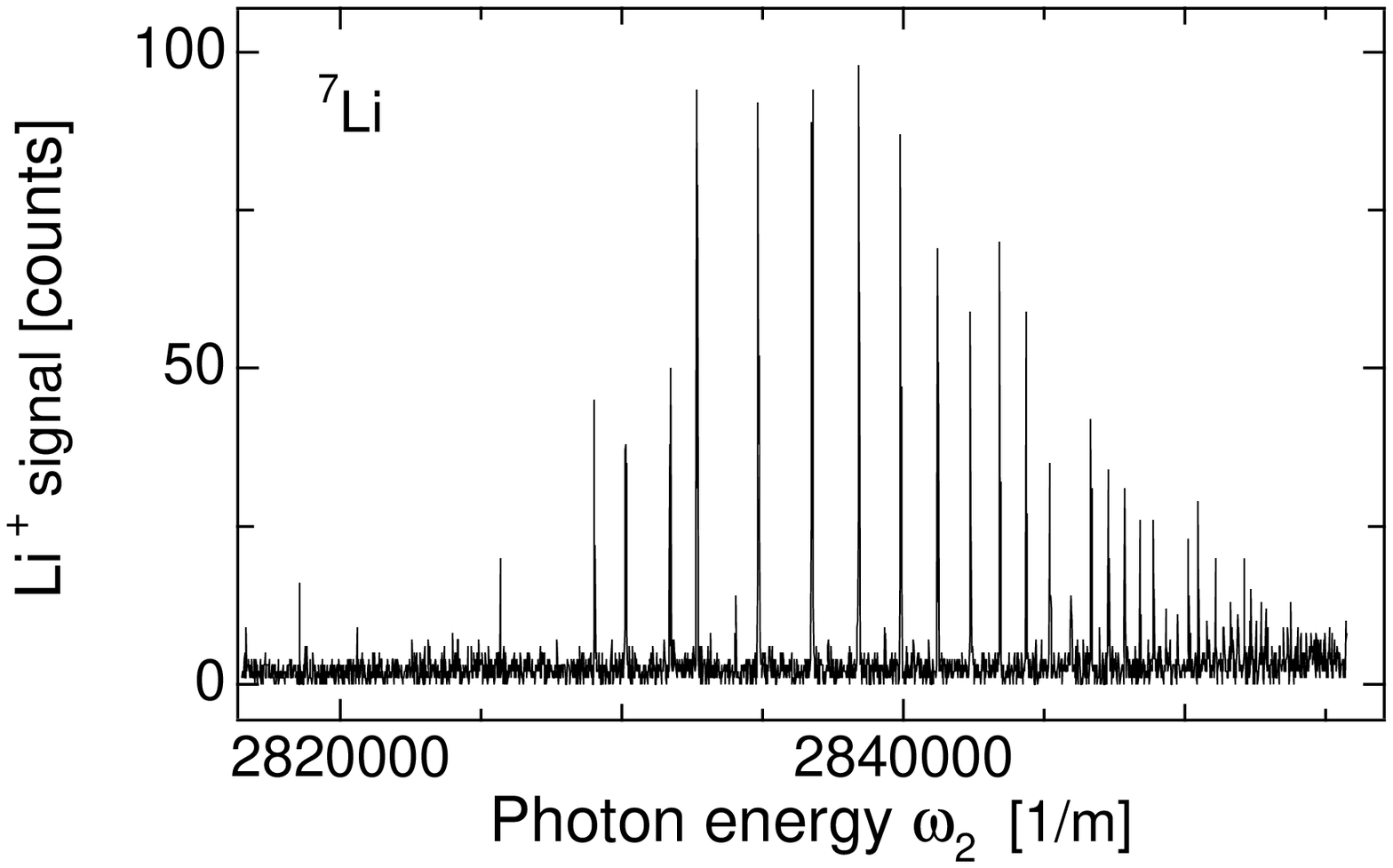,
width=0.495\textwidth}}\hfill
\subfigure[Lithium fine structure]{\epsfig{file=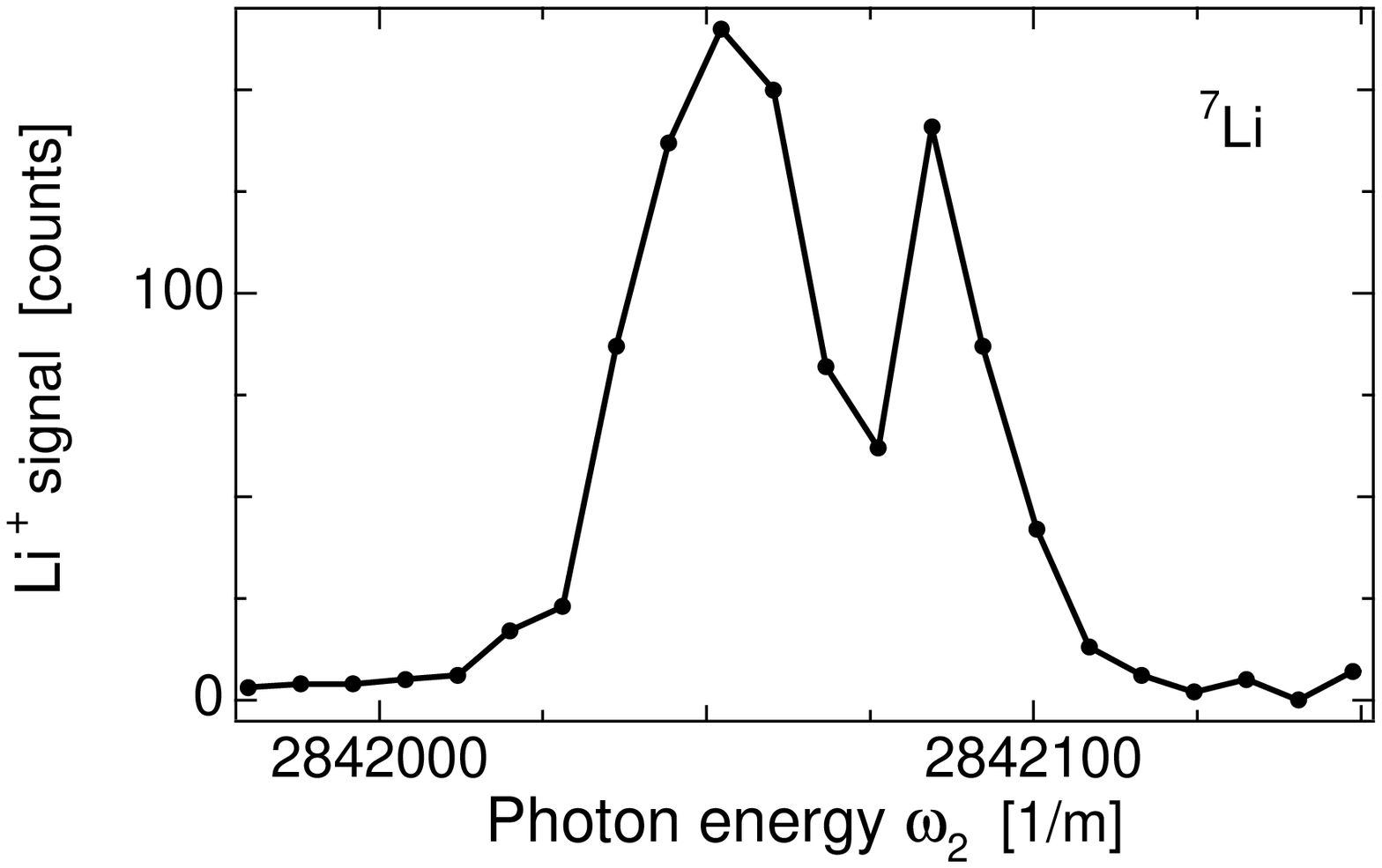,
width=0.495\textwidth}}
\protect\caption[Lithium Rydberg series and fine
structure]{\label{bi470}\sloppy Lithium Rydberg series and fine
structure: In part (a) a survey of the Rydberg states below the
ionisation limit of the Li atom.  Note how the series are terminated at
about 2\,827\,000~m$^{-1}$. The lines below that limit we interpret as
being due to photo-ionisation. In part (b) the spectrum of the fine
structure of the $2p\,^{2}\!P$ term is a magnification of the spectrum
depicted in the figure in part (a). The left peak is due to the
transition from the $J=3/2$ level and the right one is due to the
transition from the $J=1/2$ level.  For the threshold measurements,
$\omega_{2}$ was tuned to the peak of the $J=1/2$ component. The energy
scale is Doppler-shifted by 3100~m$^{-1}$ to the red.}
\end{figure}

To find an appropriate wavelength for the ionisation laser $\omega_{2}$
we tuned $\omega_{1}$ above the threshold and then tuned $\omega_{2}$
from the ionisation limit down. In figure \ref{bi470} the result of such
a scan is shown. Below a certain principal quantum number $n$ the field
ionisation ceases to be effective. In part (b) of the figure a
magnification of the subsequently used Rydberg transition is
displayed. The doublet structure stems from the fine structure of the
Li(2p) state with a splitting of 33.7~m$^{-1}$ \cite{Moo-71-1}. We found
from our data a value of 33(4)~m$^{-1}$ in agreement with the known
value. This, in addition, demonstrates the resolution of our machine to
only be limited by the laser bandwidth of 20~m$^{-1}$.

Light of frequency $\omega_{1}$ was generated by a dye laser operated
with Coumarin 307 and light of frequency $\omega_{2}$ was produced by a
dye laser operated with BMQ. Both dye lasers were pumped by a common
XeCl excimer laser (figure \ref{bi300}(a)).  The maximum energy in a
laser pulse delivered into the interaction region was 1.5~mJ for the
radiation of frequency $\omega_{1}$ and 200~$\mu$J for the radiation of
frequency $\omega_{2}$.  During the experiment both lasers were
attenuated, as will be discussed below.

\begin{figure}
\begin{minipage}{\textwidth}
\parbox[b]{0.64\textwidth}{
\epsfig{file=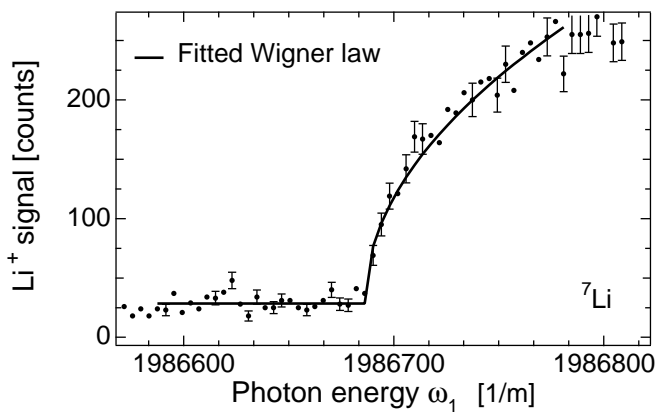, width=0.62\textwidth}
} \hfill
\parbox[b]{0.350\textwidth}{
\protect\caption[Li(2p+$\epsilon$s)
threshold]{\label{bi490}\sloppy\small Li(2p)+$\epsilon$s threshold:
Measurement of the partial relative photodetachment cross section of
Li$^{-}$ around the Li(2p) threshold. The use of counter-propagating
laser ($\omega_{1}$) and ion beams leads to a Doppler shift of
2\,169~m$^{-1}$ to the red. The solid line is a fit of the Wigner law
\eqref{eqch416} to the data in the range of the line. The error bars on
selected data points represent the shot noise. Each data point is
obtained from 100 laser pulses.}  }
\end{minipage}
\end{figure}

A typical measurement of the Li(2p) photodetachment threshold with
counter-propagating laser and ion beams is shown in figure
\ref{bi490}. The laser frequency $\omega_{2}$ was held constant and the
intensity was set to saturate the transition to the Rydberg state. The
frequency $\omega_{1}$ was then scanned over the Li(2p) threshold.

It was established that two processes contribute to the background,
namely (a),
\begin{equation}
\text{Li}^{-}(\text{2s}^{2}\,^{1}\text{S}) + \hbar\omega_{2} \mapsto
\text{Li}(2\text{p}\,^{2}\text{P}) + \text{e}^{-}
\end{equation}
followed by resonance ionisation according to \eqref{eqch64b},
\eqref{eqch64c} and (b) two-electron collisional ionisation. Process (a)
produced nearly 80\% of the background, even though the intensity of
laser $\omega_{2}$ was attenuated. Process (b), two-electron ionisation,
contributed the remaining background at the operating pressure of
5$\times$10$^{-7}$~Pa. This contribution was found to be proportional to
the pressure and is thus dominated by double detachment, as has been
discussed previously by Bae \cite{Bae-88}. The laser intensities were
too low for other processes, such as direct two-electron multi-photon
detachment, to influence the experiment.

To find the photodetachment threshold energy we fitted \eqref{eqch416}
to our data and obtained the threshold energy $E_{0}$ according to
\eqref{eqch51}. With aid of O'Malley's correction
\eqref{eqch418} we estimated the range of
validity of the Wigner law \eqref{eqch416}. For an electric dipole
polarisability of $\alpha_{\text{D}}$=152~a.u. for the Li(2p\,$^{2}$P)
state
\cite{Man-75} we found the second term in \eqref{eqch418} to be much
less than unity up to 100~m$^{-1}$ above the threshold, and consequently
the Wigner law should be applicable over this range. This, of course,
assumes that there are no resonances in this region, a condition that
has previously been established by Dellwo \cite{Del-92-2}. 

\begin{figure}
\begin{minipage}{\textwidth}
\parbox[b]{0.62\textwidth}{
\epsfig{file=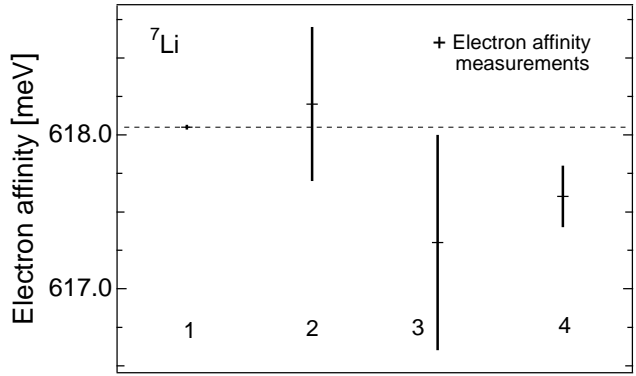, width=0.60\textwidth}
} \hfill
\parbox[b]{0.37\textwidth}{
\protect\caption[Lithium electron affinity]{\label{bi500}\sloppy
Lithium
electron affinity: Graphical comparison of the different experimental
lithium electron affinity values with their respective
uncertainties. From left to right the values are: \textsf{1} this work,
\textsf{2} Feldmann \cite{Fel-76}, \textsf{3} Bae \cite{Bae-85}, 
and \textsf{4} Dellwo \cite{Del-92-2}. See also table
\ref{tach62}.}
}
\end{minipage}
\end{figure}

Our value for the threshold energy $E_{0}$ is
\begin{equation}
E_{0} = 1\,988\,855(16)\; \text{m}^{-1}\quad .
\end{equation}
From this the lithium electron affinity is determined by subtracting the
well known 2p\,$^{2}$P$_{1/2}$\,$\mapsto$\,$2\text{s}\,^{2}$S transition
of 1\,490\,364.8130(14)~m$^{-1}$ \cite{San-95-2}. We find a lithium
electron affinity of 498\,490(17)~m$^{-1}$. In figure \ref{bi500} this
value is compared to previous measurements and in table \ref{tach62} we
also compiled a number of theoretical results.

\begin{table}[ht]
\begin{center}
\caption[Lithium electron affinity]{\label{tach62}
\slshape Lithium
electron affinity: Comparison of the different measurements and
calculations. The value in m$^{-1}$ is only given if the publication
contains the value.}\medskip
\begin{tabular}{lr@{.}lr@{.}l}
\hline\hline
Author & \multicolumn{2}{c}{Affinity in m$^{-1}$} &
\multicolumn{2}{c}{Affinity in meV} \\ \hline
\textsl{Experiment} & \multicolumn{4}{c}{}\\
Feldman (1976) \cite{Fel-76} & \multicolumn{2}{c}{} & 618&2(5) \\
Bae (1985) \cite{Bae-85} &\multicolumn{2}{c}{} & 617&3(7)\\ 
Dellwo (1992) \cite{Del-92-2} & 4&980\,9(16)$\times 10^{5}$ &
617&6(2) \\
This work \cite{Hae-96-1} & 4&984\,90(17)$\times 10^{5}$ & 618&049(20) \\
\textsl{Theory (after 1992)} & \multicolumn{4}{c}{}\\
Chung (1992) \cite{Chu-92-2} & \multicolumn{2}{c}{} & 617&4(2) \\
Froese Fischer (1993) \cite{Fis-93} & \multicolumn{2}{c}{} & 617&64\\
\hline\hline
\end{tabular}
\end{center}
\end{table}

We have demonstrated that photodetachment spectroscopy combined with
resonance ionisation is a powerful method for studying partial
photodetachment cross sections of negative ions.  The collinear beam
geometry simultaneously provides high sensitivity, due to the large
interaction volume, and excellent resolution, due to velocity
compression.  In addition, removal of the Doppler shift to all orders
can be achieved by use of two separate measurements involving co- and
counter-propagating laser and ion beams. Both merits can be fully
retained for channel-specific photodetachment investigations with the
excitation scheme used in this experiment. Our improved Li electron
affinity reveals the potential of this method, which, in principle, can
be extended to essentially all elements.

\chapter{Autodetachment resonances}
\label{ch63}
The study of resonance structure in the photodetachment cross section of
lithium is the first in a series of experiments
\cite{Han-97,Ber-95-1,Kli-97,Lju-96} aiming to deepen our knowledge of
doubly excited states in negative ions. In this experiment we used
neutral particle detection to unveil resonance structure near the Li(3p)
threshold. The experimental work was inspired and accompanied by
calculations of Pan \cite{Pan-96} and Lindroth
\cite{Ber-95-1,Lin-95}.

Doubly excited states allow us to investigate the correlated motion of a
pair of electrons and therefore extend our knowledge of the limits of
the independent particle approximation. The two prototype two-electron
systems, H$^{-}$ and He, have been studied for the last 30 years and a
great deal of insight in the nature of two-electron correlation has been
gained. With the work performed on autodetaching doubly excited states
of Li$^{-}$ and He$^{-}$ we studied systems where two highly excited
electrons interact with a rather simple singly charged core. Comparison
of H$^{-}$ and He with Li$^{-}$ and He$^{-}$ reveals similarities as
well as differences related to the finite core of Li$^{-}$ and
He$^{-}$. These simple negative ions are fairly tractable to theory, and
therefore supply valuable insight in the strive of understanding the
more general many-body problem.

\section{Li$^{-}$ doubly excited states}
\label{ch631}
Here we report on the first observation of resonance structure in the
total cross section for the photodetachment of Li$^-$. The resonances
are of $^1$P$^{\text{o}}$ final state symmetry and are optically coupled
to the ground state.  They arise from the photo excitation and
subsequent autodetaching decay of doubly-excited states of Li$^-$ that
are embedded in continua representing an excited Li atom and a free
electron.  The resonances were observed to lie in the energy region
between the Li(3s) and Li(3p) thresholds.  Recently Pan, Starace and
Greene used an eigenchannel R-matrix method to predict the shape of the
photodetachment cross section in the vicinity of the Li($n$=3) threshold
\cite{Pan-96} and the Li($n$=4,5,6) thresholds
\cite{Pan-94}.
The measurements reported here present us with an opportunity to compare
experimental data on Li$^-$ with corresponding data on H$^-$
\cite{Ham-79}.  Differences in the spectra of Li$^-$ and H$^-$
associated with the lifting of the degeneracy characteristic of the
hydrogen atom are apparent as are certain similarities \cite{Lin-95}.  I
also present calculations on Li$^-$ and H$^-$ by Lindroth
\cite{Ber-95-1} which are used to
explain the origin of the observed resonances.  The method accounts for
full correlation between the outer electrons. The autodetaching decay of
doubly excited states is treated using complex rotation.

\begin{figure}
\begin{minipage}{\textwidth}
\parbox[b]{0.754\textwidth}{
\epsfig{file=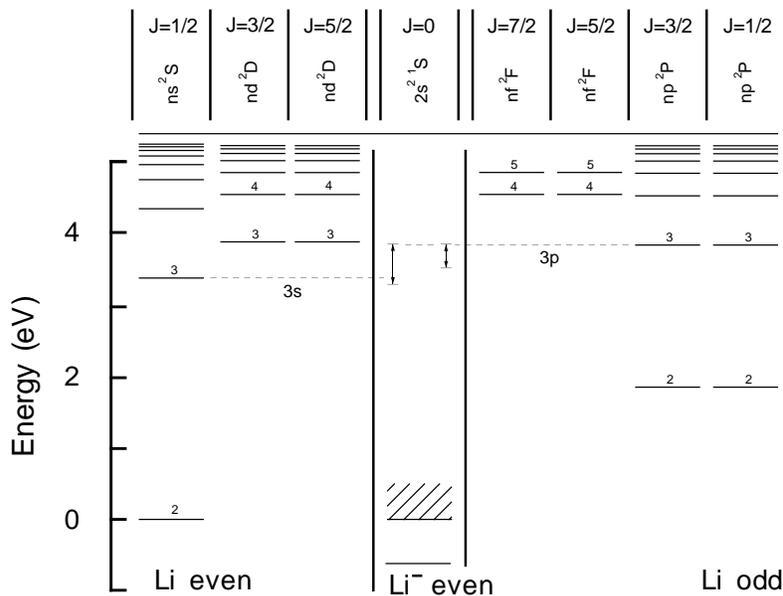, width=0.74\textwidth}
} \hfill
\parbox[b]{0.236\textwidth}{
\protect\caption[Li$^{-}$ excitation scheme 
\textsc{II}]{\label{bi510}\sloppy
\footnotesize
Li$^{-}$ excitation scheme \textsc{II}: Selected bound states 
of Li/Li$^{-}$
grouped according to their parity and total angular momentum. The left
arrow indicates the range of the theoretical curve in figure \ref{bi520}
and the right one the range of our data in figure \ref{bi520}.}  }
\end{minipage}
\end{figure}

The Li$^{-}$ ions were generated in the plasma ion source, but, in
contrast to the normal procedure, the negative ions were \emph{directly}
extracted from the discharge because the charge exchange cell was not
yet operative.  This is less efficient than the usual method of charge
exchange of positive ions and consequently gave a current not exceeding
1~nA in the interaction region.

The light required for the photodetachment was generated by frequency
doubling the output of our dye laser operated with
Rhodamine~6G and Coumarin 153. A normalisation to the laser power was
performed where ever necessary.

In figure \ref{bi510} the tuning range of the experiment is
indicated. The excitation process can be summarised as
\begin{equation}\label{eqch66}
\text{Li$^{-}$(1s$^{2}$2s$^{2}$\,$^{1}$S)} + \hbar\omega \mapsto
\text{Li} + \text{e}^{-}\quad .
\end{equation}

\begin{figure}\centering
\subfigure[Measured and calculated photodetachment cross section near
the Li(3p) threshold]{\epsfig{file=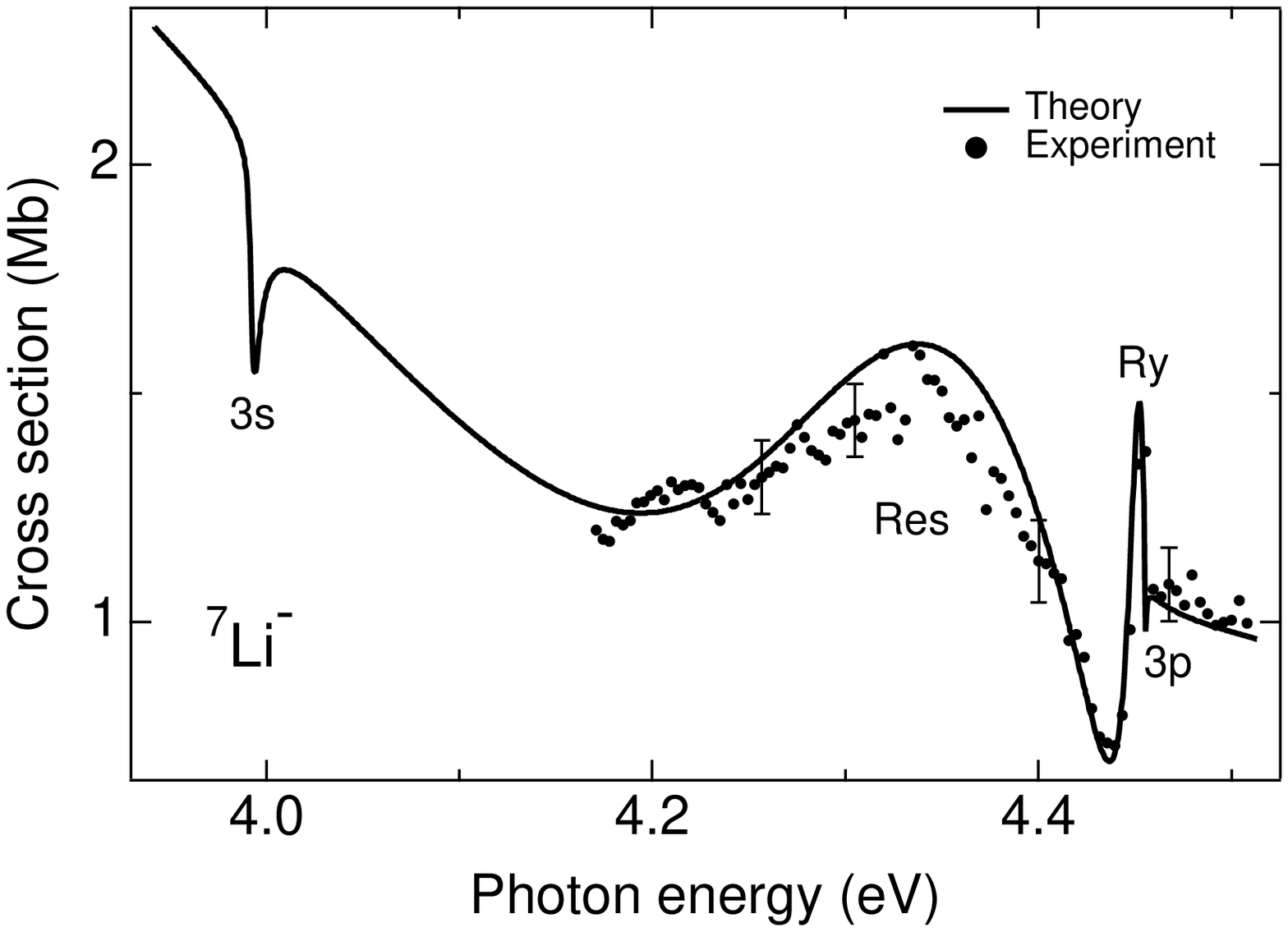,
width=0.49\textwidth}}\hfill
\subfigure[Magnification near the Li(3p) cusp]{\epsfig{file=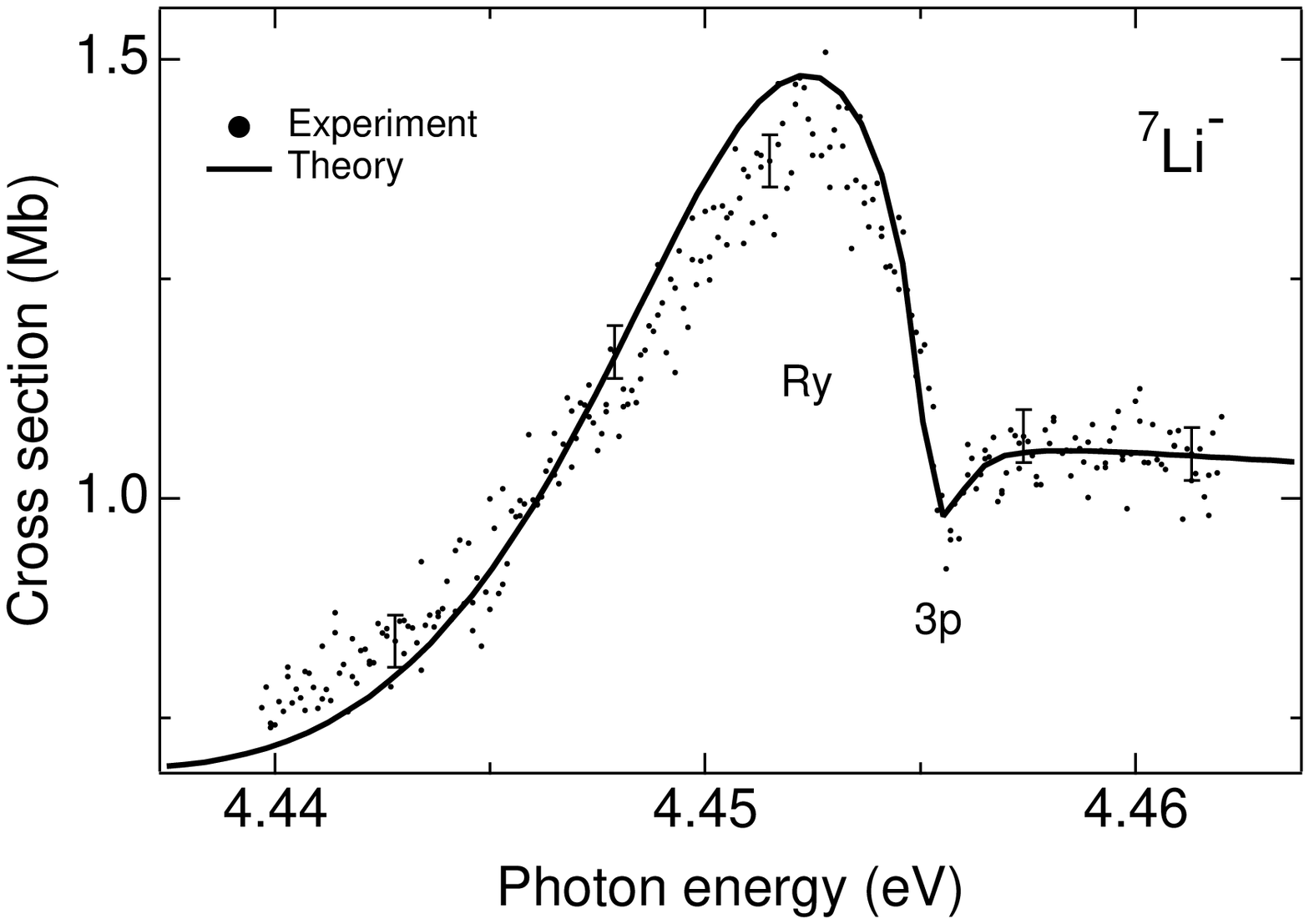,
width=0.49\textwidth}}
\protect\caption[Photodetachment cross section near the
Li(3p) threshold]{\label{bi520}\sloppy Lithium total photodetachment
cross section near the Li(3p) threshold: 
The solid line is a calculated absolute photodetachment cross
section and the dots are the measured relative photodetachment cross
section, which is scaled, but not fitted, to match the calculated value
at the Li(3p) threshold. The labels \textsf{3s} and \textsf{3p} mark the
Li(3s) and Li(3p) thresholds, and \textsf{Res,Ry} mark a broad
resonance and a Rydberg like resonance (see tabular \ref{tach63}). The
error bars represent one estimated standard deviation. Note how the data
follow the sharp Wigner cusp.}
\end{figure}
The experimentally determined total photodetachment cross section is
shown as the dots in figure \ref{bi520}. All data points have been
normalised to the theoretically calculated value at the Li(3p)
threshold. Also the energy scale has been shifted so that the Li(3p)
threshold of the data and the calculation coincide.  The scatter in the
data is primarily caused by the small but unavoidable change in the
overlap of the ion and laser beams as the wavelength of the laser was
scanned. The normalisation procedure applied in this experiment was
unable to completely account for beam overlap variations. We estimated,
however, that this effect causes a maximum error in the relative cross
section over the whole spectrum of less than 10\%. The statistical
scatter, mainly caused by the shot noise in counting of the neutral
particles, is less than 3\%.

The experimental data exhibit three significant features. First there is
the Wigner cusp, \textsf{3p} figure \ref{bi520}, at the Li(3p)
threshold, second there is a narrow resonance structure \textsf{Ry}
and finally a very broad resonance \textsf{Res} occupies a large part of
the region between the Li(3s) and Li(3p) thresholds. 

The calculated cross section is shown as a solid line and is in close
agreement with another recent calculation \cite{Pan-96,Lin-95}.  The
presented calculation predicts a resonant state of 0.36~eV width to lie
4.32~eV above the ground state of Li$^-$ as shown in table
\ref{tach63}.
\begin{table}
\begin{center}
\caption[Li$^{-}$ doubly excited states]{\label{tach63}\slshape
Li$^{-}$ doubly excited states: Comparison of theoretical results for
intra-shell $^{1}$P$^{\text{o}}$ states of H$^{-}$ below the H($n$=3)
threshold and of Li$^{-}$ below the Li(3p) threshold. The binding energy
$E_{\text{bind}}$ is relative to the double detachment limit and
$E_{\text{trans}}$ is the transition energy from the ground
state. Calculated resonance parameters for the two Rydberg-like
resonances, marked \textsf{Ry} in figure \ref{bi520}(b), below the
Li(3p) threshold are given. Note the strong overlap of these two
resonances.}\medskip
\begin{tabular}{lcr@{.}lr@{.}lr@{.}l}
\hline\hline
Element & Label & \multicolumn{2}{c}{$E_{\text{bind}}$ (eV)} & 
\multicolumn{2}{c}{$\Gamma$ (eV)} 
 & \multicolumn{2}{c}{$E_{\text{trans}}$ (eV)} \\ \hline
H$^{-}$ & & $-1$&706 & 0&033 & 12&647\\
Li$^{-}$ & \textsf{Res} & $-1$&688 & 0&36  & 4&322 \\
Li$^{-}$ & \textsf{Ry} & $-1$&556 & 0&020 & 4&453 \\
Li$^{-}$ & \textsf{Ry} & $-1$&555 & 0&016 & 4&454 \\
\hline\hline
\end{tabular}
\end{center}
\end{table}
This state appears to be analogous to the symmetrically excited
intra-shell $(_3\{0\}^+_3)$\,$^1$P$^{\text{o}}$
\cite{Hei-93}
state in H$^-$. It has a similar binding energy relative to the double
detachment limit, but is about one order of magnitude broader.  The
broadening of the resonance in Li$^-$ arises from the strong coupling to
the 3s$\varepsilon\text{p}$ continuum, which is not available below
H($n$=3) in H$^-$.  The resonant state in Li$^-$ is dominated by the
configurations 3p3d and 4s3p and there appears to be no significant
contribution to the localised part of the wave function from
configurations with one electron in the 3s orbital, which is also in
contrast to the case of H$^-$.

While the H$^-$ intra-shell resonance is well described by a Fano
profile the width of the resonant state in Li$^-$ is too broad for this
to be possible. The calculated width of 0.36~eV is overlapping the 3p
threshold as well as the narrow resonance seen just below it in figure
\ref{bi520}. The latter resonance is due to asymmetrically excited
Rydberg-like states.  In these states one of the electrons in the doubly
excited pair occupies a more loosely bound orbital than the other.  The
presence of the threshold and the interference between the resonances
affects the shape of the cross section curve significantly.  The
assumption necessary to obtain a Fano profile, that only one state
determines the shape, is thus not valid.  The interference results in a
narrowed structure in the spectrum, especially on the high energy side,
compared to that which would arise from a hypothetically isolated doubly
excited state of width 0.36~eV.  Rydberg-like resonant states are also
apparent in the calculated spectrum of H$^-$.  These states are bound
relative to the H($n$=3) limit by the strong dipolar field between the
two electrons.  The strength of this field is due to the degeneracy of
the H(3$\ell$) states which results in nearly equal admixture of 3snp
and 3pns (or 3pns and 3dnp) configurations in the composition of the
Rydberg states.

The Rydberg-like states \textsf{Ry} in Li$^-$, however, do not have this
character.  They are completely dominated by 3pn$\ell$ configurations,
with $n \gg 3$, and the dipolar field is in this case insignificant. The
explanation for the existence of the Rydberg states in Li$^-$ is the
inability of the monopole part of the electron-electron interaction to
screen the singly-charged core completely. The residual nuclear
attraction binds the states below the Li(3p) threshold.

We have studied the total photodetachment cross section of the Li$^{-}$
ion in the energy region between the Li(3s) and Li(3p) thresholds. This
region is dominated by a broad resonance. A narrower resonance structure
lies just below the Li(3p) threshold. By comparing the data with
calculated photodetachment cross sections for Li$^{-}$ and H$^{-}$ we
have been able to identify the broad resonance in the Li$^{-}$ spectrum
as being associated with the presence of a symmetrically excited
intra-shell doubly excited state, analogous to, but much broader than,
the intra-shell state in H$^{-}$. The sharper resonance structure is
identified with asymmetrically excited Rydberg like states.

\section{He$^{-}$ doubly excited states}
\label{ch632}
Another, and perhaps even more exciting, aspect of the state selective
detection method is the possibility to seperately monitor partial cross
sections. This is especially useful to investigate high lying
resonances. The modulation of the total photodetachment cross section
due to autodetaching states can become very small \cite{Pan-96,Xi-96-2},
but in certain partial cross sections there may still occur quite
pronounced modulation. Therefore this state selective detection scheme
has rapidly become an indispensible tool for studies of negative ions.

The expriments on He doubly excited states, in particular, profited from
both the backgound suppression and enhanced modulation of partial cross
sections. The background suppression proved to be truly vital since
the decay of the He$^{-}$ ground state already leads to a background of
neutral helium, thus rendering any investigation with neutral particle
detection impossible. Earlier ill fated attempts to measure resonance
structure near the He($n$=3) thresholds in this laboratory suffered from
saturation of the detection system caused by excessive neutral particle
background. As is further elaborated in this section, we fard much
better with the state selective detection now at our disposal.

The metastable He$^{-}$ ion has received considerable attention since
its discovery by Hiby in 1939 \cite{Hib-39}. This simple three
electron ion is the prototype of an unusual class of negative ions that
are not stable, but rather metastable, against autodetachment. The
lowest lying state is the 1s2s2p\,$^{4}$P$^{\mathrm{o}}$ state, which is
bound by 77.524(5)~meV relative to the 1s2s\,$^{3}$S state of He
\cite{Pet-96-3} in agreement with the calculated value of 77.51(4) by
Bunge \cite{Bun-79}. The He$^{-}$ ion in this spin aligned quartet state
cannot radiate and, since it is embedded in a doublet continuum, it can
only autodetach via the relatively weak magnetic interactions (table
\ref{tach31}). The varying strengths of these spin dependent
interactions result in a differential metastability among the three fine
structure levels. The longest lived $J$=$5/2$ level which can only decay
via the spin-spin interaction (table \ref{tach31}) has, for example, a
lifetime of 350(15)~$\mu$s \cite{And-93}. Metastable He$^{-}$ ions are
therefore sufficiently long lived to pass through a typical apparatus
with only minor depletion by autodetachment.

Excited states of He$^{-}$, on the other hand, decay rapidly via Coulomb
autodetachment. Their presence is manifested as resonance structure in
scattering cross section close to thresholds for new channel openings,
i.~e.~the excited state energies of the He atom. Many doublet resonances
have been observed, for example, as transient intermediate states in
studies of electron impact on atomic He targets \cite{Buc-94}. Excited
quartet states of He$^{-}$, however, have hitherto received far less
attention. Such states appear as resonances in the photodetachment cross
section and, to a lesser extent, in cross section for detachment via
heavy particle collisions. Selection rules on photo-excitation from the
1s2s2p\,$^{4}$P$^{\mathrm{o}}$ ground state of He$^{-}$ allow
transitions only to $^{4}$S, $^{4}$P and $^{4}$D excited states. There
have been several experimental \cite{Com-80,Hod-81,Peg-90} and
theoretical studies \cite{Haz-81,Sah-90,Dou-90} of the photodetachment
cross section of He$^{-}$. Most recently, Xi and Froese Fischer
\cite{Xi-96} have calculated the position and width of quartet states of
He$^{-}$ up to the He($n=4$) thresholds.

\begin{figure}\centering
\subfigure[Monitoring on 1s2s\,$^{3}$S]{\epsfig{file=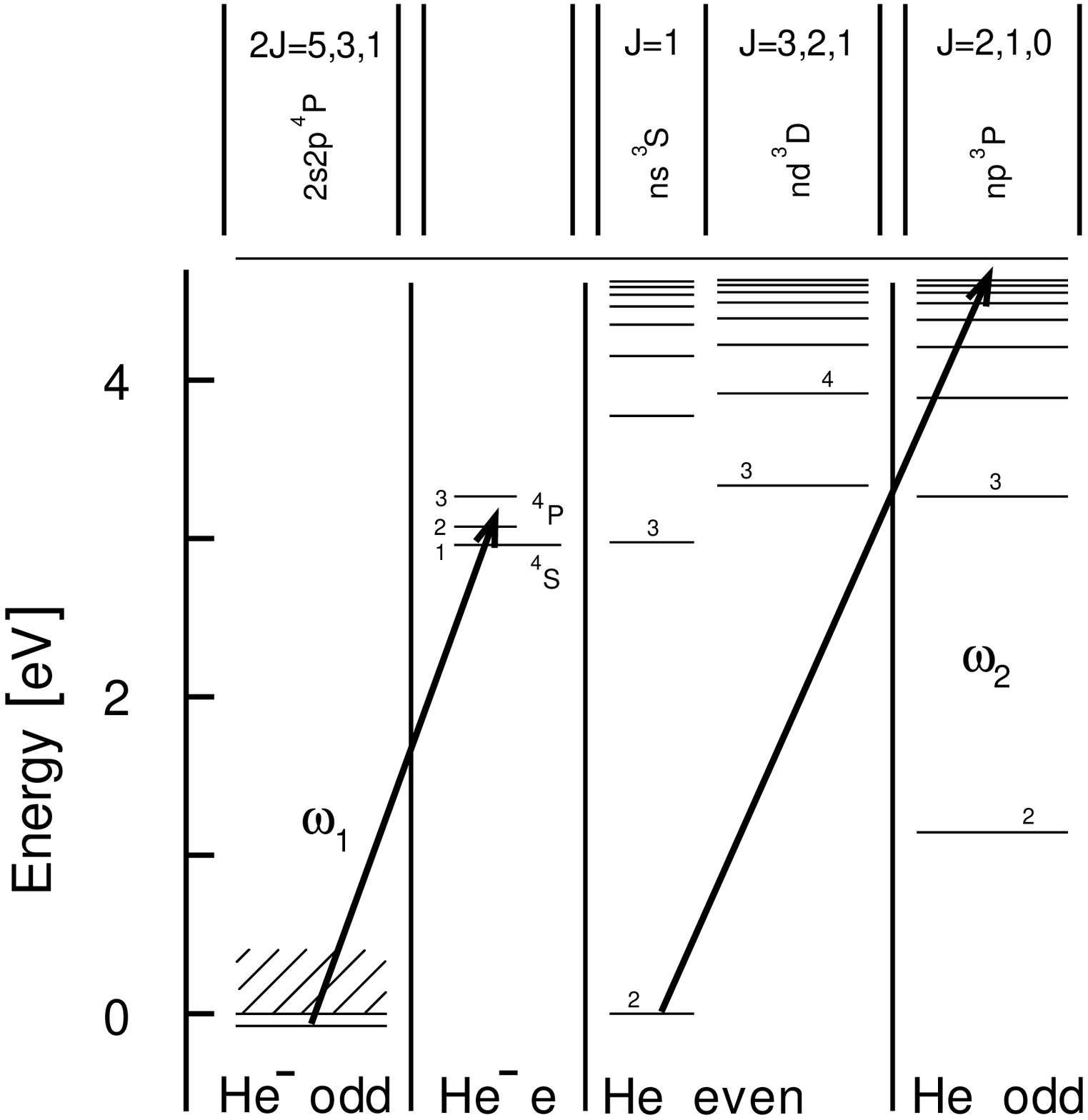,
width=0.48\textwidth}}\hfill
\subfigure[Monitoring on 1s2p\,$^{3}$P$^{o}$]{\epsfig{file=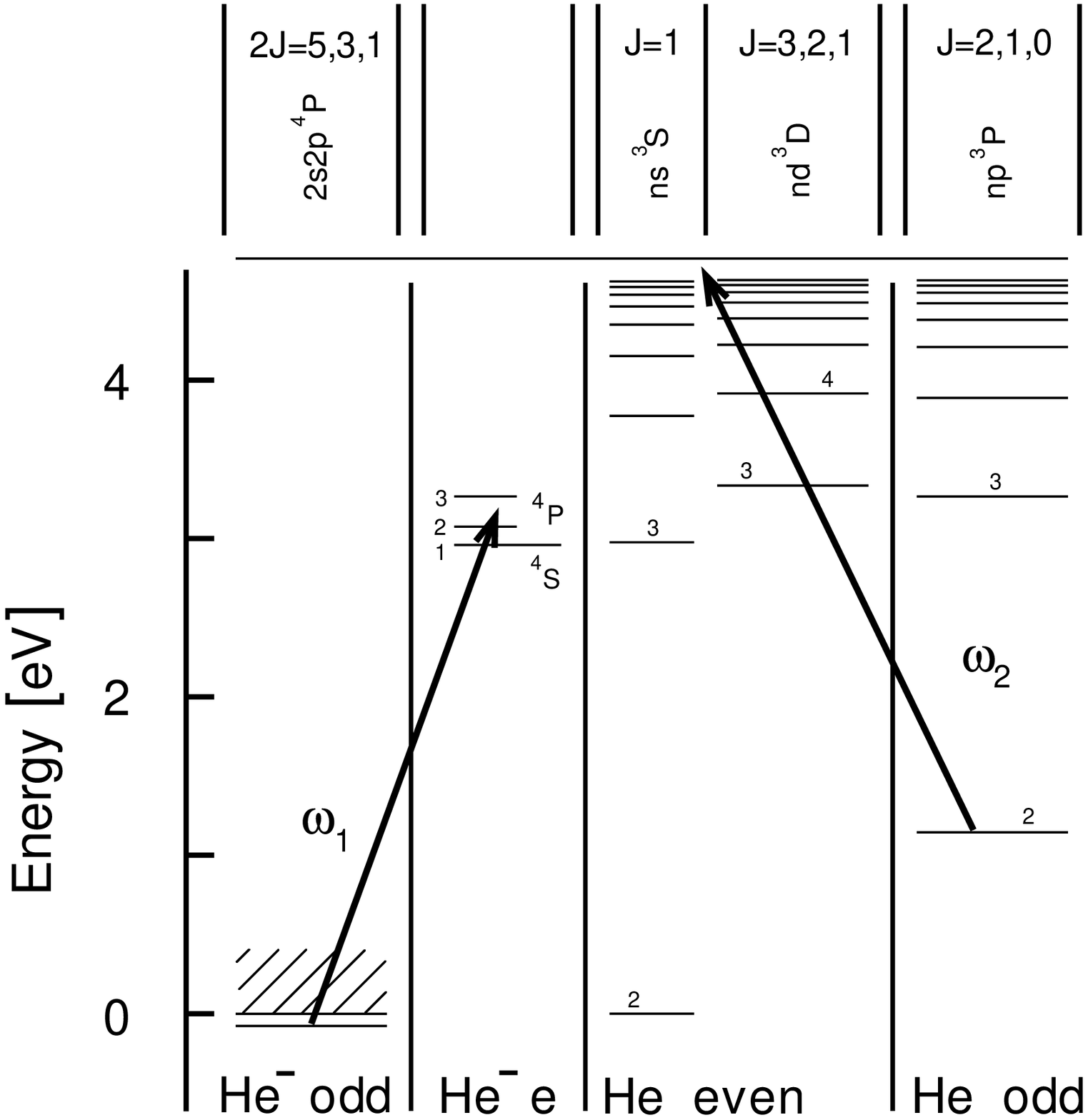,
width=0.48\textwidth}}
\protect\caption[He$^{-}$ excitation scheme]{
\label{bi560}\sloppy He$^{-}$ excitation scheme: 
Selected states of He/He$^{-}$ grouped according to their parity
and total angular momentum. The solid arrows represent the transitions
induced in the measurements, see also \eqref{eqch67a} to \eqref{eqch67f}
and figure \ref{bi580}. The designation of the He$^{-}$ levels is,
\textsf{1}, 1s3s4s\,$^{4}$S; \textsf{2}, 1s3p$^{2}$\,$^{4}$P;
\textsf{3}, 1s3p4p\,$^{4}$P; according to \cite{Xi-96}.}
\end{figure}

To investigate the He$^{-}$ photodetachment cross section we use the
previously introduced (section \ref{ch622}) two-colour state selective
excitation scheme. The specific steps for the excitations are depicted
in figure \ref{bi560}. For all resonances the yield of He$^{+}$ ions in
the state selective excitation was recorded as a function of
$\omega_{1}$, while the frequency $\omega_{2}$ was held constant on the
transition to the Rydberg state. During a scan we assume the laser power
of $\omega_{2}$ to be constant. If the resonance is narrow also
$\omega_{1}$ is assumed to stay constant, otherwise we account for the
varying light intensity by a normalisation to the measured laser power.

The detection scheme based on the selective detection of residual atoms
was effective in eliminating a potential background source unique to
measurements involving metastable negative ions. Since He$^{-}$ ions are
non-stable, the ion beam will contain a fraction of He atoms produced,
in flight, by autodetachment. These helium atoms will, however, be in
the He(1s$^{2}$\,$^{1}$S) ground state and will therefore not be
resonantly photo-ionised.

The state selectivity allows us to concentrate our measurement efforts
on the partial cross sections that actually contain resonance
structure. The autodetachment selection rules, given in table
\ref{tach31} on page \pageref{tach31}, impose restrictions on
which partial cross sections can be expected to contain resonance
structure.

The $^{4}$He$^{-}$ beam was produced from a mass selected He$^{+}$ ion
beam via charge exchange in a cesium vapour cell.  The beam energy was
3.1 keV. A current of typically 1 nA was obtained in the interaction
region.

\begin{figure}
\begin{minipage}{\textwidth}
\parbox[b]{0.50\textwidth}{
\epsfig{file=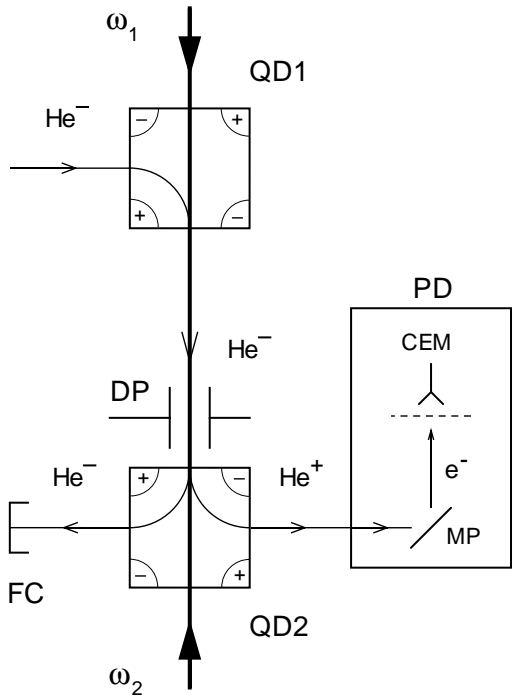, width=0.48\textwidth}
} \hfill
\parbox[b]{0.49\textwidth}{
\protect\caption[Modified interaction-detection
chamber]{\label{bi550}\sloppy Modified interaction-detec\-tion chamber:
\textsf{QD1,QD2}, electrostatic quadrupole deflectors; \textsf{CEM},
channel electron multiplier; \textsf{DP}, deflection plates;
\textsf{PD}, positive ion detector; \textsf{FC}, Faraday cup;
\textsf{MP}, metal plate. The deflection plates \textsf{DP} serve to
sweep collisionally created positive ions out of the beam, thereby
improving the signal to noise ratio. They are periodically grounded to
monitor the negative ion current. Ion- and laser beams were merged in
the 0.5~m long interaction region between the quadrupole deflectors. }
}
\end{minipage}
\end{figure}

The apparatus was designed to reduce the background of He$^{+}$ ions
produced by double collisional detachment by installing a pair of
deflection plates (\textsf{DP} figure \ref{bi550}) just before the
second quadrupole deflector (\textsf{QD2}).  The transverse electric
field between the deflection plates was insufficient to field ionise the
Rydberg atoms, but strong enough to sweep collisionally created He$^{+}$
ions out of the beam.
%

\minisec{The 1s3s4s\,$^{\mathbf{4}}$S state}
To study the 1s3s4s\,$^{4}$S resonance we used an excitation scheme,
depicted in figure \ref{bi560}, that can be summarised by the following
equations for state selective detection on He(1s2s\,$^{3}$S), 
\begin{align}
\text{He$^{-}$(1s2s2p\,$^{4}$P$^{\mathrm{o}}$)} 
+ \hbar \omega_{1}
& \mapsto
\text{He(1s2s\,$^{3}$S)} + \text{e}^{-}(\epsilon\text{s})\quad ,
\label{eqch67a} 
\\ 
\label{eqch67b} \text{He(1s2s\,$^{3}$S)} + 
\hbar 
\omega_{2} & \mapsto  \text{He(1s24p\,$^{3}$P$^{\mathrm{o}}$)}\quad
, \\ \label{eqch67c} \text{He(1s24p\,$^{3}$P$^{\mathrm{o}}$)} &
\leadsto
\text{He$^{+}$(1s\,$^{2}$S)} + \text{e}^{-}\quad ,\\
\intertext{and for
state selective detection of He(1s2p\,$^{3}$P$^{\mathrm{o}}$),}
\text{He}^{-} \text{(1s2s2p\,$^{4}$P$^{\mathrm{o}}$)} + \hbar \omega_{1}
& \mapsto \text{He(1s2p\,$^{3}$P$^{\mathrm{o}}$)} +
\text{e}^{-}(\epsilon\text{p})\quad , \label{eqch67d}\\ \label{eqch67e}
\text{He(1s2p\,$^{3}$P$^{\mathrm{o}}$)} + 
\hbar 
\omega_{2} & \mapsto  \text{He(1s25d\,$^{3}$D)}\quad
, \\ \label{eqch67f} \text{He(1s25d\,$^{3}$P)} & \leadsto
\text{He$^{+}$(1s\,$^{2}$S)} + \text{e}^{-}\quad ,
\end{align}
where $\leadsto$ in both cases represents field ionisation.

The calibration of the energy scale is done as described in section
\ref{ch52} using the four transitions in argon given in table
\ref{tach64}.

\begin{table}
\begin{minipage}{\textwidth}
\protect\caption[He$^{-}$(1s3s4s\,$^{4}$S) calibration lines]{
\label{tach64}\sloppy He$^{-}$(1s3s4s\,$^{4}$S) calibration lines
\cite{Min-73}:  
The argon levels are designated in Paschen notation.  For
counter-propagating laser ($\omega_{1}$) and ion beams we used the lines
a, b and c and for co-propagating laser ($\omega_{1}$) and ion beam we
used the lines b, c and d for calibration.}
\begin{center}
\medskip
\begin{tabular}{lcc}\hline\hline
\# & Transition & Line (m$^{-1}$) \\
\hline
a  & 3p$_{5}$ $\rightarrow$ 1s$_{4}$ & 2\,381\,235.9(2) \\
b  & 3p$_{4}$ $\rightarrow$ 1s$_{3}$ & 2\,385\,376.7(2) \\
c  & 3p$_{8}$ $\rightarrow$ 1s$_{5}$ & 2\,385\,556.5(2) \\
d  & 3p$_{2}$ $\rightarrow$ 1s$_{3}$ & 2\,390\,593.3(2) \\
\hline\hline
\end{tabular}
\end{center}
\end{minipage}
\end{table}

\begin{figure}\centering
\subfigure[Measured He(1s2s\,$^{3}$S) and
He(1s2p\,$^{3}$P$^{\text{o}}$) partial cross
sections]{\epsfig{file=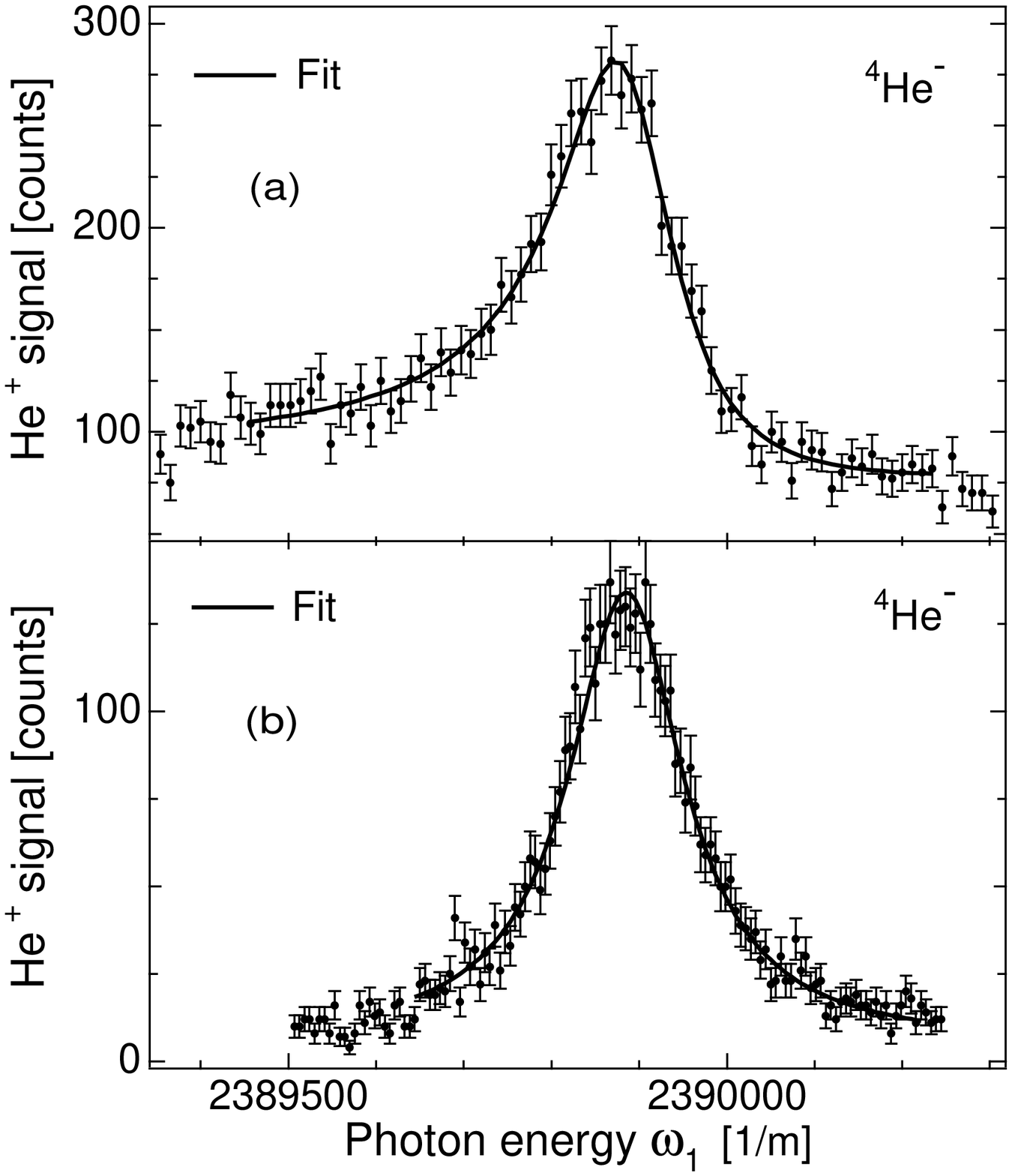, width=0.52\textwidth}}\hfill
\subfigure[Calculated partial cross
section]{\parbox[b]{0.42\textwidth}{\epsfig{file=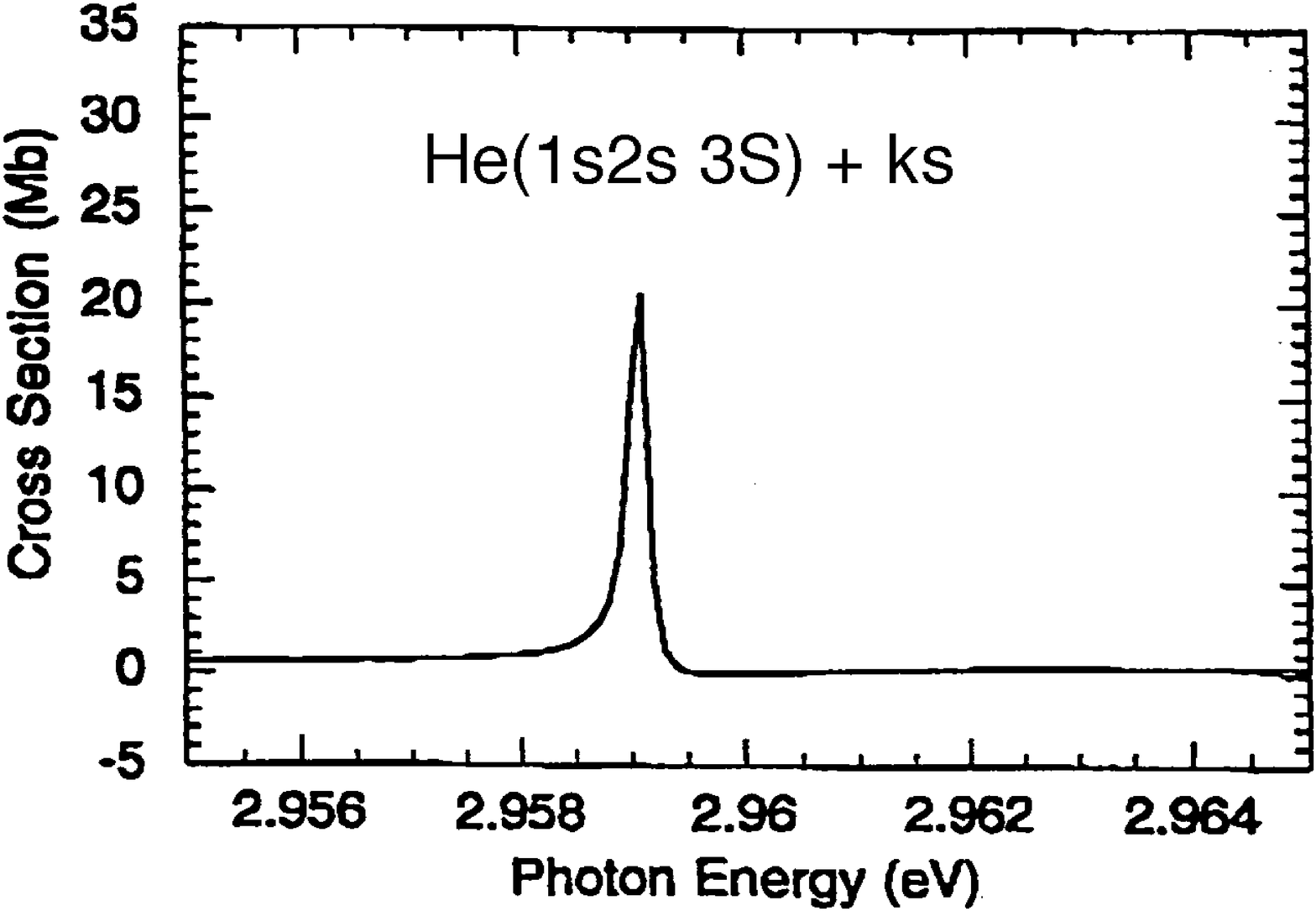,
width=0.42\textwidth}\vspace*{0.35ex}
\epsfig{file=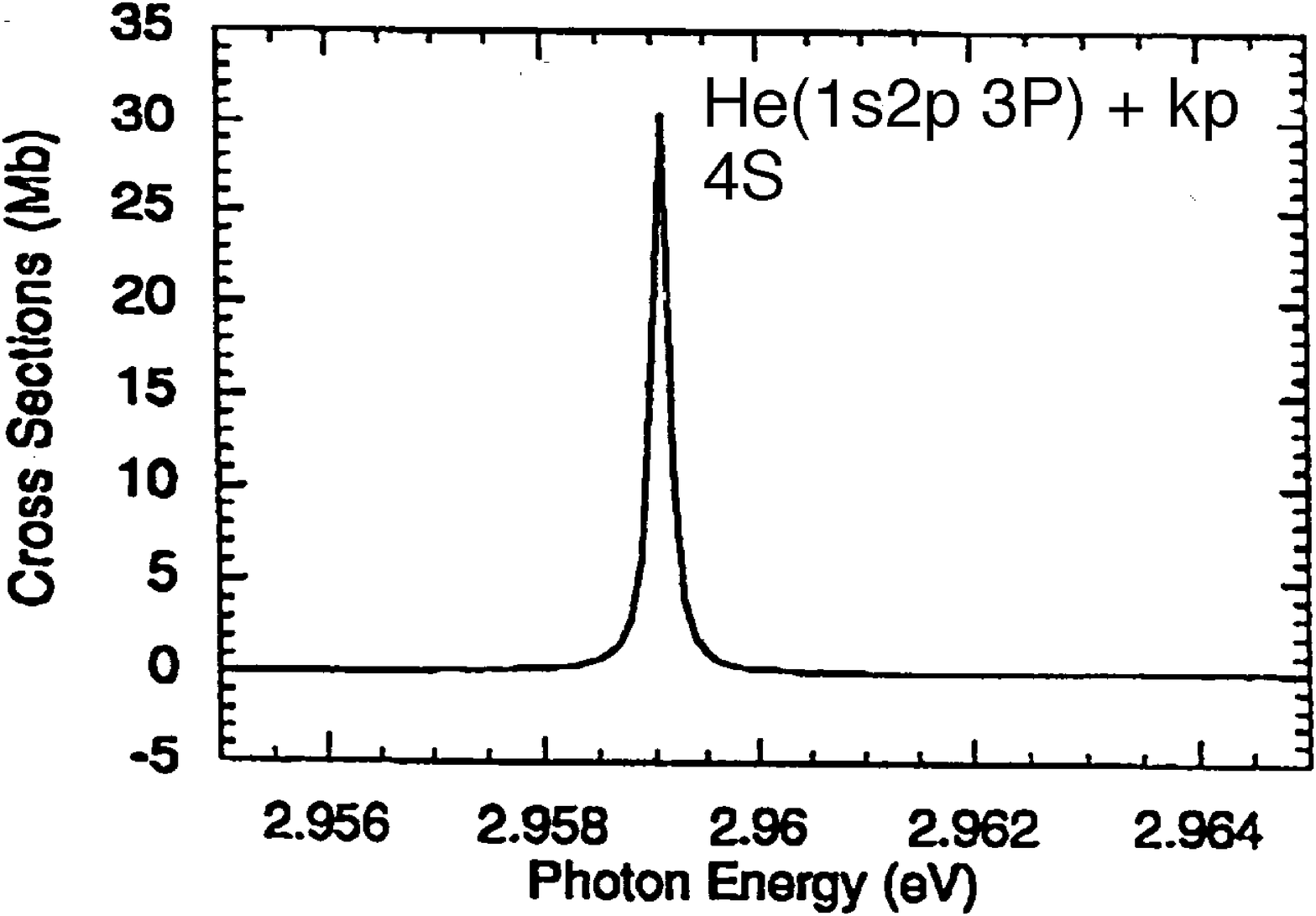, width=0.42\textwidth}}}
\protect\caption[Partial photodetachment cross sections near
the He$^{-}$(1s3s4s\,$^{4}$S) state]{\label{bi580}\sloppy Partial
photodetachment cross section near the He$^{-}$(1s3s4s\,$^{4}$S) state:
In the upper graph of part (a) we show a measurement of the
He(1s2s\,$^{3}$S)+e$^{-}$($\epsilon$s, $\epsilon$d) partial
photodetachment cross sections of He$^{-}$ in the vicinity of the
1s3s4s\,$^{4}$S resonance, and in the lower graph of part (a) a
measurement of the He(1s2p\,$^{3}$P$^{\text{o}}$)+e$^{-}$($\epsilon$p)
partial photodetachment cross sections of the same state. The use of
co-propagating laser ($\omega_{1}$) and ion beams causes the resonance
to be Doppler shifted by 3\,089 m$^{-1}$ to the blue. The solid line is
a fit to the data using equation
\eqref{eqch421}.  The error bars represent the shot noise.  Each
data point represents 200 laser pulses in the upper graph of part (a)
and 100 laser pulses in the lower graph of part (a). For a comparison of
width and position see figure \ref{bi650},\ref{bi670} and table
\ref{tach65}. The partial cross sections in part
(b) are calculated by Xi \cite{Xi-96-2}.}
\end{figure}

A typical measurement of the 1s3s4s\,$^{4}$S resonance in the
He(1s2s\,$^{3}$S) partial cross sections and the
He(1s2p\,$^{3}$P$^{\mathrm{o}}$) partial cross sections is shown in
figure \ref{bi580}. For the He(1s2s\,$^{3}$S) cross sections we found
three processes to significantly contribute to the background: firstly
(a), collisional detachment leaving the He atom in the 1s2s\,$^{3}$S
state, and secondly (b), photodetachment by the laser $\omega_{2}$
\begin{equation}\label{eqch68}
\text{He$^{-}$(1s2s2p\,$^{4}$P$^{\mathrm{o}}$)} + \hbar
\omega_{2} \mapsto  \text{He(1s2s\,$^{3}$S)} +
\text{e}^{-}(\epsilon\text{s},\epsilon\text{d})\quad ,
\end{equation}
followed by resonance ionisation indicated \eqref{eqch67b} and
\eqref{eqch67c}. Thirdly (c), non-resonant photodetachment
by the laser of frequency $\omega_{1}$,
\begin{equation}\label{eqch69}
\text{He$^{-}$(1s2s2p\,$^{4}$P$^{\mathrm{o}}$)} + \hbar
\omega_{1} \mapsto  \text{He(1s2s\,$^{3}$S)} +
\text{e}^{-}(\epsilon\text{d})\quad ,
\end{equation}
and then proceeding as in \eqref{eqch67b} and \eqref{eqch67c}.

To reduce the collisional detached contribution (a) we maintained a
pressure of $5\times 10^{-7}$~Pa  in the
interaction chamber. To reduce contribution (b) we attenuated the
output of laser $\omega_{2}$.

The selection rule $\Delta L$=$0$ for Coulomb autodetachment (table
\ref{tach31} page \pageref{tach31}) in
LS-coupling forbids the 1s3s4s\,$^{4}$S resonance from appearing in the
He(1s2s\,$^{3}$S)+e$^{-}$($\epsilon$d) partial cross section. Thus, the
process (c), as represented by \eqref{eqch69}, contributes a constant
background over the region shown in the upper part of figure
\ref{bi580}(a).  Xi and Froese Fischer \cite{Xi-96-2} predict a d-wave
photodetachment cross section of 6 Mb across the resonance, figure
\ref{bi690}(c) page \pageref{bi690}, and a peak
s-wave cross section of approximately 20 Mb, upper part of figure
\ref{bi580}(b).

For the He(1s2p\,$^{3}$P$^{\mathrm{o}}$) cross sections we found two
processes to significantly contribute to this background: firstly (a),
collisional detachment leaving the He atom in the
1s2p\,$^{3}$P$^{\mathrm{o}}$ state, and secondly (b), photodetachment by
the laser $\omega_{2}$
\begin{equation}\label{eqch610}
\text{He$^{-}$(1s2s2p\,$^{4}$P$^{\mathrm{o}}$)} + \hbar
\omega_{2} \mapsto  \text{He(1s2p\,$^{3}$P$^{\mathrm{o}}$)} +
\text{e}^{-}(\epsilon\text{p}),
\end{equation}
followed by resonance ionisation indicated \eqref{eqch67e} and
\eqref{eqch67f}. 

To reduce the collisional detached contribution (a) we maintained a
pressure of $5\times 10^{-7}$~Pa  in the
interaction chamber and  to reduce contribution (b) we attenuated the
output of laser $\omega_{2}$.

\begin{figure}
\begin{minipage}{\textwidth}
\parbox[b]{0.5\textwidth}{
\epsfig{file=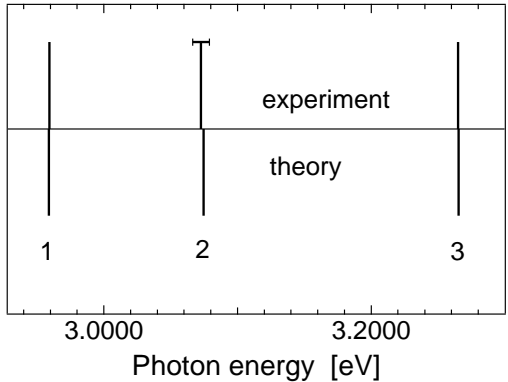, width=0.48\textwidth}
} \hfill
\parbox[b]{0.49\textwidth}{
\protect\caption[Energies of He$^{-}$ doubly
excited states near the He($n$=3) thresholds]{\label{bi650}\sloppy
Energies of He$^{-}$ doubly excited states near the He($n$=3)
thresholds: A very good agreement of the predicted \cite{Xi-96} and
measured energies is found. If no error bar on the experimental value
(upward) is indicated then it is less or equal to the line width. No
error bars for the theoretical values are avaliable. For numerical
values see table \ref{tach74} page \pageref{tach74}.} }
\end{minipage}
\end{figure}

\begin{figure}
\begin{minipage}{\textwidth}
\parbox[b]{0.665\textwidth}{
\epsfig{file=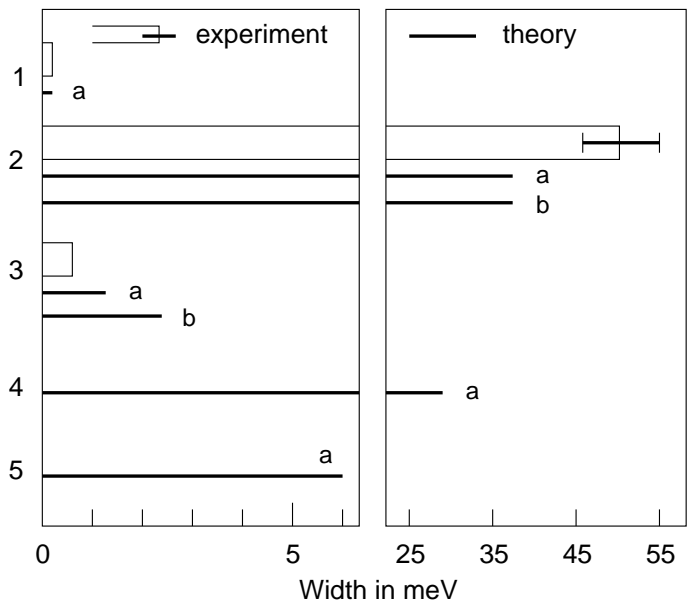, width=0.645\textwidth}
} \hfill
\parbox[b]{0.321\textwidth}{
\protect\caption[Widths of He$^{-}$ doubly
excited states near the He($n=3$) thresholds]{\label{bi670}
\sloppy\small Widths of He$^{-}$
doubly excited states near the He($n=3$) thresholds: The states with 
odd numbers are as
narrow or narrower as predicted. These state are of $A$=$-1$ character,
see section \ref{ch32} and the states with even numbers are all broader
than predicted. These states should be about 100 to 10\,000 times
broader than the adjacent $A$=$-1$ resonances.  If no error bar is given
then it is less or equal to the line width.  For numerical values see
tables \ref{tach65}, \ref{tach66}. The \textsf{a} marks values from
Xi's calculation \cite{Xi-96} and \textsf{b} values of Bylicki
\cite{Byl-97}. }}
\end{minipage}
\end{figure}

The Doppler free resonance energy $E_{0}^{\text{4s}}$, obtained as
outlined in \eqref{eqch51} on page \pageref{eqch51}, is
\begin{align}\label{eqch611}
E_{0}^{\text{4s}} &= (2\,386\,803.1\pm 4.2)\; \text{m}^{-1}\quad ,\\
\intertext{and the resonance width $\Gamma^{\text{4s}}$ is,}
\Gamma^{\text{4s}} &= 160(16)\;\text{m}^{-1}
\end{align}
This result is an average obtained from 11 spectra taken with
co-propagating laser and ion beams and 14 spectra taken with
counter-propagating beams. There are two major contributions to the
quoted uncertainty: 1~m$^{-1}$ is due to the calibration uncertainty,
and the remainder is due to statistical scatter of the fitted resonance
parameters. In table \ref{tach65} and in figure
\ref{bi650},\ref{bi670} the values are compared with recently calculated
resonance parameters.

\begin{table}
\begin{center}
\caption[He$^{-}$(1s3s4s\,$^{4}$S)
state]{\label{tach65}He$^{-}$(1s3s4s\,$^{4}$S) state: Comparison of the
experimental resonance energy and width with the theoretical prediction
of Xi \cite{Xi-96}. The energy position agrees, within the limited
accuracy of the calculation, with the experimental value.}\medskip
\begin{tabular}{lr@{.}lr@{.}l}
\hline\hline
&
\multicolumn{4}{c}{1s3s4s\,$^{4}$S}\\
\cline{2-5}
&
\multicolumn{2}{c}{$E_{0}$ (eV)} & \multicolumn{2}{c}{$\Gamma$ (meV)}\\
\hline
\emph{Experiment}: & \multicolumn{4}{c}{}\\
This work & 2&959\,260(6) & 0&20(2)\\
\emph{Theory}: &  \multicolumn{4}{c}{}\\
Xi \cite{Xi-96} (1996)& 2&959\,07 & 0&19\\
\hline\hline
\end{tabular}
\end{center}
\end{table}

The measured resonance parameters agree with those calculated by Xi and
Froese Fischer \cite{Xi-96} within the limited precision of the
latter. Our measurement is, however, almost two orders of magnitudes
more precise and should stimulate further theoretical work.

\minisec{The 1s3p$^{\mathbf{2}}$\,$^{\mathbf{4}}$P and 
1s3p4p\,$^{\mathbf{4}}$P state}

To study the 1s3p$^{2}$\,$^{4}$P and 1s3p4p\,$^{4}$P resonances we used
an excitation scheme, depicted in figure \ref{bi560}(b), that can be
summarised by
\eqref{eqch67d}, \eqref{eqch67e} and \eqref{eqch67f}. 
In this case a calibration for the 1s3p$^{2}$\,$^{4}$P was achieved with
the known energy of the 1s3s4s\,$^{4}$S resonance, lying close to the
investigated structure, and as relative scale we used the wavelength
read from the laser. Seperate scans over the 1s3s4s\,$^{4}$S resonance
had to be taken since the step length in the scans of the
1s3p$^{2}$\,$^{4}$P resonance was too long to even find the
1s3s4s\,$^{4}$S resonance. Owing to the broadness of the
1s3p$^{2}$\,$^{4}$P resonance a more precise calibration was not
neccessary. For the 1s3p4p\,$^{4}$P we calibrated with several argon
lines (table \ref{ta01}) surrounding the resonance. We did not take
Fabry Perot fringes but as a relative wavelength scale we used the
internal scale of the laser. Since we took four calibration lines, we
could check the wavelength dependency of the laser wavelength scale
offset. Over the range of the measurement the offset was linearly
dependent on the wavelength.

\begin{figure}\centering
\subfigure[The He$^{-}$(1s3p$^{2}$\,$^{4}$P) (upper graph) and
He$^{-}$(1s3p4p\,$^{4}$P) resonance (lower graph) both monitored on the
He(1s2p\,$^{3}$P$^{\text{o}}$)
state.]{\parbox[b]{0.52\textwidth}{\epsfig{file=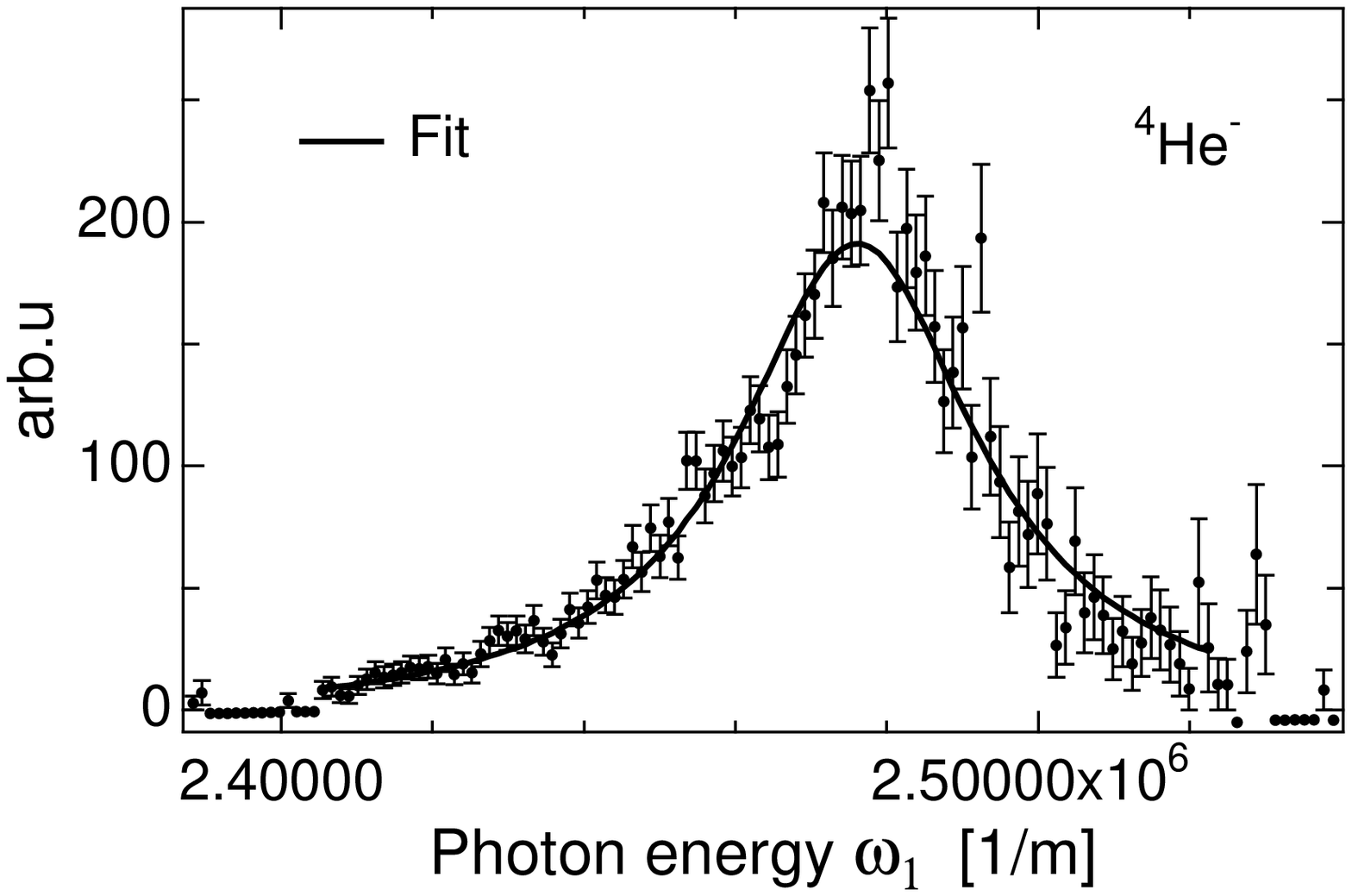,
width=0.52\textwidth}\vspace*{1ex}
\epsfig{file=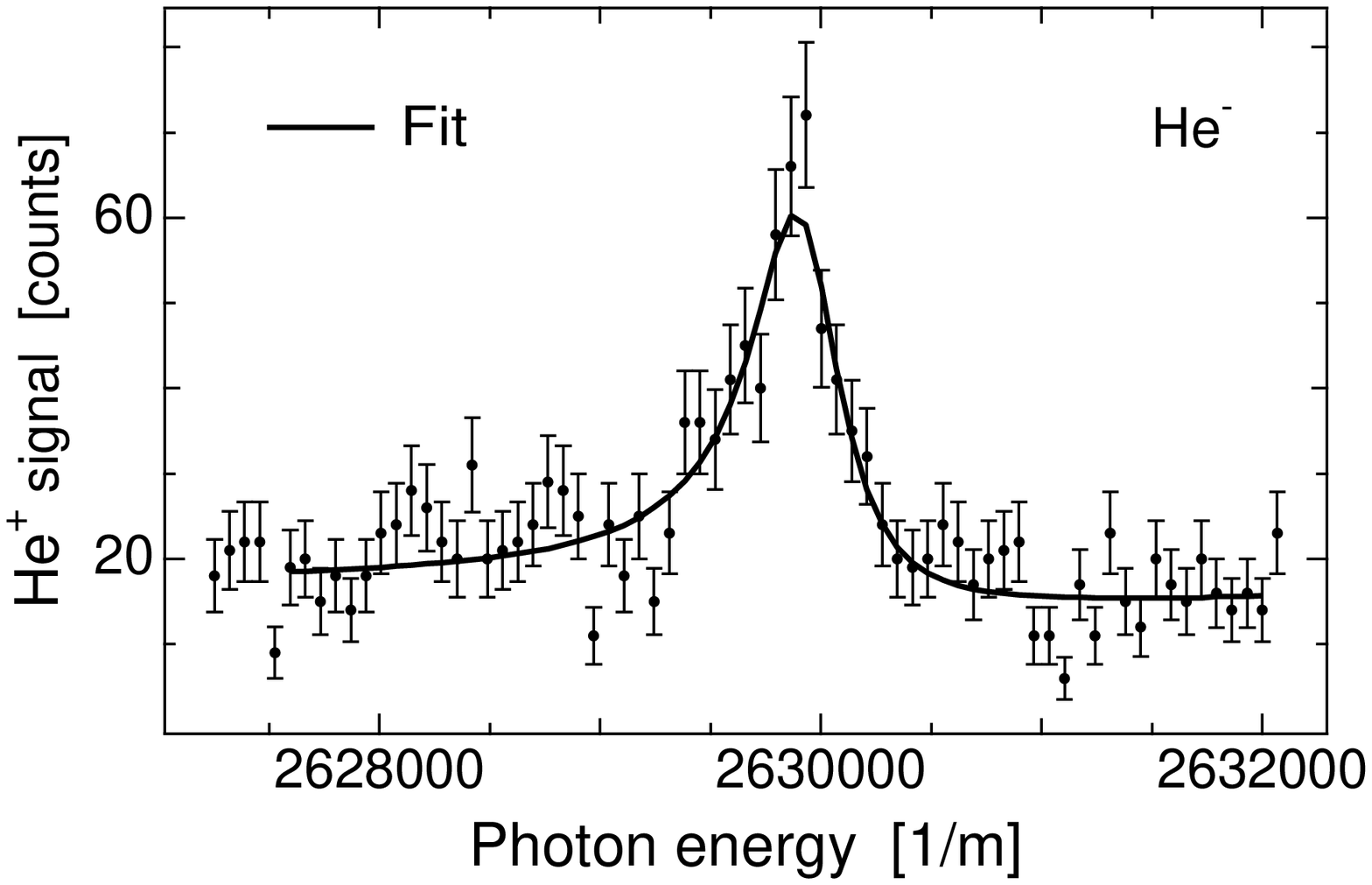, width=0.52\textwidth}}}\hfill
\subfigure[Calculated He(1s2p\,$^{3}$P$^{\text{o}}$)+$\epsilon$p
absolute partial cross section for He$^{-}$(1s3p$^{2}$\,$^{4}$P) (left
peak) and He$^{-}$(1s3p4p\,$^{4}$P) (right
peak)]{\epsfig{file=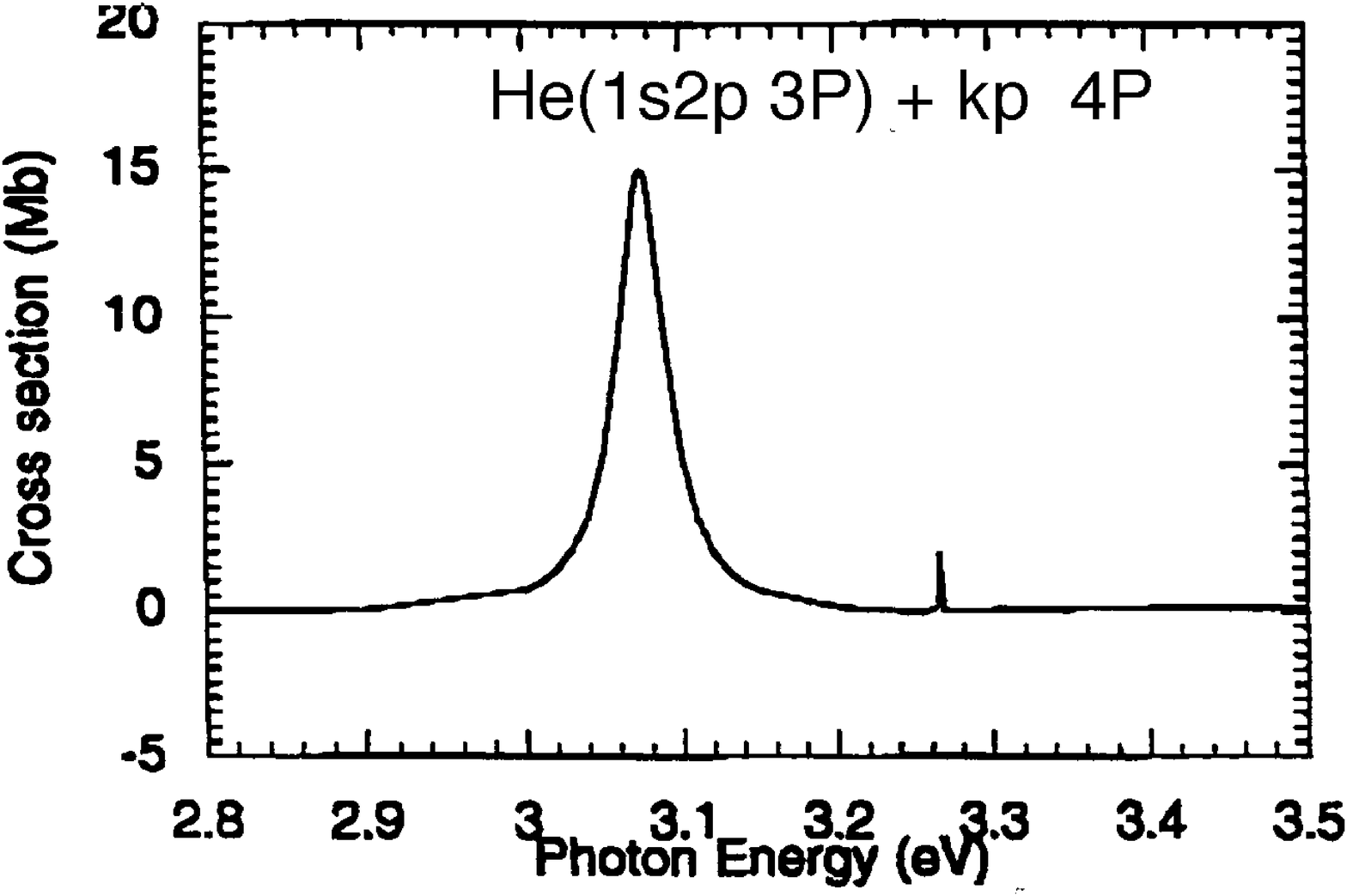, width=0.47\textwidth}}
\protect\caption[Photodetachment cross section near the 
He$^{-}$($n$=3 $^{4}$P) states]{\label{bi630}\sloppy Photodetachment
cross section near the He$^{-}$($n$=3 $^{4}$P) states: In the part (a)
we show a measurement of the He(1s2p\,$^{3}$P$^{\text{o}}$)+$\epsilon$p
partial photodetachment cross section in He$^{-}$ near the
1s3p$^{2}\,^{4}$P and 1s3p4p\,$^{4}$P state. The solid line is a fit to
the data using \eqref{eqch421}.  In the part of (a) each data point
represents 100 laser pulses. The curve in the upper part of (a) is
normalised to the laser power with the small background subtracted.  The
error bars indicate the shot noise.  The calculated partial
photodetachment cross section in part (b) is calculated by Xi
\cite{Xi-96-2}. For a comparison of width and position of the resonances
see table \ref{tach66} and figure \ref{bi650} and
\ref{bi670}.} 
\end{figure}

A typical measurement of both resonances in the
He(1s2p\,$^{3}$P$^{\mathrm{o}}$) partial cross sections is shown in
figure \ref{bi630}(a). Three processes contribute to the background of
both measurements: firstly (a) collisional detachment leaving the helium
atom in the 1s2p\,$^{3}$P$^{\mathrm{o}}$ state, and, secondly (b)
photodetachment by $\omega_{2}$ summarised in \eqref{eqch610}. Thirdly
(c) photodetachment creating a He(1s3s\,$^{3}$S) atom, which decays with
a lifetime of 36~ns \cite{Wie-66-1} to the 1s2p\,$^{3}$P$^{\mathrm{o}}$,
\begin{align}\label{eqch612}
\text{He$^{-}$(1s2s2p\,$^{4}$P$^{\mathrm{o}}$)} + \hbar\omega_{1,2}
&\mapsto \text{He(1s3s\,$^{3}$S)} + \text{e}^{-}\quad ,\\
\text{He(1s3s\,$^{3}$S)} &\mapsto
\text{He(1s2p\,$^{3}$P$^{\mathrm{o}}$)} + \hbar\omega\quad ,
\end{align}
proceeding further as in \eqref{eqch67e} and \eqref{eqch67f}. The short
lifetime makes this an efficient process.  This background contribution
is unfortunately proportional to the intensity of the laser light
$\omega_{1}$.  Therefore we could not improve the signal to noise ratio
much by attenuating $\omega_{2}$. We found a resonance energy
$E_{0}^{\text{3p}}$ for the 1s3p$^{2}$\,$^{4}$P resonance of,
\begin{align}\label{eqch613}
E_{0}^{\text{3p}} &= 2.478\,2(55)\times 10^{6}\; \text{m}^{-1}\quad ,\\
\intertext{and a width $\Gamma^{\text{3p}}$,}
\Gamma^{\text{3p}} &= 40(3)\times 10^{3}\; \text{m}^{-1}\quad .
\end{align}
The uncertainty of the resonance energy is due to the calibration
uncertainty of 40~m$^{-1}$ and the statistical scatter of the fitted
parameters of 36~m$^{-1}$. For the 1s3p4p\,$^{4}$P resonance we found an
energy of,
\begin{align}
E_{0}^{\text{4p}} &= 2\,633\,297(40)\;\text{m}^{-1}\quad ,\\
\intertext{and a width $\Gamma^{\text{4p}}$of,}
\Gamma^{\text{4p}} &= 492(35)\;\text{m}^{-1}
\end{align}
A comparison of the measured parameters with recent calculations is
given in table \ref{tach66}. 
\begin{table}
\begin{center}
\caption[He$^{-}$($n=3$\,$^{4}$P)
states]{\label{tach66}He$^{-}$($n=3$\,$^{4}$P) states: Comparison of
experimental resonance energies and widths with theoretically predicted
ones. The energy positions agree well with the calculations. The widths
deviate from the predicted values. }\medskip
\begin{tabular}{lr@{.}llr@{.}lr@{.}l}
\hline\hline
& \multicolumn{3}{c}{1s3p$^{2}$\,$^{4}$P} &
\multicolumn{4}{c}{1s3p4p\,$^{4}$P}\\
\cline{2-4}\cline{5-8}
& \multicolumn{2}{c}{$E_{0}$ (eV)} & \multicolumn{1}{c}{$\Gamma$ (meV)} 
&
\multicolumn{2}{c}{$E_{0}$ (eV)} & \multicolumn{2}{c}{$\Gamma$ (meV)}\\
\hline
%
\emph{Experiment}: & \multicolumn{7}{c}{}\\
This work & 3&072(7) & 50(5) & 3&264\,87(5) & 0&61(5)\\
\emph{Theory (after 1995)}: &  \multicolumn{7}{c}{}\\
Bylicki \cite{Byl-97} (1997)& 3&074\,24 & 37 & 3&264\,78 & 2&45\\
Xi \cite{Xi-96} (1996)& 3&074\,70 & 37.37 & 3&265\,54 & 1&30\\
Themelis \cite{The-95} (1995) & 3&096\,6 & 34.6 & \multicolumn{2}{c}{--}
& \multicolumn{2}{c}{--}\\
\hline\hline
\end{tabular}
\end{center}
\end{table} 

\begin{table}
\protect\caption[He$^{-}$(1s3p4p\,$^{4}$P) calibration lines]{ 
\label{ta01}
He$^{-}$(1s3p4p\,$^{4}$P) calibration lines: The tansition in
Ar~\textsc{i} is labled in Paschen notation and the other in the usual
spectroscopic notation. }\medskip
\begin{center}
\begin{tabular}{lr@{$\mapsto$}lr@{.}ll}
\hline\hline \# & \multicolumn{2}{c}{Transition} &
\multicolumn{2}{c}{Energy (m$^{-1}$)}  & Element \\
\hline 
a \cite{Min-73} & 4p$_{5}$&1s$_{2}$ & \multicolumn{2}{c}{2\,607\,041(1)}
& Ar
\textsc{i}\\ 
b \cite{Meg-75} & [Rn]?$J$=9&[Rn]?$J$=8 &
\multicolumn{2}{c}{2\,643\,532(1)} & U \textsc{i}
\\ 
c \cite{Kur-75} & [Mg]3p$^{4}$6f\,$J$=5/2&%
[Mg]3p$^{4}$($^{3}$P)4d\,$^{4}$D$_{3/2}$ &
\multicolumn{2}{c}{2\,647\,870(1)} & Ar \textsc{ii}
\\ 
d \cite{Kur-75} &
[Mg]3p$^{4}$($^{3}$P)5s\,$^{4}$P$_{3/2}$&%
[Mg]3p$^{4}$($^{3}$P)4p\,$^{4}$P$^{\text{o}}_{1/2}$ &
\multicolumn{2}{c}{2\,651\,401(1)} & Ar
\textsc{ii}\\ \hline\hline
\end{tabular}
\end{center}
\end{table}

\minisec{Discussion}
We have investigated three doubly excited states of He$^{-}$ below the
double detachment limit. The lowest lying resonance is of $^{4}$S
symmetry and all others of $^{4}$P symmetry, all with even parity. The
resonance structure in different partial cross sections due to these
autodetaching states has been studied.

In the case of the He$^{-}$(1s3s4s\,$^{4}$S) state \emph{all} possible
decay channels have been measured. The shape of both partial cross
sections (figure \ref{bi580}) qualitatively agrees with the predictions
by Xi \cite{Xi-96-2}. In addition, the width of this resonance is in
perfect agreement with the predicted value. The amplitude ratio of the
peak resonance cross section (figure \ref{bi580}(b) upper part) to the
flat background (figure \ref{bi690}(c) page \pageref{bi690}) is
predicted \cite{Xi-96-2} to be 3.0(3), whereas we find this ratio to be
about 3.6(3) which is not quite significantly higher. 

For the 1s3p$^{2}$\,$^{4}$P state we also find good agreement between
the predicted resonance shape (figure \ref{bi630}) and our measurement.
A similar comparison is unfortunately not possible for the
1s3p4p\,$^{4}$P resonance due to the poor resolution of the avaliable
graphs.  In all cases the energy position of the resonance coincides
well with the experimentally determined value, shown in table
\ref{tach65},\ref{tach66}  and figure \ref{bi650} within the limited
accuracy of the calculation. 

The 1s3s4s\,$^{4}$S state has also been theoretically investigated by Le
Dourneuf \cite{Dou-90}. They analysed He$^{-}$ doubly excited states
from a hyperspherical point of view and find a good correspondence
between doubly excited states of H$^{-}$ and He$^{-}$. Specifically the
charge density of doubly excited states of He$^{-}$ very closely
resembles the charge density distribution of the analog states in
H$^{-}$. It is remarkable how little the core pertubation disturbes the
correlation pattern of the two excited electrons. From the quantum
numbers given by Le Dourneuf we find that the 1s3s4s\,$^{4}$S state has
$A=-1$.

The widths of doubly-excited states of Li$^{-}$ does not vary
systematically with the value of $A$. The resonance of Li$^{-}$ cannot
be classified as narrow or broad as for He$^{-}$. The core appears to
have a more profound effect on the excited pair of electrons than in
He$^{-}$. This seems surprising to us because the cores are rather
similar with respect to their first excitation energy and dipole
polarisability.

The 1s3s4s\,$^{4}$S resonance has $A=-1$ and should accordingly be
narrow. This is confirmed by our experiment and the theory by Xi. For
the 1s3p$^{2}$\,$^{4}$P and 1s3p4p\,$^{4}$P resonances the measured
widths deviates from the predicted ones
\cite{Byl-97,Xi-96,The-95,The-94,Dav-90}. Taking the widths as guidance
I suggest the 1s3p$^{2}$\,$^{4}$P state has $A=+1$ and the
1s3p4p\,$^{4}$P state has $A=-1$. It appears more difficult to
calculate the widths of higher lying $A=-1$ resonances than for
resonances with $A=+1$. With our experiments we hope to inspire
theorists to refined calculations on He$^{-}$.

Xi \cite{Xi-96} also predicts two states of He$^{-}$ near the He($n=4$)
thresholds. Here, just as near the He($n=3$) thesholds, the lower lying
1s4p$^{2}$\,$^{4}$P state is predicted to be much broader than the
higher lying 1s4p5p\,$^{4}$P state. In the near future the study
presented here will be amended with measurements of the
1s4p$^{2}$\,$^{4}$P and 1s4p5p\,$^{4}$P states.

%

%
\chapter{Summary and conclusion}
\label{ch7}
In this chapter we compile all results of this thesis, mainly in form of
tables. A very brief description of the experiment, based on the
abstracts of the respective publication, and an assessment of the
accuracy is accompanied by a table to give the reader convenient access
to the main results of this thesis. I present the experiments in the
same order as they appear in the table of contents.

\section{Summary of Results}
\label{ch71}
%
\minisec{The electron affinity of tellurium}
The electron affinity of tellurium has been determined to be
1\,589\,618(5)~m$^{-1}$. The threshold for photodetachment of
Te$^{-}$($^{2}$P$_{3/2}$) forming neutral Te in the ground state was
investigated by measuring the total photodetachment cross section
using a collinear laser-ion beam geometry. The electron affinity was
obtained from a fit to the Wigner law in the threshold region.

The error is dominated by the calibration uncertainty of 4~m$^{-1}$ due
to the unknown laser line profile, the remainder is due to statistical
scatter of the fitted threshold parameters.

\setlongtables
\begin{longtable}[c]{lr@{.}lr@{.}l}
\caption[Tellurium electron affinity]{\label{tach71}\slshape Tellurium
electron affinity: Comparison of the different measurements. The value
in m$^{-1}$ is only given if the publication contains the value.}\\[-1.4ex]
\hline\hline
Author & \multicolumn{2}{c}{Affinity in m$^{-1}$} &
\multicolumn{2}{c}{Affinity in meV} \\ \hline
\endfirsthead
\multicolumn{5}{r}{continued from previous page}\\
\hline\hline
Author & \multicolumn{2}{c}{Affinity in m$^{-1}$} &
\multicolumn{2}{c}{Affinity in meV} \\ \hline
\endhead
\hline\hline
\multicolumn{5}{r}{continued on next page}
\endfoot
\hline\hline
\endlastfoot
Slater (1977) \cite{Sla-77} & \multicolumn{2}{c}{} & 1970&8(3) \\
Th{\o}gersen (1996) \cite{Tho-96-1} & 1&589\,69(4)$\times 10^{6}$ &
1970&965(50)\\ 
This work (1996) \cite{Hae-96-2} & 1&589\,618(5)$\times 10^{6}$ &
1970&876(7) \\
\end{longtable}

\minisec{The electron affinity of lithium}
We have investigated the photodetachment threshold of Li$^{-}$ leading
to the formation of the residual Li atom in the 2p\,$^{2}$P state. The
excited residual atom was selectively photo-ionised via an intermediate
Rydberg state and the resulting Li$^{+}$ ion was detected. A collinear
laser-ion beam geometry enabled both high resolution and sensitivity to
be attained. We have demonstrated the potential of this state selective
photodetachment spectroscopic method by improving the accuracy of Li
electron affinity measurements one order of magnitude. From a fit to the
Wigner law in the threshold region, we obtained a Li electron affinity
of 498\,490(17)~m$^{-1}$.

The uncertainty of 17~m$^{-1}$ is mainly due to a calibration
uncertainty of 13~m$^{-1}$, and the remainder is due to statistical
scatter of the fitted threshold values.

\setlongtables
\begin{longtable}[c]{lr@{.}lr@{.}l}
\caption[Lithium electron affinity]{\label{tach72}\slshape Lithium
electron affinity: Comparison of the different measurements and
calculations. The value in m$^{-1}$ is only given if the publication
contains the value.} \\[-1.4ex]
\hline\hline
Author & \multicolumn{2}{c}{Affinity in m$^{-1}$} &
\multicolumn{2}{c}{Affinity in meV} \\ \hline
\endfirsthead
\multicolumn{5}{r}{continued from previous page}\\
\hline\hline
Author & \multicolumn{2}{c}{Affinity in m$^{-1}$} &
\multicolumn{2}{c}{Affinity in meV} \\ \hline
\endhead
\hline\hline
\multicolumn{5}{r}{continued on next page}
\endfoot
\hline\hline
\endlastfoot
\textsl{Experiment} & \multicolumn{4}{c}{}\\
Feldman (1976) \cite{Fel-76} & \multicolumn{2}{c}{} & 618&2(5) \\
Bae (1985) \cite{Bae-85} &\multicolumn{2}{c}{} & 617&3(7)\\ 
Dellwo (1992) \cite{Del-92-1} & 4&980\,9(16)$\times 10^{5}$ &
617&6(2) \\
This work \cite{Hae-96-1} & 4&984\,90(17)$\times 10^{5}$ & 618&049(20) \\
\textsl{Theory (after 1992)} & \multicolumn{4}{c}{}\\
Chung (1992) \cite{Chu-92-2} & \multicolumn{2}{c}{} & 617&4(2) \\
Froese Fischer (1993) \cite{Fis-93} & \multicolumn{2}{c}{} & 617&64\\
\end{longtable}

\minisec{Doubly excited states of Li$^{-}$}
We report on the first observation of resonance structure in the total
cross section for the photodetachment of Li$^-$. The structure arises
from the autodetaching decay of doubly excited $^1$P$^{\text{o}}$ states
of Li$^-$ that are bound with respect to the Li(3p) state.  Calculations
have been performed for both Li$^-$ and H$^-$ to assist in the
identification of these resonances. The lowest lying resonance is
analogous to the previously observed $(_3\{0\}_3^+)$ symmetrically
excited intra-shell resonance in H$^-$ but it is much broader. Higher
lying resonant states are observed to converge on the Li(3p)
limit. These Rydberg-like resonances are much narrower and correspond to
asymmetrically excited inter-shell states.

\setlongtables
\begin{longtable}[c]{lcr@{.}lr@{.}lr@{.}l}
\caption[Li$^{-}$ doubly excited states]{\label{tach73}\slshape
Li$^{-}$ doubly excited states: Comparison of theoretical results for
intra-shell $^{1}$P$^{\text{o}}$ states of H$^{-}$ below the H($n$=3)
threshold and of Li$^{-}$ below the Li(3p) threshold. The binding energy
$E_{\text{bind}}$ is relative to the double detachment limit and
$E_{\text{trans}}$ is the transition energy from the ground
state. Calculated resonance parameters for the two Rydberg like
resonances, marked \textsf{Ry} in figure \ref{bi520} (b) page
\pageref{bi520}, below the Li(3p) threshold are given. Note the strong
overlap of these two resonances.}\\[-1.4ex]
\hline\hline
Element & Label & \multicolumn{2}{c}{$E_{\text{bind}}$ (eV)} & 
\multicolumn{2}{c}{$\Gamma$ (eV)} 
 & \multicolumn{2}{c}{$E_{\text{trans}}$ (eV)} \\ \hline
\endfirsthead
\multicolumn{8}{r}{continued from previous page}\\
\hline\hline
Element & Label & \multicolumn{2}{c}{$E_{\text{bind}}$ (eV)} & 
\multicolumn{2}{c}{$\Gamma$ (eV)} 
  & \multicolumn{2}{c}{$E_{\text{trans}}$ (eV)} \\ \hline
\endhead
\hline\hline
\multicolumn{8}{r}{continued on next page}
\endfoot
\hline\hline
\endlastfoot
H$^{-}$ & & $-1$&706 & 0&033 & 12&647\\
Li$^{-}$ & \textsf{Res} & $-1$&688 & 0&36  & 4&322 \\
Li$^{-}$ & \textsf{Ry} & $-1$&556 & 0&020 & 4&453 \\
Li$^{-}$ & \textsf{Ry} & $-1$&555 & 0&016 & 4&454 \\
\end{longtable}

The uncertainty in this experiment is caused by the small but
unavoidable change in the overlap of the ion and laser beams as the
wavelength of the laser was scanned. The normalisation procedure applied
in this experiment was unable to completely account for the beam overlap
variation. We estimated, however, that this effect caused a maximum
error in the relative cross section over the whole spectrum of less than
10\%. The statistical scatter, mainly caused by the shot noise in
counting of the neutral particles, is less than 3\%.

\minisec{Doubly excited states of He$^{-}$}
We investigated Feshbach resonances at the He($n$=3) thresholds.  The
lowermost resonance has been identified to be of $A=-1$ chararcter
\cite{Dou-90}. The resonance energies are in good agreement with the
calculation. The widths have been found to deviate systematically from
the predictions.

In all experiments we have combined laser photodetachment and resonance
ionisation to study Feshbach resonances associated with the
autodetaching decay of doubly excited states of He$^{-}$. The
measurements were made under the simultaneous conditions of high
sensitivity and high energy resolution using a collinear beam apparatus.

\begin{center}\footnotesize
\setlongtables
\begin{longtable}{cr@{.}lr@{.}lr@{.}lr@{.}lr@{.}lr@{.}lr@{.}l}
\caption[He$^{-}$ doubly excited
states]{\label{tach74}\normalsize\slshape
He$^{-}$ doubly excited states: Comparison of the experimental resonance
energies and width with the theoretically predicted ones of Xi
\cite{Xi-96}. The energy positions agree within the limited accuracy of
the calculation with the experimental values. Some widths deviate from
the predicated values. The numbers in the outermost column on the left
refer to the following designations:
\textsf{1}, 1s3s4s\,$^{4}$S; \textsf{2}, 1s3p$^{2}$\,$^{4}$P;
\textsf{3}, 1s3p4p\,$^{4}$P. 
} \\[-1.4ex]
\hline\hline
\# &
\multicolumn{6}{c}{Experiment} &
\multicolumn{8}{c}{Theory}  \\
\cline{8-15}
& \multicolumn{2}{c}{} &
\multicolumn{2}{c}{} &
\multicolumn{2}{c}{} & \multicolumn{4}{c}{length from} &
\multicolumn{4}{c}{velocity form} \\
&
\multicolumn{2}{c}{$E_{\text{r}}$ (eV)} &
\multicolumn{2}{c}{$E_{\text{r}}$ ($10^{6}$ m$^{-1}$)} &
\multicolumn{2}{c}{$\Gamma$ (meV)} &\multicolumn{2}{c}{$E_{\text{r}}$
(eV)} &
\multicolumn{2}{c}{$\Gamma$ (meV)} & \multicolumn{2}{c}{$E_{\text{r}}$
(eV)} &
\multicolumn{2}{c}{$\Gamma$ (meV)} \\
\hline
\endfirsthead
\multicolumn{15}{r}{continued from previous page}\\
\hline\hline
\# &
\multicolumn{6}{c}{Experiment} &
\multicolumn{8}{c}{Theory}  \\
\cline{8-15}
& \multicolumn{2}{c}{} &
\multicolumn{2}{c}{} &
\multicolumn{2}{c}{} & \multicolumn{4}{c}{length from} &
\multicolumn{4}{c}{velocity form} \\
&
\multicolumn{2}{c}{$E_{\text{r}}$ (eV)} &
\multicolumn{2}{c}{$E_{\text{r}}$ ($10^{6}$ m$^{-1}$)} &
\multicolumn{2}{c}{$\Gamma$} (meV) &\multicolumn{2}{c}{$E_{\text{r}}$
(eV)} &
\multicolumn{2}{c}{$\Gamma$ (meV)} & \multicolumn{2}{c}{$E_{\text{r}}$
(eV)} &
\multicolumn{2}{c}{$\Gamma$ (meV)}\\
\hline
\endhead
\hline\hline
\multicolumn{15}{r}{continued on next page}
\endfoot
\hline\hline
\endlastfoot
\textsf{1} &
2&959\,260(6) & 2&386\,803\,1(42) &   0&20(2) & 2&959\,07 & 0&19 &
2&959\,08 & 0&18 \\
\textsf{2} &
3&072(7) & 2&478\,2(55) & \multicolumn{2}{l}{50(5)} & 3&074\,70 & 37&37
& 3&074\,71 & 37&37 \\
\textsf{3} &
3&264\,87(4) & 2&633\,297(40) & 0&61(5) & 3&265\,54 & 1&30 & 3&265\,47
& 1&31 \\
\end{longtable}
\end{center}

For the He$^{-}$(1s3s4s\,$^{4}$S) state, number \textsf{1} in table
\ref{tach74}, there are two major contributions to the quoted
uncertainty: 1~m$^{-1}$ is due to the calibration uncertainty, and the
remainder is due to statistical scatter of the fitted resonance
parameters. The error of the width reflects the statistical scatter of
the fitted width parameter. 

For the He$^{-}$(1s3p$^{2}$\,$^{4}$P) state, number \textsf{2} in table
\ref{tach74}, there are two major contributions to the error, namely a
calibration uncertainty of 40~m$^{-1}$ and the statistical scatter of
the fitted resonance parameters of 36~m$^{-1}$. The uncertainty of the
width is entirely due to statistical scatter of the fitted width
parameter.  

For the He$^{-}$(1s3p4p\,$^{4}$P) state, number \textsf{3} in table
\ref{tach74}, we found two dominant source of uncertainty: a calibration
uncertainty of 30~m$^{-1}$ and the remainder is due to statistical
scatter of the fitted resonance parameters. The uncertainty of the width
stems only from statistical scatter of the fitted values.

\section{Conclusion}
\label{ch72}
In the beginning of the nineties, about fifteen years after the boost
from the advent of tunable lasers, high resolution studies of negative
ions was a mature field. The most precise method in negative ion studies
by that time was laser photodetachment threshold spectroscopy in a
collinear geometry. With this methode we studied the photodetachment
threshold of tellurium. We attained a value for the electron affinity
that ranks among the most accurately known in the periodic system. All
the very precise electron affinities had so far been obtained in a
collinear geometry with neutral particle detection, but this method is
limited to elements with s-wave photodetachment thresholds and has been
applied to most suitable elements. The majority of the elements, cannot
be investigted with this method.

Using neutral particle detection it is also possible to investigate
resonance structure in the photodetachment cross section of negative
ions. We were able to measure resonance structure between the Li(3s) and
Li(3p) thresholds. The signal contains a large fraction of
non-resonantly detached ions, which contribute only to the background,
thus masking the sought resonance structure. For higher lying resonances
the modultation of the cross section rapidly dwindle, hence aggravating
the previously mentioned problems. Furthermore we found that background
caused by photo-electrons set a definite upper limit to the usable
photon energies.  The experiment on the Li$^{-}$ doubly excited states
therefore revealed some limitations of neutral particle detection: to
study resonances with very small photodetachment cross sections other
more sensitive methods are necessary.

The shortcommings of laser photodetachment spectroscopy with neutral
particle detection are eliminated by a novel state selective detection
scheme, based on resonance ionisation of the residual atom. The scheme
retains all advantages of the collinear geometry, and, at the same time
fundamentally augments experimental possibilities. The technique enables
seperate studies of partial cross sections with little or no
background. To demonstrate the feasibilty of the state selective
detection we studied the Li(2p) threshold and obtained a very precise
electron affinity. With this state selective detection it will be
possible to measure the electron affinities of essentially all metals
using favourable s-wave thresholds. It is consequently for the first
time possible to determine the electron affinities of almost any
elements with the high precision previously only attainable for the few
with s-wave photodetachment thresholds. In addition, with the now
available tunable infra-red laser light sources investigations of the so
far largely unexplored rare earth element negative ions have become
possible. Most notably a number of calculations
\cite{Vos-91-1,Vos-91-2,Gri-92,Din-94,Dat-94,Din-95-1,Din-95-2,Din-96}
predict electron affinities and the fine structure of numerous rare
earth negative ions.

This state selective detection was also applied to study three doubly
excited state of He$^{-}$. Investigations of this metastable
three-electron system with neutral particle detection suffered from
excessive backgound of neutrals created, in flight, by autodetachment of
the He$^{-}$ ground state. State selective detection remedied this
severe problem by completely eliminating this source of background,
because the neutrals resulting from autodetachment are in the ground
state and are thus not resonantly ionised. Prior to these investigations
studies of doubly excited states of negative ions had essentially been
limited to the negative hydrogen ion.  The results presented here are a
first step towards a systematic study of doubly excited states in
another fundamental negative ion.

There is at present a large theoretical activity in the field of doubly
excited states in negative ions.  Calculations on Li$^{-}$ showed a good
agreement with the experiment presented here.  In He$^{-}$ a recent
calculation agrees with the experimentally determined energy and width
for the lowest lying doubly excited state 1s3s4s\,$^{4}$S. The energies
of the two higher lying 1s3p$^{2}$\,$^{4}$P and 1s3p4p\,$^{4}$P states
agree with calculations, but their width deviates from these
predictions.  The calculations on Li$^{-}$ and He$^{-}$ had a direct
impact on our investigations. It is interesting to note that in this
research field theory and experiment currently are developing
parallelly.  We hope and believe the measurements on He$^{-}$ will
inspire theorists to further refine their methods.

One of the new perspectives opened up by the state selective detection
is to improve the electron affinities of a large number of
elements. Investigations of fine structure of negative ion ground states
can often be made in conjunction with affinity measurements. The studies
of doubly excited states of negative ions has just begun, and the
investigations of He$^{-}$ and Li$^{-}$ will be extended towards the
double detachment limit. This type of studies can also be pursued
further with negative ions with more complex cores like heavier alkaline
or alkaline earth metals.

\begin{appendix}
\chapter{Glossary}
\label{ch8}
\setlongtables
\begin{longtable}[c]{p{0.17\textwidth}p{0.4\textwidth}}
\caption[Abbreviations]{\label{tach81}\slshape Abbreviation: Acronym
expansion of abbreviations used in this thesis.} \\
\hline\hline
Abbreviation & Expansion \\ \hline
\endfirsthead
\multicolumn{2}{r}{continued from previous page}\\
\hline\hline
Abbreviation & Expansion \\ \hline
\endhead
\hline\hline
\multicolumn{2}{r}{continued on next page}
\endfoot
\hline\hline
\endlastfoot
\textsf{a.u.} & (Hartree) atomic units\\
\textsf{BNC} & Banana normal contact\\
\textsf{CEM} & channel electron multiplier\\
\textsf{CG} & conducting glass plate\\
\textsf{CX} & charge exchange chamber\\
\textsf{DL} & dye laser\\
\textsf{DP} & deflection plate\\
\textsf{EA} & electron affinity\\
\textsf{EC} & extractor of the sputter ion source\\
\textsf{EX} & excimer laser\\
\textsf{FC} & Faraday cup\\
\textsf{FP} & Fabry Perot etalon\\
\textsf{GPC} & gated photon counter\\
\textsf{GPIB} & general purpose interface bus\\
\textsf{HL} & hollow cathode lamp\\
\textsf{ID} & interaction detection chamber\\
\textsf{IO} & ioniser of the sputter ion source\\
\textsf{L} & laser beam\\
\textsf{LM} & liquid metal\\
\textsf{LPT} & laser photodetachment threshold spectroscopy\\
\textsf{M} & sector magnet\\
\textsf{MCDF} & multi-configuration Dirac Fock\\
\textsf{MCHF} & multi-configuration Hartree Fock\\
\textsf{MCP} & micro channel plate\\
\textsf{MP} & metal plate in \textsf{PID} and \textsf{NPD}\\
\textsf{NPD} & neutral particle detector\\
\textsf{P} & photodiode\\
\textsf{PD} & positive ion detector\\
\textsf{PID} & positive ion detector\\
\textsf{PIS} & plasma ion source\\
\textsf{QD} & electric quadrupole deflector\\
\textsf{QED} & Quantum electrodynamics\\
\textsf{RIS} & resonance ionisation spectroscopy\\
\textsf{SIS} & sputter ion source\\
\textsf{T} & sputter target\\
\textsf{TC} & thermo couple\\
\textsf{TOF} & time of flight\\
\end{longtable}

%
\chapter{Symbols}
\label{ch9}
I have tried as much as possible to use the standard symbols and
notations. Here I compile first the conventions used for the different
mathematical objects, and then a table with all symbols in this thesis
follows.

\setlongtables
\begin{longtable}[c]{lll}
\caption[Symbol conventions]{\label{tach91}\slshape Symbol conventions:
Compilation of how I indicate which type of mathematical object is
denoted by the typography.}\\
\hline\hline
Item & typeface & example\\ \hline
\endfirsthead
\multicolumn{3}{r}{continued from previous page}\\
\hline\hline
Item & typeface & example\\ \hline
\endhead
\hline\hline
\multicolumn{3}{r}{continued on next page}
\endfoot
\hline\hline
\endlastfoot
vector & upper and lower case bold italic & $\boldsymbol{r}$\\
variable & italic & $t$\\
matrix & uppercase italic & $H$\\
Hamilton operator & calligraphic & $\mathcal{H}$\\
vector operators & bold roman & $\mathbf{L}$\\
scalar operator & italic & $L^{2}$\\
quantum number & italic & $l$\\
\end{longtable}

\setlongtables
\begin{longtable}[c]{ll}
\caption[Symbols in the thesis]{\label{tach92}\slshape Symbols in the
thesis: Explanation of symbols and variables that appear in this
thesis. } \\
\hline\hline
Symbol & Explanation \\ \hline
\endfirsthead
\multicolumn{2}{r}{continued from previous page}\\
\hline\hline
Symbol & Explanation\\ \hline
\endhead
\hline\hline
\multicolumn{2}{r}{continued on next page}
\endfoot
\hline\hline
\endlastfoot
$\mathcal{H}$ & Hamilton operator \\
$\hbar$ & Planck's constant \\
$m$ & electron mass\\
$\boldsymbol{\nabla}$ & nabla operator \\
$Z$ & nuclear charge\\
$q$ & electron charge, Beutler-Fano asymmetry parameter\\
$\epsilon_{0}$ & dielectric constant\\
$\boldsymbol{r}$ & point in space\\
$\boldsymbol{r}_{i,j}$ & difference vector in space\\
$\mathbf{L}$ & multi-electron angular momentum operator\\
$\mathbf{l}$ & one-electron angular momentum operator\\
$\mathbf{S}$ & multi-electron spin operator\\
$\mathbf{s}$ & one-electron spin operator\\
$\mathbf{J}$ & multi-electron total angular momentum operator\\
$\mathbf{j}$ & one-electron total angular momentum operator\\
$\psi$ & one-electron orbital\\
$\Psi$ & multi-electron wave function\\
$R_{nl}(r)$ & Laguerre polynomials\\
$Y_{lm}(\Omega)$ & spherical harmonics\\
$\Omega$ & solid angle\\
$\xi$ & spinor\\
$m_{s}$ & magnetic spin quantum number\\
$\lambda_{i}$ & i$^{\text{th}}$ set of spatial and spin coordinates\\
$c_{i}$ & expansion coefficients\\
$E_{\text{int}}$ & interaction energy\\
$E_{\text{exp}}$ & experimental energy\\
$E_{\text{c}}$ & correlation energy\\
$E_{\text{rel}}$ & relativistic energy\\
$V_{\text{pol}}$ & polarisation potential\\
$\alpha_{\text{D}}$ & electric dipole polarisability\\
$\Phi_{\text{cont}}$ & continuum wavefunction\\
$n$ & principal quantum number\\
$n_{\text{eff}}$ & effective principal quantum number\\
$K$ & two-electron quantum number\\
$T$ & two-electron quantum number\\
$A$ & two-electron quantum number\\
$\pi$ & parity\\
$\alpha$ & fine structure constant\\
$\sigma (\omega)$, $\sigma (E)$ & photodetachment cross section\\
$\sigma_{\text{W}}$ & Wigner law fit function\\
$\sigma_{\text{OM}}$ & Wigner law with O'Malley's correction\\
$\sigma_{\text{BF}}$ & Beutler-Fano profile fit function\\
$\sigma_{\text{Sh}}$ & Shore fit function for Beutler-Fano profile\\
$\sigma_{\text{SF}}$ & Starace-Fano fit function\\
$\sigma_{\text{P}}$ & generalised Beutler-Fano fit function\\
$\sigma_{\text{Cu}}$ & Wigner cusp fit function\\
$\rho^{2}$ & Beutler-Fano correlation parameter\\
$M_{0\nu}$ & transition matrix element\\
$\mathbf{e}$ & polarisation vector operator\\
$\mathbf{P}_{q}$ & electron momentum operator\\
$\omega$ & frequency of light in SI; energy of light in a.u.\\
$V(r)$ & potential energy\\
$\boldsymbol{k}$ & wave vector\\
$f_{i,j}(\Omega)$ & scattering amplitude\\
j$_{l}$ & spherical Bessel function\\
$\delta_{ij}$ & Kronecker-Delta symbol\\
$F(q;\epsilon)$ & Beutler-Fano function\\
$\Gamma$ & resonance width\\
$E_{0}$ & resonance position\\
$E_{0}^{\text{r,(b)}}$ & red (blue) shifted resonance position\\
\end{longtable}

\chapter{He$^{-}$ partial cross sections}
\label{ch10}
Here I compile partial photodetachment cross sections for He$^{-}$. Some
of them have been presented earlier in this thesis, others are not
present because the may not contain resonance structure. Even the
non-resonant cross sections can provide valuable information, for
example about possible contributions to background. All partial cross
section presented here are calculated by Xi and Froese Fischer
\cite{Xi-96-2} and kindly provided to aid the interpretation of our
data.

\begin{figure}
\begin{minipage}{\textwidth}
\parbox[b]{0.5\textwidth}{
\epsfig{file=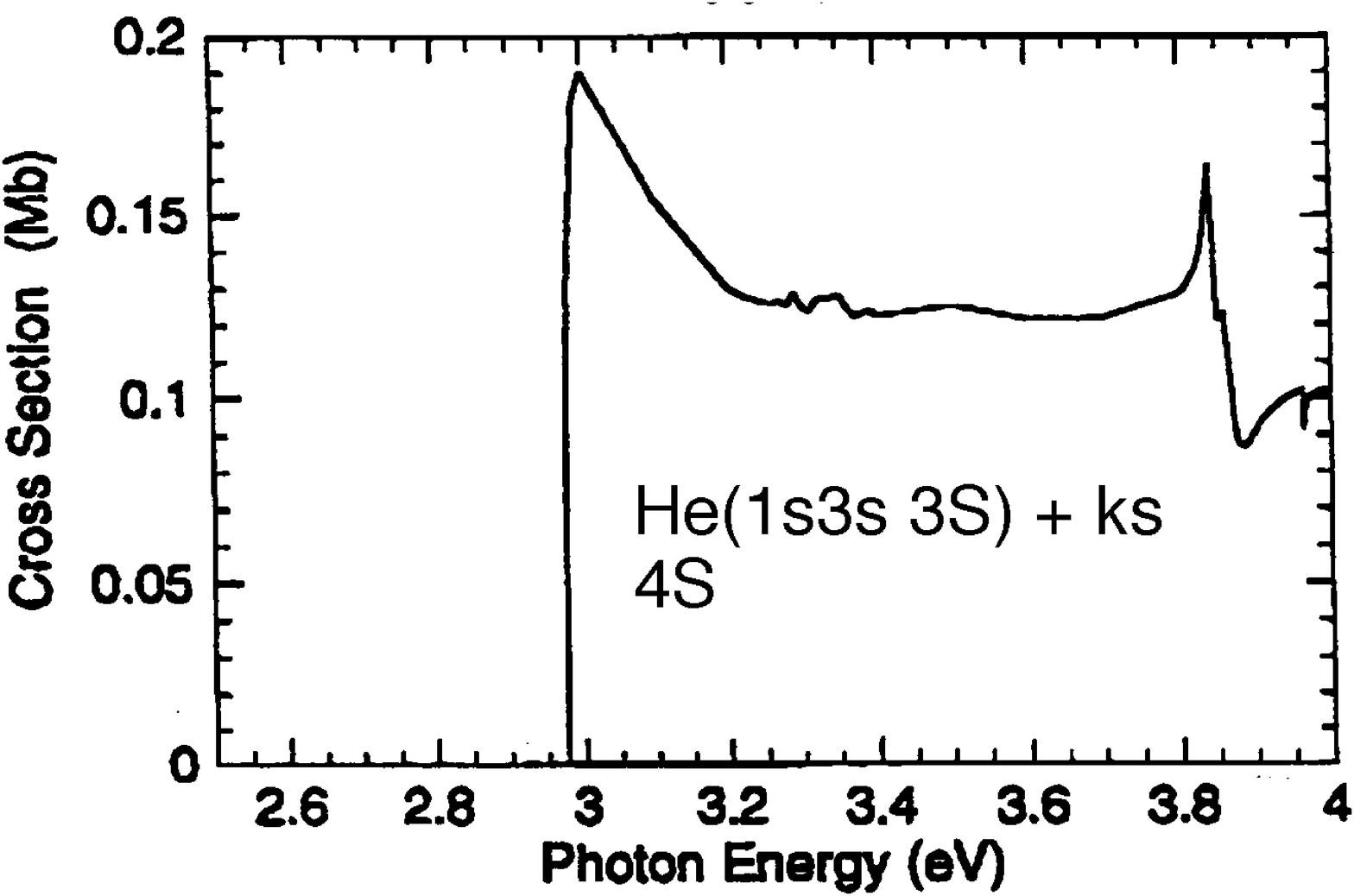, width=0.49\textwidth}
} \hfill
\parbox[b]{0.49\textwidth}{
\protect\caption[He(1s3s\,$^{3}$S) partial cross
section of $^{4}$S symmetry]{\label{bi740}\sloppy He(1s3s\,$^{3}$S)
partial cross section of $^{4}$S symmetry: The parital photodetachment
cross section monitored by selectively detecting He atoms in the
1s3s\,$^{3}$S state \cite{Xi-96-2}.}  }
\end{minipage}
\end{figure}
%

\begin{figure}
\begin{minipage}{\textwidth}
\parbox[b]{0.5\textwidth}{
\epsfig{file=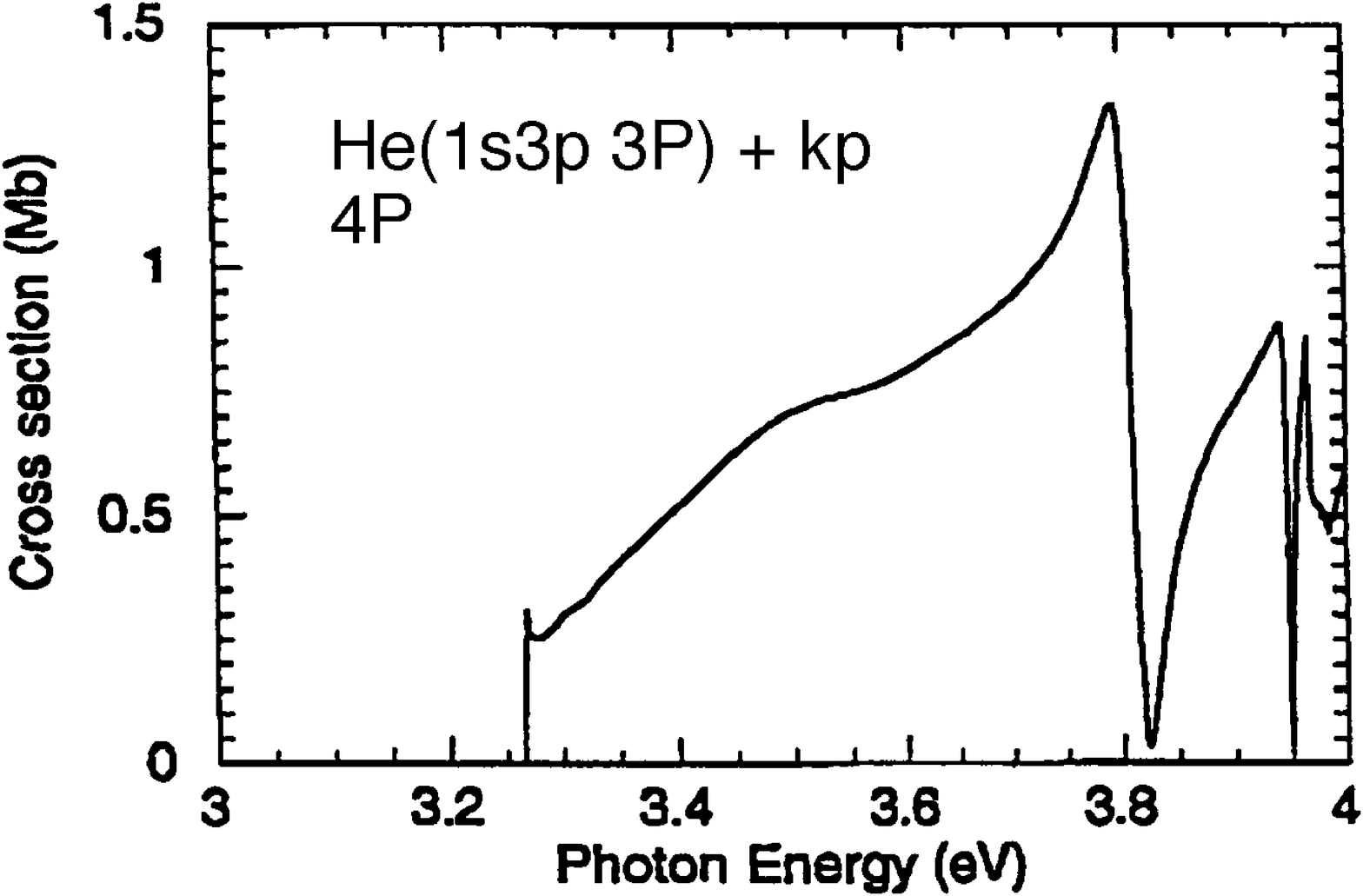, width=0.49\textwidth}
} \hfill
\parbox[b]{0.49\textwidth}{
\protect\caption[He(1s3p\,$^{3}$P$^{\text{o}}$) partial cross
section of $^{4}$P symmetry]{\label{bi680}\sloppy
He(1s3p\,$^{3}$P$^{\text{o}}$) partial cross section of $^{4}$P
symmetry: The partial photodetachment cross section monitored by
selectively detecting He atoms in the 1s3p\,$^{3}$P$^{\mathrm{o}}$ state
\cite{Xi-96-2}.} }
\end{minipage}
\end{figure}
%

\begin{figure}
\subfigure[Overview of He(1s2s\,$^{3}$S) partial cross
sections of $^{4}$S symmetry]{\epsfig{file=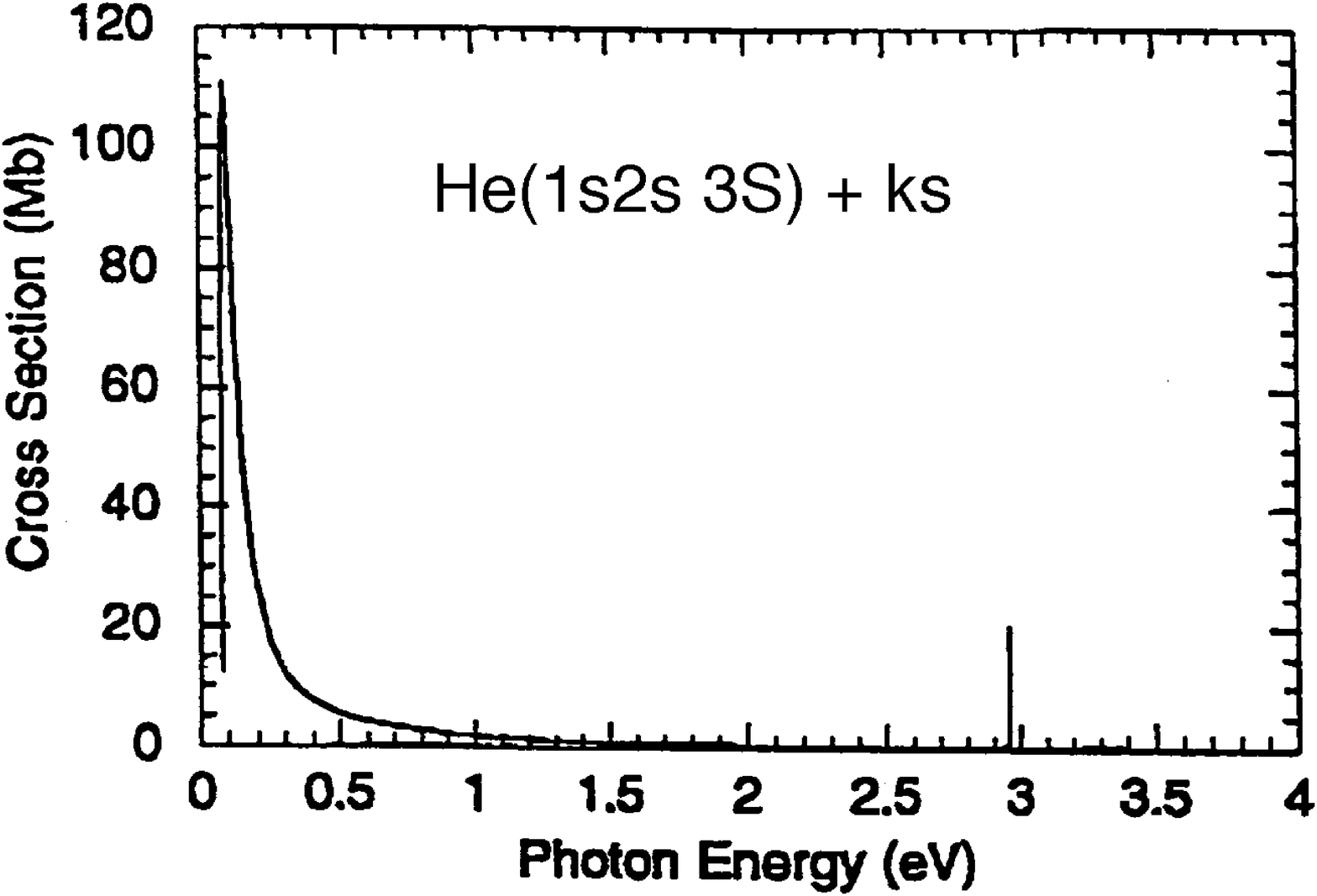 ,
width=0.49\textwidth}}\hfill
\subfigure[He(1s2s\,$^{3}$S) partial cross
sections of $^{4}$S symmetry near the 1s3s4s\,$^{4}$S state]{
\epsfig{file=bi590.eps, width=0.49\textwidth}}
\subfigure[Overview of He(1s2s\,$^{3}$S) partial cross
sections of $^{4}$D symmetry]{\epsfig{file=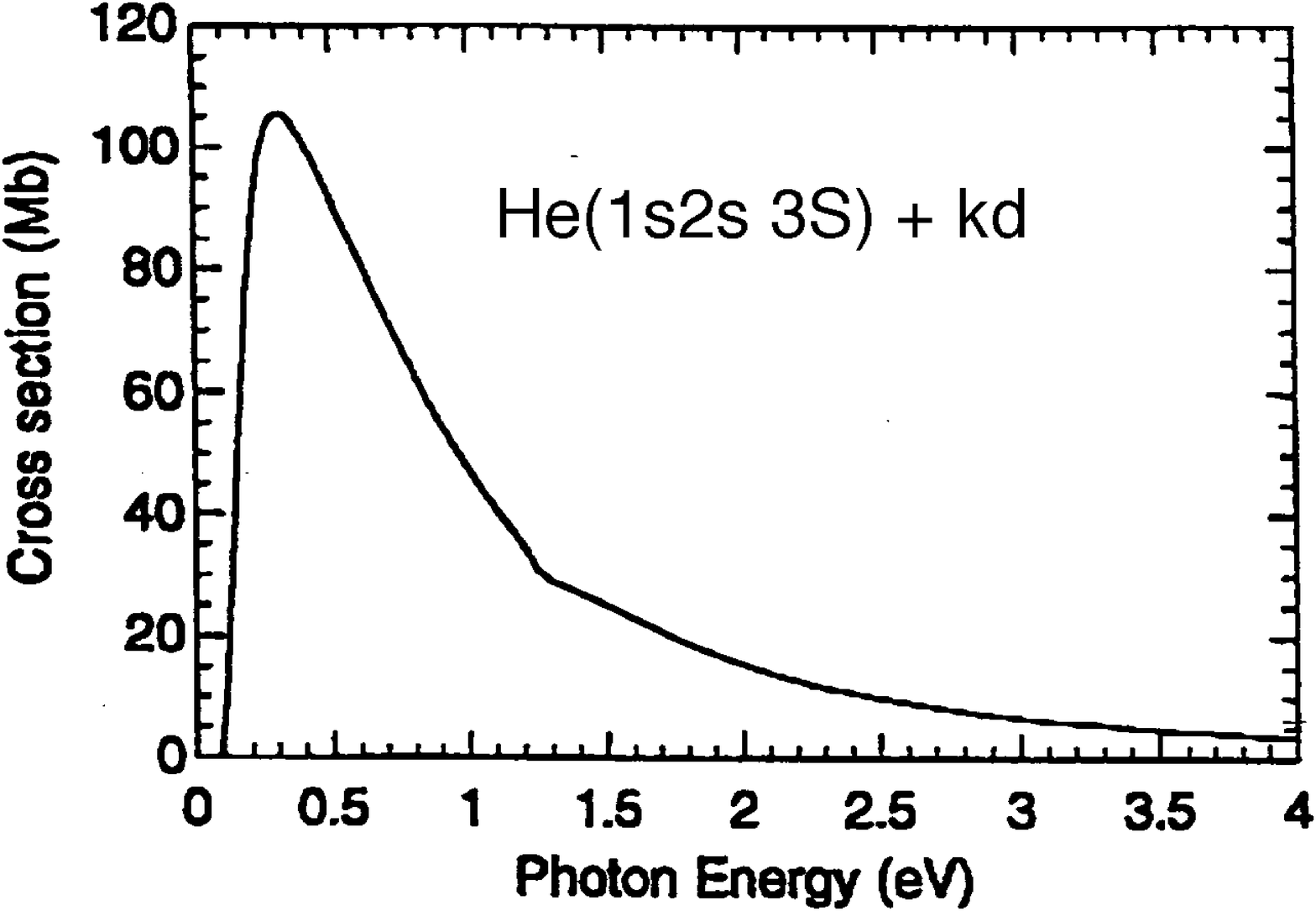,
width=0.49\textwidth}} 
\caption[He(1s2s\,$^{3}$S) partial cross
sections]{\label{bi690}\sloppy He(1s2s\,$^{3}$S) partial cross sections:
The partial photodetachment cross sections monitored by selectively
detecting He atoms in the 1s2s\,$^{3}$S state \cite{Xi-96-2}.}
\end{figure}

\begin{figure}
\subfigure[Overview of He(1s2p\,$^{3}$P$^{\text{o}}$) partial cross
sections of $^{4}$S symmetry]{\epsfig{file=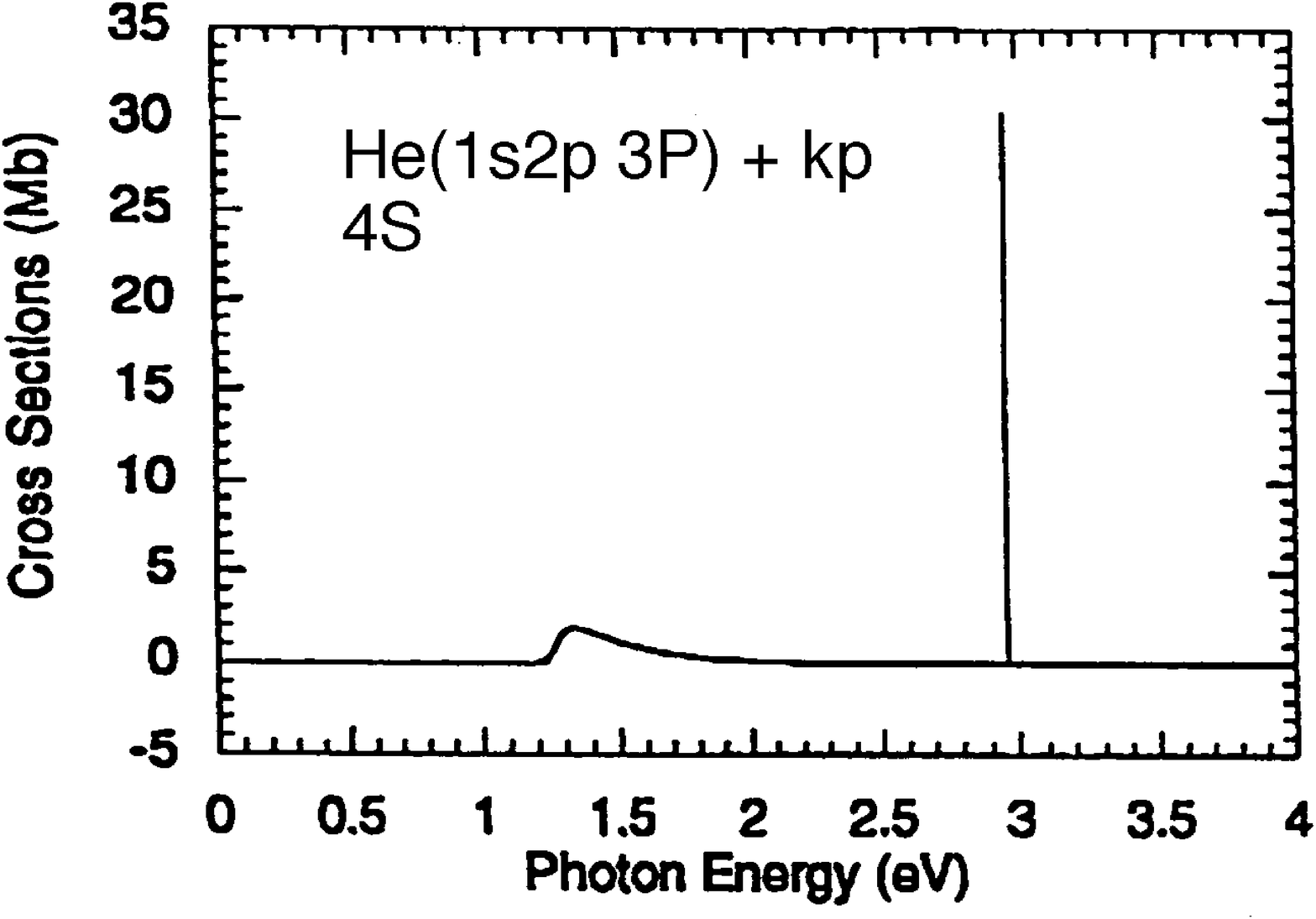,
width=0.49\textwidth}}
\subfigure[He(1s2p\,$^{3}$P$^{\text{o}}$) partial cross
sections of $^{4}$S symmetry near the 1s3s4s\,$^{4}$S
state]{\epsfig{file=bi600.eps, width=0.49\textwidth}}
\subfigure[Overview of He(1s2p\,$^{3}$P$^{\text{o}}$) partial cross
sections of $^{4}$P symmetry]{\epsfig{file=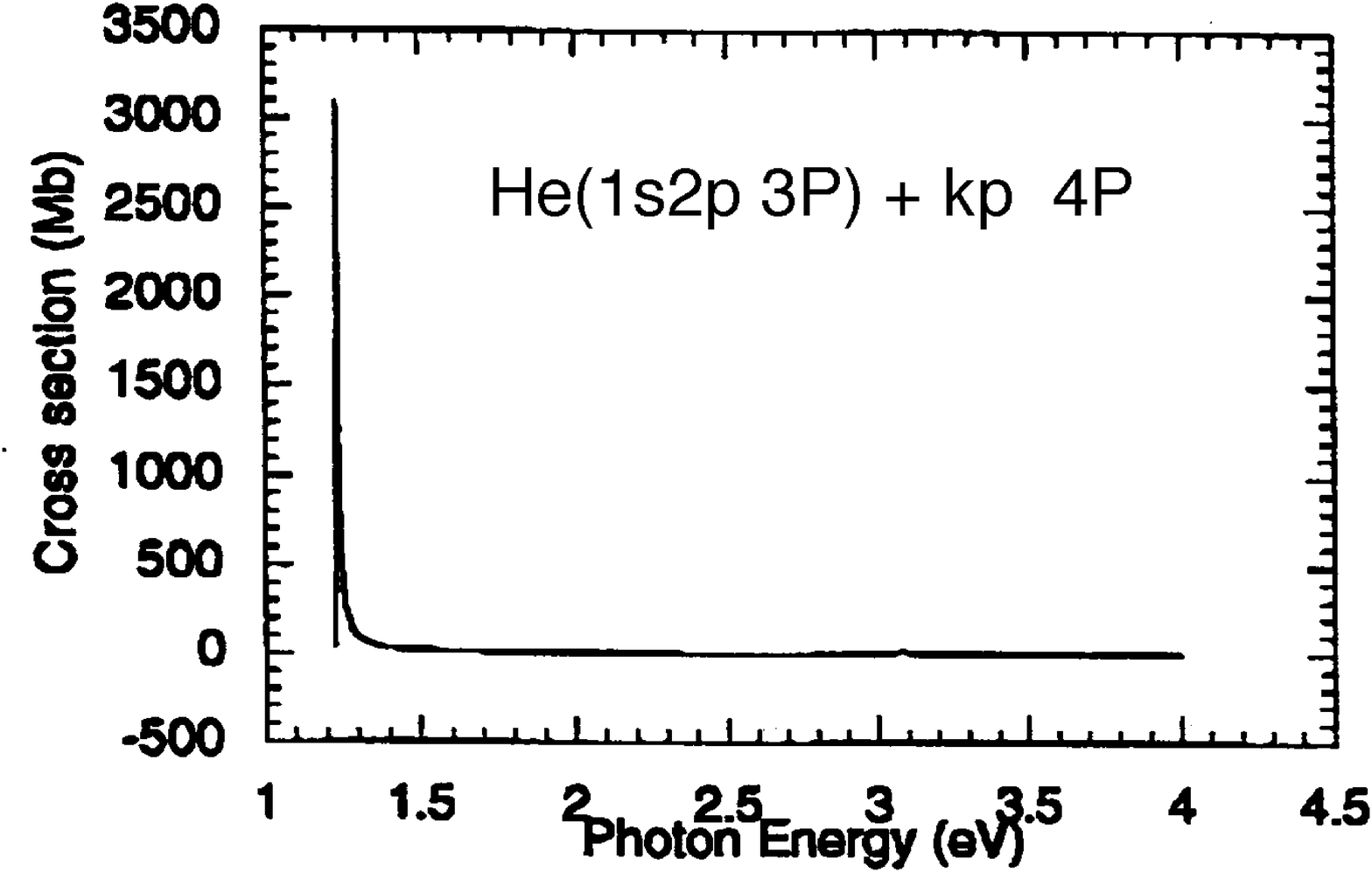,
width=0.49\textwidth}}
\subfigure[He(1s2p\,$^{3}$P$^{\text{o}}$) partial cross
sections of $^{4}$P symmetry near the 1s3p$^{2}$\,$^{4}$P and
1s3p4p\,$^{4}$P state]{\epsfig{file=bi640.eps, width=0.49\textwidth}}
\subfigure[Overview of He(1s2p\,$^{3}$P$^{\text{o}}$) partial cross
sections of $^{4}$D symmetry]{\epsfig{file=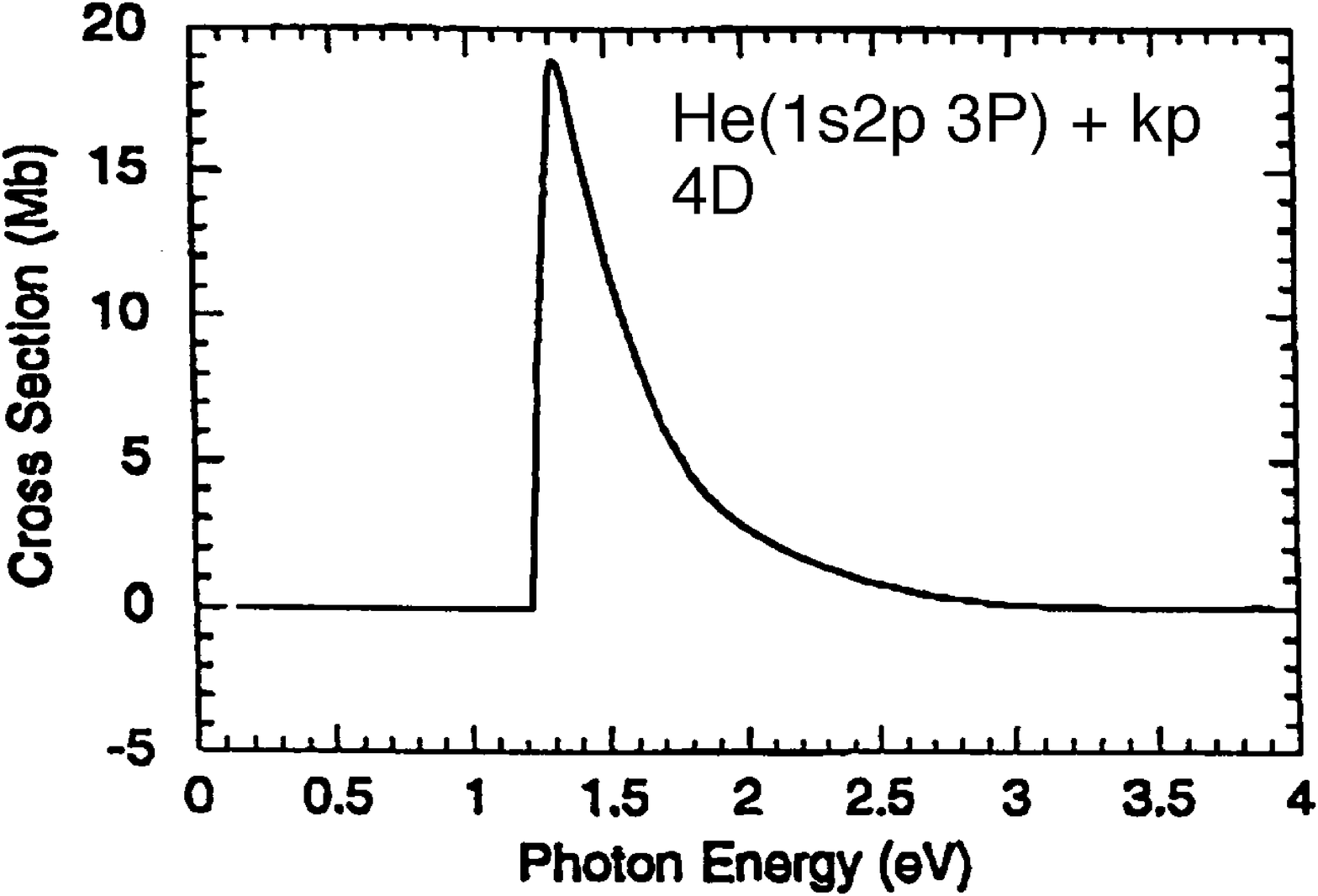,
width=0.49\textwidth}}
\protect\caption[He(1s2p\,$^{3}$P$^{\text{o}}$) partial cross
sections]{\label{bi710}\sloppy The He(1s2p\,$^{3}$P$^{\text{o}}$)
partial cross sections: The partial photodetachment cross sections
monitored by selectively detecting He atoms in the
1s2p\,$^{3}$P$^{\text{o}}$ state \cite{Xi-96-2}. }
\end{figure}

\end{appendix}
\cleardoublepage
\addcontentsline{toc}{chapter}{Bibliography}

\begin{thebibliography}{100}

\bibitem{Lju-94}
U.~{Ljungblad}, A.~{Klinkm{\"u}ller}, and D.~{Hanstorp}.
\newblock A new apparatus for studies of negative ions.
\newblock In Mike {Hopkins} and Samantha {Fahy}, editors, {\em Fifth european
  workshop on the production and application of light negative ions}, pages
  35--40, Glasnevin, Dublin 9, Ireland, March 1994. Dublin City University.

\bibitem{Ber-95-3}
U.~{Berzinsh}, G.~{Haeffler}, D.~{Hanstorp}, A.~{Klinkm{\"u}ller},
  E.~{Lindroth}, U.~{Ljungblad}, and D.~J. {Pegg}.
\newblock Resonance structure in the {L}i$^{-}$ photodetachment cross section.
\newblock In R.~C. {Thompson}, editor, {\em 5th EPS conference on atomic and
  molecular physics}, page~65, Edinburgh, UK, April 1995. European physical
  society.

\bibitem{Kli-96}
A.~E. {Klinkm{\"u}ller}, G.~{Haeffler}, U.~{Berzinsh}, D.~{Hanstorp}, I.~Yu.
  {Kiyan}, U.~{Ljungblad}, D.~J. {Pegg}, and J.~{Rangell}.
\newblock Electron affinities of {As}, {Te} and {Li}: {A} comparison of three
  different measurements.
\newblock In {\em RIS-96, Program and Abstracts}, page 131. Eighth
  interantional symposium on resonance ionisation spectroscopy and its
  applications, June 1996.

\bibitem{Hae-96-4}
G.~{Haeffler}, A.~E. {Klinkm{\"u}ller}, D.~{Hanstorp}, I.~Yu. {Kiyan}, C.~W.
  {Ingram}, D.~J. {Pegg}, and J.~R. {Peterson}.
\newblock Photodetachment study of doubly excited states in {He}$^{-}$.
\newblock {\em Bull. Am. Phys. Soc.}, 42(2):1136, April 1996.

\bibitem{Hae-96-5}
G.~{Haeffler}, U.~{Berzinsh}, D.~{Hanstorp}, I.~Yu. {Kiyan} A.~E.
  {Klinkm{\"u}ller}, U.~{Ljungblad}, and J.~{Rangell}.
\newblock Electron affinities of {As}, {Te} and {Li}: {A} comparison of three
  different measurements.
\newblock {\em Bull. Am. Phys. Soc.}, 41(3):1086, May 1996.

\bibitem{Kli-97-2}
Andreas~E. {Klinkm{\"u}ller}, Gunnar {Haeffler}, Dag {Hanstorp}, Chris~W.
  {Ingram}, Igor~Yu. {Kiyan}, Uldis {Berzinsh}, David~J. {Pegg}, and James~R.
  {Peterson}.
\newblock Photodetachment study of {H}e$^{-}$ resonances.
\newblock In H.-D. {Kronfeldt}, editor, {\em 29. EGAS Berlin abstracts},
  europhysics conference abstracts, pages 16--17. European group for atomic
  spectroscopy, European physical society, July 1997.

\bibitem{Kli-97-3}
Andreas~E. {Klinkm{\"u}ller}, Gunnar {Haeffler}, Dag {Hanstorp}, Chris~W.
  {Ingram}, Igor~Yu. {Kiyan}, Uldis {Berzinsh}, David~J. {Pegg}, and James~R.
  {Peterson}.
\newblock Photodetachment study of {H}e$^{-}$ resonances.
\newblock In F.~{Aumayr}, G.~{Betz}, and H.~P. {Winter}, editors, {\em XX.
  ICPEAC}, volume~1, page WE047, July 1997.

\bibitem{UIP-78}
L.~{Kerwin}, editor.
\newblock {\em Symbols, units and nomenclature in physics}.
\newblock {U.I.P.} 20. Union internationale de physique pure et
  appliqu{\'{e}}e, 1978.

\bibitem{Tay-95}
Barry~N. {Taylor}.
\newblock Guide for the use of the international system of units {(SI)}.
\newblock NIST Special publication 811, United {S}tates {D}epartment of
  {C}ommerce {T}echnology {A}dministration, National {I}nstitute of {S}tandards
  and {T}echnology, Gaithersburg, MD-20\,899-0001, April 1995.

\bibitem{Coh-88}
E.~Richard {Cohen} and Barry~N. {Taylor}.
\newblock The 1986 {CODATA} recommended values of the fundamental physical
  constants.
\newblock {\em J. Phys. Chem. Ref. Data}, 17(4):1795--1803, 1988.

\bibitem{Mas-38}
Sir~Harrie Massey.
\newblock {\em Negative ions}.
\newblock Cambridge University Press, The Pitt building, Trumpington street,
  Cambridge CB2 1RP, third edition, 1976.

\bibitem{Smi-76}
B.~M. {Smirnov}.
\newblock {\em Negative Ions}.
\newblock McGraw-Hill, New York, 1976.

\bibitem{Buc-94}
Stephen~J. {Buckman} and Charles~W. {Clark}.
\newblock Atomic negative-ion resonances.
\newblock {\em Rev. Mod. Phys.}, 66(2):539--655, April 1994.

\bibitem{Esa-86}
A.~V. {Esaulov}.
\newblock Electron detachment from atomic negative ions.
\newblock {\em Ann. Phys. Fr.}, 11(5):493--592, October 1986.

\bibitem{And-91}
T.~{Andersen}.
\newblock Spectroscopy of negativ ions.
\newblock {\em Phys. Scr.}, T34:23--35, 1991.

\bibitem{Opa-74}
V.~A. {Oparin}, R.~N. {II'in}, I.~T. {Serenkov}, and E.~S. {Solov'ev}.
\newblock Negative ion exicted states and the determination of their binding
  energy by electron detachment by an electric field.
\newblock {\em Sov. Phys.-JETP}, 39(6):989--994, 1974.

\bibitem{Nad-92}
M.-J. {Nadeau}, X.-L. {Zhao}, M.~A. {Garwan}, and A.~E. {Litherland}.
\newblock Ca negative ion binding energy.
\newblock {\em Phys. Rev.~A}, 46(7):R3588--R3590, October 1992.

\bibitem{Dra-96-2}
Gordon W.~F. {Drake}, editor.
\newblock {\em Atomic molecular \& optical physics \textsc{handbook}},
  chapter~58, pages 681--689.
\newblock American Institute of Physics, 1996.

\bibitem{Dra-96-4}
Gordon W.~F. {Drake}, editor.
\newblock {\em Atomic molecular \& optical physics \textsc{handbook}},
  chapter~59, pages 690--700.
\newblock American Institute of Physics, 1996.

\bibitem{Hot-85}
H.~{Hotop} and W.~C. {Lineberger}.
\newblock Binding energies in atomic negative ions.
\newblock {\em J. Phys. Chem. Ref. Data}, 14(3):731--750, 1985.

\bibitem{Blo-95}
C.~{Blondel}.
\newblock Recent experimental achievements with negative ions.
\newblock {\em Phys. Scr.}, T58:31--42, 1995.

\bibitem{Pet-95}
V.~V. {Petrunin}, J.~D. {Voldstad}, P.~{Balling}, P.~{Kristensen}, and
  T.~{Andersen}.
\newblock Resonant ionization spectroscopy of {B}a$^{-}$: Metastable and stable
  ions.
\newblock {\em Phys. Rev. Lett.}, 75(10):1911--1914, September 1995.

\bibitem{Hae-96-1}
Gunnar {Haeffler}, Dag {Hanstorp}, Igor {Kiyan}, Andreas~E. {Klinkm{\"u}ller},
  Ulric {Ljungblad}, and David~J. {Pegg}.
\newblock Electron affinity of {L}i: {A} state-selective measurement.
\newblock {\em Phys. Rev.~A}, 53(6):4127--4131, June 1996.
\newblock e-print: physics/9703013.

\bibitem{Har-90}
P.~G. {Harris}, H.~C. {Bryant}, A.~H. {Mohagheghi}, R.~A. {Reeder},
  H.~{Sharifian}, C.~Y. {Tang}, J.~B. {Donahue}, C.~R. {Quick}, D.~C. {Rislov},
  W.~W. {Smith}, and J.~E. {Stewart}.
\newblock Observation of high lying resonances in the {H}$^{-}$ ion.
\newblock {\em Phys. Rev. Lett.}, 65(3):309--312, 1990.

\bibitem{And-93}
T.~{Andersen}, L.~H. {Andersen}, P.~{Balling}, H.~K. {Haugen}, P.~{Hvelplund},
  W.~W. {Smith}, and K.~{Taulbjerg}.
\newblock Metastable-ion lifetime studies utilizing a heavy-ion storage ring:
  Measurements on {H}e$^{-}$.
\newblock {\em Phys. Rev.~A}, 47(2):890--896, February 1993.

\bibitem{Lar-88}
D.~J. {Larson}, C.~J. {Edge}, R.~E. {Elmquist}, N.~B. {Mansour}, and
  R.~{Trainham}.
\newblock Physics with negative ions in ion traps.
\newblock {\em Phys. Scr.}, T22:183--190, 1988.

\bibitem{Lar-85}
D.~J. {Larson} and R.~{Stoneman}.
\newblock Photodetachment of atomic negative ions near threshold in a magnetic
  field.
\newblock {\em Phys. Rev.~A}, 31(4):2210--2214, 1985.

\bibitem{Cra-87}
M.~{Crance}.
\newblock Multiphoton detachment from the negtive ion of fluorine.
\newblock {\em J. of Phys.~B}, 20:L411--L416, February 1987.

\bibitem{Blo-93}
C.~{Blondel} and C.~{Delsart}.
\newblock Two-photon detachment from the negative ions {F}$^{-}$ and {I}$^{-}$:
  {A}ngular distributions with elliptically polarised light.
\newblock {\em Nucl. Instrum. Methods B}, 79:156--158, June 1993.

\bibitem{Sta-94}
H.~{Stapelfeldt}, P.~{Kristensen}, U.~{Ljungblad}, T.~{Andersen}, and H.~K.
  {Haugen}.
\newblock Autoionizing states of negative ions in strong resonant laser fields:
  The negative rubidium ion.
\newblock {\em Phys. Rev.~A}, 50:1618--1626, August 1994.

\bibitem{Kri-93}
P.~{Kristensen}, H.~{Stapelfeldt}, P.~{Balling}, and T.~{Andersen}.
\newblock Spectroscopy of negative ions utilizing multiphoton detachment in a
  {R}aman coupling regime.
\newblock {\em Phys. Rev. Lett.}, 71(21):3435--3438, November 1993.

\bibitem{Tho-96-2}
J.~{Th{\o}gersen}, L.~D. {Steele}, M.~{Scheer}, H.~K. {Haugen},
  P.~{Kristensen}, P.~{Balling}, H.~{Stapelfeldt}, and T.~{Andersen}.
\newblock Fine structure measurements for negative ions: Studies of {S}e$^{-}$
  and {T}e$^{-}$.
\newblock {\em Phys. Rev.~A}, 53(5):3023--3028, May 1996.

\bibitem{Wei-78}
M.~Weissbluth.
\newblock {\em Atoms and Molecules}.
\newblock Academic Press, New York, 1986.

\bibitem{Cow-81-2}
Robert~Duane {Cowan}.
\newblock {\em The theory of atomic structure and spectra}, chapter~4, pages
  93--141.
\newblock Los Alamos series in basic and applied sciences. University of
  {C}alifornia {P}ress, Berkeley and Los Angeles, California, 1981.

\bibitem{Amu-90-1}
M.~Ya. {Amusia}.
\newblock {\em Atomic {P}hotoeffect}, chapter 1.3, pages 6--7.
\newblock Physics of atoms and molecules. Plenum press, New {Y}ork and
  {L}ondon, 1990.

\bibitem{Alo-96}
J.~A. {Alonso} and N.~A. {Cordero}.
\newblock Variation of the ground state correlation energy of atoms with atomic
  number.
\newblock {\em J. of Phys.~B: At., Mol. Opt.}, 29:1629--1636, 1996.

\bibitem{Sal-96}
Sten {Salomonson}, H{\aa}kan {Warston}, and Ingvar {Lindgren}.
\newblock Many-body calculations of the electron affinity for {C}a and {S}r.
\newblock {\em Phys. Rev. Lett.}, 76(17):3092--3095, April 1996.

\bibitem{Bra-91}
T.~{Brage} and C.~{Froese Fischer}.
\newblock Autodetachment of negative ions.
\newblock {\em Phys. Rev.~A}, 44:72--79, 1991.

\bibitem{Dra-96-3}
Gordon W.~F. {Drake}, editor.
\newblock {\em Atomic molecular \& optical physics \textsc{handbook}},
  chapter~21, pages 243--257.
\newblock American Institute of Physics, 1996.

\bibitem{Bet-57}
Hans~A. {Bethe} and Edwin~E. {Salpeter}.
\newblock {\em Quantum mechanics of one- and two-electron atoms}.
\newblock Springer Verlag, Berlin--G{\"o}ttingen--Heidelberg, third edition,
  1957.
\newblock published by Plenum Publishing Corporation.

\bibitem{Coo-63}
J.~W. {Cooper}, U.~{Fano}, and F.~{Pratts}.
\newblock Classification of two electron excitations of helium.
\newblock {\em Phys. Rev. Lett.}, 10(12):518--521, 1963.

\bibitem{Mad-63}
R.~P. Madden and K.~Codling.
\newblock New auto-ionizing atomic energy levels in {H}e, {N}e, and {A}r.
\newblock {\em Phys. Rev. Lett.}, 10(12):516--518, 1963.

\bibitem{Reh-78-2}
Paul {Rehmus}, Michael~E. {Kellman}, and Stephen {Berry}.
\newblock Spatial correlation of atomic electrons: He.
\newblock {\em Cont. Phys.}, 31:239--262, 1978.

\bibitem{Rau-90}
A.~R.~P. {Rau}.
\newblock Group theoretical treatment of strongly correlated atomic dynamics.
\newblock {\em Rep. on Prog. in Phys.}, 53:181--220, 1990.

\bibitem{Ber-89-2}
R.~S. Berry.
\newblock How good is {N}iels {B}ohr's atomic model?
\newblock {\em Cont. Phys.}, 30(1):1--19, 1989.

\bibitem{Ros-91}
J.~M. {Rost} and J.~S. {Briggs}.
\newblock Saddle structure of the three body {C}oulomb problem; symmetries of
  doubly excited states and propensity rules for transitions.
\newblock {\em J. of Phys.~B: At., Mol. Opt.}, 24:4293--4322, 1991.
\newblock Topical review.

\bibitem{Win-94}
Dieter {Wintgen} and Klaus {Richter}.
\newblock Semiclassical nature of planetary atom states.
\newblock {\em Comments At. Mol. Phys.}, 29(5):261--274, January 1994.

\bibitem{Lin-93-2}
C.~D. {Lin}.
\newblock {\em Review of fundamental processes and applications of atoms and
  ions}, book~8., pages 357--401.
\newblock World Scientific Co.~Pte.~Ltd., P O Box 128, Farrer Road, Singapore
  91\,28, 1993.

\bibitem{Lin-87-2}
C.~D. {Lin}.
\newblock Classification of two-electron doubly excited states.
\newblock {\em Comments At. Mol. Phys.}, 19(2):89--98, 1987.

\bibitem{SiH-75}
O.~Sinano\u{g}lu and D.~R. Herrick.
\newblock Group theoretic prediction of configuration mixing effects due to
  {C}oulomb repulsions in atoms with applications to doubly-excited spectra.
\newblock {\em J. Chem. Phys.}, 62:886--892, 1975.

\bibitem{SiH-76}
O.~Sinano\u{g}lu and D.~R. Herrick.
\newblock Erratum: {G}roup theoretic prediction of configuration mixing effects
  due to {C}oulomb repulsions in atoms with applications to doubly excited
  spectra {[J. Chem. Phys. 62, 886 (1975)]}.
\newblock {\em J. Chem. Phys.}, 65:850, 1976.

\bibitem{HeS-75}
D.~R. Herrick and O.~Sinano\u{g}lu.
\newblock Comparison of doubly excited helium energy levels, isoelectronic
  series, auto-ionization lifetimes, and group theoretical configuration mixing
  predictions with large configuration interaction calculations and
  experimental spectra.
\newblock {\em Phys. Rev.~A}, 11:97--110, 1975.

\bibitem{Lin-76}
C.~D. Lin.
\newblock Properties of resonance states in {H$^-$}.
\newblock {\em Phys. Rev.~A}, 14:30--35, 1976.

\bibitem{KeH-78}
M.~E. Kellman and D.~R. Herrick.
\newblock Rotor-like spectra for some doubly excited two-electron states.
\newblock {\em J. of Phys.~B: At. and Mol.}, 11:L755--L759, 1978.

\bibitem{HeK-80}
D.~R. Herrick and M.~E. Kellman.
\newblock Novel supermultiplet energy levels for doubly excited {H}e.
\newblock {\em Phys. Rev.~A}, 21:418--425, 1980.

\bibitem{HKP-80}
D.~R. Herrick, M.~E. Kellman, and R.~D. Poliak.
\newblock Supermultiplet classification of higher intrashell doubly excited
  states of {H$^-$} and {H}e.
\newblock {\em Phys. Rev.~A}, 22(4):1517--1535, October 1980.

\bibitem{Lin-82-2}
C.~D. Lin.
\newblock Properties of high-lying doubly excited states of {H}$^-$.
\newblock {\em Phys. Rev.~A}, 25(3):1535--1545, March 1982.

\bibitem{Lin-82-1}
C.~D. {Lin}.
\newblock Radial and angular correlations of doubly excited electrons.
\newblock {\em Phys. Rev.~A}, 25(1):76--87, January 1982.

\bibitem{FeB-86}
J.~M. Feagin and J.~S. Briggs.
\newblock Molecular descripion of two-electron atoms.
\newblock {\em Phys. Rev. Lett.}, 57:984--987, 1986.

\bibitem{Wul-73}
Carl {Wulfman}.
\newblock Approximate dynamical symmetry of two-electron atoms.
\newblock {\em Chem. Phys. Lett.}, 23(3):370, December 1973.

\bibitem{Wul-83}
C.~{Wulfman}.
\newblock Configuration mixing within the 2s-2p shell.
\newblock {\em Phys. Rev. Lett.}, 51(13):1159--1162, September 1983.

\bibitem{Hei-93}
Thomas~A. {Heim}.
\newblock Decay properties of doubly excited ridge states.
\newblock {\em J. of Phys.~B: At., Mol. Opt.}, 26:4343--4365, 1993.

\bibitem{Lin-85}
C.~D. {Lin}.
\newblock Classification and supermultiplet structure of doubly excited states.
\newblock {\em Nucl. Instrum. Methods B}, 240:572--576, 1985.

\bibitem{RoB-90}
J.~M. Rost and J.~S. Briggs.
\newblock Propensity rules for radiative and non-radiative decay of
  doubly-excited states.
\newblock {\em J. of Phys.~B: At., Mol. Opt.}, 23:L339 -- L346, 1990.

\bibitem{Tyk-78}
P.~{Tykesson}.
\newblock The production of negative heavy ion beams through charge exchange
  processes.
\newblock Symposium of Northeastern Accelerator Personnel Oak Ridge, Tennessee,
  USA., October 1978.

\bibitem{Amu-90-3}
M.~Ya. {Amusia}.
\newblock {\em Atomic {P}hotoeffect}, chapter 2.1, pages 13--16.
\newblock Physics of atoms and molecules. Plenum press, New {Y}ork and
  {L}ondon, 1990.

\bibitem{Dem-88-3}
Yu.~N. {Demkov} and V.~N. {Ostrovskii}.
\newblock {\em Zero-range potentials and their applications in atomic physics},
  chapter 1.3, pages 8--14.
\newblock Physics of atoms and molecules. Plenum Press, 233 Spring Street, New
  York, N.Y. 10\,013, 1988.

\bibitem{Amu-90-4}
M.~Ya. {Amusia}.
\newblock {\em Atomic {P}hotoeffect}, chapter 3.6, pages 72--74.
\newblock Physics of atoms and molecules. Plenum press, New {Y}ork and
  {L}ondon, 1990.

\bibitem{Lin-94-2}
Eva {Lindroth}.
\newblock private communication, 1994.

\bibitem{Wig-48}
Eugene~P. {Wigner}.
\newblock On the behavior of cross sections near thresholds.
\newblock {\em Phys. Rev.}, 73(9):1002--1009, 1948.

\bibitem{Bae-86}
Y.~K. {Bae} and J.~R. {Peterson}.
\newblock Modified photodetachment threshold behavior near resonances.
\newblock In J.~R.~{Peterson} D.~C.~{Lorents}, W. E.~{Meyerhof}, editor, {\em
  Electronic and atomic collisions}, pages 799--806. Elsevier Science
  Publisher, 1986.

\bibitem{Fri-90}
H.~Friedrich.
\newblock {\em {Theoretische Atomphysik}}.
\newblock {Springer--Lehrbuch}. Springer--Verlag, Berlin, 1990.

\bibitem{Joa-75-1}
C.~J. Joachain.
\newblock {\em Quantum Collision Theory}, volume~1.
\newblock North Holland Publishing Company, Amsterdam, 1975.

\bibitem{Arf-85}
George {Arfken}.
\newblock {\em Mathematical methods for physicists}.
\newblock Academic Press, INC., 1985.

\bibitem{OMa-65}
Thomas~F. {O'Malley}.
\newblock Effect of long-range final state forces on the negative ion
  photodetachment cross section near threshold.
\newblock {\em Phys. Rev.}, 137(6A):A1668--A1672, March 1965.

\bibitem{Hot-73}
H.~{Hotop} and W.~C. {Lineberger}.
\newblock Dye-laser photodetachment studies of $\mbox{Au}^-$, $\mbox{Pt}^-$,
  $\mbox{PtN}^{-}$, and $\mbox{Ag}^{-\ast }$.
\newblock {\em J. Chem. Phys.}, 58(6):2379--2387, March 1973.

\bibitem{Hot-73-2}
H.~{Hotop}, T.~A. {Patterson}, and W.~C. {Lineberger}.
\newblock High resolution photodetachment study of se$^{\text{-}}$ ions.
\newblock {\em Phys. Rev.~A}, 8(2):762--774, August 1973.

\bibitem{Far-89}
John~W. {Farley}.
\newblock Photodetachment cross section of negative ions: {T}he range of
  validity of the {W}igner threshold law.
\newblock {\em Phys. Rev.~A}, 40(11):6286--6292, December 1989.

\bibitem{Pet-85}
J.~R. {Peterson}, Y.~K. {Bae}, and D.~L. {Huestis}.
\newblock Threshold behavior near an electronic shape resonance: Analysis of
  the $\mbox{{H}e}(^{3}\!{P})$ threshold in $\mbox{{H}e}^{-}$ photodetachment
  and determination of the $\mbox{{H}e}(2^{3}\!{S})$ electron affinity.
\newblock {\em Phys. Rev. Lett.}, 55(7):692--695, August 1985.

\bibitem{Wal-94-4}
C.~W. {Walter}, J.~A. {Seifert}, and J.~R. {Peterson}.
\newblock Re-examination of the {H}e$^{-}$ $1s2p^{2}\,^{4}\!{P}^{\mathrm{e}}$
  shape resonance: Details of its properties and a precise electron affinifty
  for {H}e $2\;^{3}\!{S}$.
\newblock {\em Phys. Rev.~A}, 50(3):2257--2262, September 1994.

\bibitem{Fan-61}
U.~Fano.
\newblock Effects of configuration interactions on intensities and phase
  shifts.
\newblock {\em Phys. Rev.}, 124:1866--1878, 1961.

\bibitem{Fer-87}
T.~A. {Ferrett}, D.~W. {Lindle}, P.~A. {Heimann}, W.~D. {Brewer}, U.~{Becker},
  H.~G. {Kerkhoff}, and D.~A. {Shirley}.
\newblock Lithium 1$s$ main line and satellite photo-emission: {R}esonant and
  nonresonant behavior.
\newblock {\em Phys. Rev.~A}, 36(7):3172--3183, October 1987.

\bibitem{Sho-68}
Bruce~W. {Shore}.
\newblock Parametrization of absorption line profiles.
\newblock {\em Phys. Rev.}, 171(1):43--54, July 1968.

\bibitem{Sta-77}
Anthony~F. {Starace}.
\newblock Behavior of partial cross sections and braching ratios in the
  neighborhood of a resonance.
\newblock {\em Phys. Rev.~A}, 16(1):231--242, July 1977.

\bibitem{Fri-90-1}
H.~Friedrich.
\newblock {\em {Theoretische Atomphysik}}, chapter 3.2.3, pages 138--143.
\newblock {Springer--Lehrbuch}. Springer--Verlag, Berlin, 1990.

\bibitem{Gre-95-2}
C.~H. {Green}.
\newblock private communication, 1995.

\bibitem{Fri-90-2}
H.~Friedrich.
\newblock {\em {Theoretische Atomphysik}}, chapter 1.4.3, pages 36--38.
\newblock {Springer--Lehrbuch}. Springer--Verlag, Berlin, 1990.

\bibitem{Bae-85}
Young~K. {Bae} and James~R. {Peterson}.
\newblock Effect of a virtual state near an $s$-wave threshold: Absolute
  {L}i$^{-}$ photodetachment cross sections near the {L}i(2\,$^{2}\!{P}$)
  threshold.
\newblock {\em Phys. Rev.~A}, 32(3):1917--1920, September 1985.

\bibitem{Amu-90-5}
M.~Ya. {Amusia}.
\newblock {\em Atomic {P}hotoeffect}, chapter 3.7, pages 74--77.
\newblock Physics of atoms and molecules. Plenum press, New {Y}ork and
  {L}ondon, 1990.

\bibitem{Kau-76}
S.~L. {Kaufman}.
\newblock High-resolution laser spectroscopy in fast beams.
\newblock {\em Opt. Comm.}, 17(3):309--312, 1976.

\bibitem{Mid-89}
Roy {Middelton}.
\newblock A negative ion cookbook, October 1989.

\bibitem{Ros-95}
J{\"o}rgen {Roslund}.
\newblock Design of a sputter negative ion source system.
\newblock Master's thesis, Chalmers tekniska h{\"o}gskola AB, S-412\,96
  G{\"o}teborg, Sweden, August 1995.
\newblock GIPR-330, ISSN 0280-2872.

\bibitem{Har-76}
E.~{Harting} and F.~H. {Read}.
\newblock {\em Electrostatic lenses}.
\newblock Elsevier scientific publishing company, Amsterdam-Oxford-New York,
  1976.

\bibitem{Hae-96-3}
Gunnar {Haeffler}.
\newblock Photodetachment spectroscopy: {N}egative ion bound states and
  resonances.
\newblock Master's thesis, G{\"o}teborgs Universitet och Chalmer tekniska
  h{\"o}gskola AB, Department of atomic physics, S--412\,96 G{\"o}teborg, 1996.
\newblock ISBN 91-7197-379-6.

\bibitem{Han-92-2}
D.~{Hanstorp}.
\newblock A secondary emission detector capable of preventing detection of the
  photoelectric effect induced by pulsed lasers.
\newblock {\em Meas. Sci. Technol.}, 3:523--527, 1992.

\bibitem{Han-89-1}
D.~{Hanstorp}, C.~{Bengtsson}, and D.~J. {Larson}.
\newblock Angular distribution in photodetachment from {O}$^{-}$.
\newblock {\em Phys. Rev.~A}, 40(2):670--675, July 1989.

\bibitem{Ran-96}
Jonas {Rangell}.
\newblock Computer control and wavelength calibration of a dye laser.
\newblock {GIPR} 337, G{\"o}terborgs universitet and Chalmers tekniska
  h{\"o}gskola AB, SE-412\,96 G{\"o}teborg, Sweden, 1996.
\newblock ISSN 0280-2872.

\bibitem{PFT-89}
W.~H. Press, B.~P. Flannery, S.~A. Teukolsky, and W.~T. Vetterling.
\newblock {\em Numerical Recipes. The Art of Scientific Computing (FORTRAN
  Version)}.
\newblock Cambridge University Press, Cambridge, 1989.

\bibitem{Han-92-1}
D.~{Hanstorp} and M.~{Gustafsson}.
\newblock Determination of the electron affinity of iodine.
\newblock {\em J. of Phys.~B: At., Mol. Opt.}, 25:1773--1783, 1992.

\bibitem{Neu-85}
D.~M. {Neumark}, K.~R. {Lykke}, T.~{Andersen}, and W.~C. {Lineberger}.
\newblock Laser photodetachment measurement of the electron affinity of atomic
  oxygen.
\newblock {\em Phys. Rev.~A}, 32(3):1890--1892, September 1985.

\bibitem{Ber-95-4}
U.~{Berzinsh}, M.~{Gustafsson}, D.~{Hanstorp}, A.~{Klinkm{\"u}ller},
  U.~{Ljungblad}, and A.-M. {M{\aa}rtensson-Pendrill}.
\newblock Isotope shift in the electron affinity of chlorine.
\newblock {\em Phys. Rev.~A}, 51(1):231--238, January 1995.

\bibitem{Blo-89}
C.~{Blondel}, P.~{Cacciani}, C.~{Delsart}, and R.~{Trainham}.
\newblock High-resolution determination of the electron affinitiy of fluorine
  and bromine using crossed ion and laser beams.
\newblock {\em Phys. Rev.~A}, 40:3698--3701, October 1989.

\bibitem{Hae-96-2}
Gunnar {Haeffler}, Andreas~E. {Klinkm{\"u}ller}, Jonas {Rangell}, Uldis
  {Berzinsh}, and Dag {Hanstorp}.
\newblock The electron affinity of tellurium.
\newblock {\em Z. Phys. D}, 38:211--214, October 1996.
\newblock e-print: physics/9703012.

\bibitem{Min-73}
Lennart {Minnhagen}.
\newblock Spectrum and the energy levels of neutral argon, {A}r {\sc i}.
\newblock {\em J. Opt. Soc. Am.}, 63(10):1185--1198, 1973.

\bibitem{Cha-94}
Edward~S. {Chang} and William~G. {Schoenfeld}.
\newblock Improved experimental and theoretical energy levels of neon {\sc i}.
\newblock {\em Phys. Scr.}, 49:26--33, 1994.

\bibitem{Kau-72}
Victor {Kaufman} and Lennart {Minnhagen}.
\newblock Accurate ground term combinations in {N}e {\sc i}.
\newblock {\em J. Opt. Soc. Am.}, 62(1):92--95, January 1972.

\bibitem{Sla-77}
J.~{Slater} and W.~C. {Lineberger}.
\newblock High resolution photodetachment studies of {P}$^{-}$ and {T}e$^{-}$.
\newblock {\em Phys. Rev.~A}, 15(6):2277--2282, June 1977.

\bibitem{Hur-79}
G.~S. {Hurst}, M.~G. {Payne}, S.~D. {Kramer}, and J.~P. {Young}.
\newblock Resonance ionization spectroscopy and one-atom detection.
\newblock {\em Rev. Mod. Phys.}, 51(4):767--819, October 1979.

\bibitem{Lju-96}
U.~{Ljungblad}, D.~{Hanstorp}, U.~{Berzinsh}, and D.~J. {Pegg}.
\newblock Observation of doubly excited states in {L}i$^{-}$.
\newblock {\em Phys. Rev. Lett.}, 77(18):3751--3754, October 1996.

\bibitem{Fel-76}
D.~{Feldmann}.
\newblock Infra-red photodetachment threshold measurements: {L}i$^{-}$ and
  {P}$^{-}$.
\newblock {\em Z. Phys. A}, 277:19--25, 1976.

\bibitem{Lan-74}
L.~D. {Landau} and E.~M. {Lifschitz}.
\newblock {\em Quantenmechanik}.
\newblock Akademie-Verlag-Berlin, 1979.

\bibitem{Fab-93}
I.~I. {Fabrikant}.
\newblock Theory of negative ion decay in an electrical field.
\newblock {\em J. of Phys.~B}, 26:2533--2541, 1993.

\bibitem{Smi-66}
B.~M. {Smirnov} and M.~I. {Chibisov}.
\newblock The breaking up of atomic particles by an electrical field and by
  electron collisions.
\newblock {\em Sov. Phys.-JETP}, 22(3):585--592, 1966.

\bibitem{Dem-2-81}
Yu.~N. {Demkov} and G.~F. {Drukarev}.
\newblock Loosely bound particle with nonzero orbital angular momentum in an
  electric or magnetic field.
\newblock {\em Sov. Phys.-JETP}, 54(4):650--656, 1981.

\bibitem{Dem-64}
Yu.~N. {Demkov} and G.~F. {Drukarev}.
\newblock Decay and polarizability of negative ions in an electric field.
\newblock {\em Sov. Phys.-JETP}, 20(3):614--618, March 1964.

\bibitem{Moo-71-1}
C.~E. Moore.
\newblock {\em Atomic Energy Levels as Derived from the Analysis of Optical
  Spectra. Volume I. $^1 \!$H to $^{23} \!$V}, volume~35 of {\em National
  Standards Reference and Data Series}.
\newblock National Bureau of Standards (U. S.), Washington, 1975.

\bibitem{Bae-88}
Y.~K. {Bae} and J.~R. {Petersson}.
\newblock Near-threshold measurements of {K}$^{-}$ two-electron
  photo-ionization cross section.
\newblock {\em Phys. Rev.~A}, 37(9):3254--3258, May 1988.

\bibitem{Man-75}
N.~L. {Manakov}, V.~D. {Ovsyannikov}, and L.~P. {Rapoport}.
\newblock Atomic calculations using pertubation theory with a model potential.
\newblock {\em Opt. Spectrosc.}, 38(2):115--117, February 1975.

\bibitem{Del-92-2}
J.~{Dellwo}, Y.~{Liu}, D.~J. {Pegg}, and G.~D. {Alton}.
\newblock Near-threshold photodetachment of the {L}i$^{-}$ ion.
\newblock {\em Phys. Rev.~A}, 45(3):1544--1547, February 1992.

\bibitem{San-95-2}
Craig~J. {Sansonetti} and Bruno {Richou}.
\newblock Precise measurements of the lithium resonance lines by doppler-free
  frequency modulation spectroscopy.
\newblock {\em Bull. Am. Phys. Soc.}, 40(4):1272--1273, 1995.

\bibitem{Chu-92-2}
Kwong~T. {Chung} and Paul {Fullbright}.
\newblock Electron affinity of lithium.
\newblock {\em Phys. Scr.}, 45:445--449, 1992.

\bibitem{Fis-93}
Charlotte {Froese Fischer}.
\newblock Convergence studies of {MCHF} calculations for {B}e and {L}i$^{-}$.
\newblock {\em J. of Phys.~B: At., Mol. Opt.}, 26(5):855--862, 1993.

\bibitem{Han-97}
D.~{Hanstorp}, G.~{Haeffler}, A.~E. {Klinkm{\"u}ller}, U.~{Ljungblad},
  U.~{Berzinsh}, I.~Yu. {Kiyan}, and D.~J. {Pegg}.
\newblock Two-electron dynamics in photodetachment.
\newblock {\em Adv. Quant. Chem.}, 29, 1997.
\newblock e-print: physics/9706013.

\bibitem{Ber-95-1}
U.~{Berzinsh}, G.~{Haeffler}, D.~{Hanstorp}, A.~{Klinkm{\"u}ller},
  E.~{Lindroth}, U.~{Ljungblad}, and D.~J. {Pegg}.
\newblock Resonance structure in the {L}i$^{-}$ photodetachment cross section.
\newblock {\em Phys. Rev. Lett.}, 74(24):4795--4798, June 1995.
\newblock e-print: physics/9703015.

\bibitem{Kli-97}
Andreas~E. {Klinkm{\"u}ller}, Gunnar {Haeffler}, Dag {Hanstorp}, Igor~Yu.
  {Kiyan}, Uldis {Berzinsh}, Christopher {Ingram}, David~J. {Pegg}, and
  James~R. {Peterson}.
\newblock Photodetachment study of the $1s3s4s\,^{4}\!{S}$ resonance in
  {He}$^{-}$.
\newblock {\em Phys. Rev.~A}, 56(4):2788--2791, October 1997.
\newblock e-print: physics/9703011.

\bibitem{Pan-96}
Cheng {Pan}, Anthony~F. {Starace}, and Chris~H. {Green}.
\newblock Photodetachment of {L}i$^{-}$ from the {L}i $3s$ threshold to the
  {L}i $6s$ threshold.
\newblock {\em Phys. Rev.~A}, 53(2):840--852, February 1996.

\bibitem{Lin-95}
Eva {Lindroth}.
\newblock Photodetachment of {H}$^{-}$ and {L}i$^{-}$.
\newblock {\em Phys. Rev.~A}, 52(4):2737--2749, October 1995.

\bibitem{Pan-94}
C.~{Pan}, A.~F. {Starace}, and C.~H. {Greene}.
\newblock Parallels between high doubly excited state spectra in {H}$^-$ and
  {L}i$^-$ photodetachment.
\newblock {\em J. of Phys.~B: At., Mol. Opt.}, 27:L137, 1994.

\bibitem{Ham-79}
M.~E. {Hamm}, R.~W. {Hamm}, J.~{Donahue}, P.~A.~M. {Gram}, J.~C. {Pratt}, M.~A.
  {Yates}, R.~D. {Bolton}, D.~A. {Clark}, H.~C. {Bryant}, C.~A. {Frost}, and
  W.~W. {Smith}.
\newblock Observation of narrow resonances in the $\mbox{H}^{-}$
  photodetachment cross section near the $n=3$ threshold.
\newblock {\em Phys. Rev. Lett.}, 43(23):1715--1718, 1979.

\bibitem{Xi-96-2}
Jinhua {Xi} and Charlotte {Froese Fischer}.
\newblock private communication, 1996.

\bibitem{Hib-39}
Julius~W. {Hiby}.
\newblock Massenspektrographische {U}ntersuchungen an {W}asserstoff- und
  {H}eliumstrahlen ({H}$^{+}_{3}$, {H}$^{-}_{2}$, {H}e{H}$^{+}$, {H}e{D}$^{+}$,
  {H}e$^{-}$).
\newblock {\em Analen der Phys.}, 34:473--487, 1939.
\newblock (5. Folge).

\bibitem{Pet-96-3}
V.~V. {Petrunin}, H.~H. {Andersen}, P.~{Balling}, P.~{Kristensen}, and
  T.~{Andersen}.
\newblock Resonance ionization spectroscopy of negative ions.
\newblock In {\em 15th international conference on atomic physics
  {Z}eeman-effect centenary}, page ThE2, Amsterdam, The Netherlands, 1996.

\bibitem{Bun-79}
A.~V. {Bunge} and C.~F. {Bunge}.
\newblock Electron affinity of helium $(1s2s)\,^{3}\!{S}$.
\newblock {\em Phys. Rev.~A}, 19(2):452--456, February 1979.

\bibitem{Com-80}
R.~N. {Compton}, G.~D. {Alton}, and D.~J. {Pegg}.
\newblock Photodetachment cross sections for
  {H}e$^{-}(^{4}\!{P}^{\mathrm{o}})$.
\newblock {\em J. of Phys.~B: At., Mol. Opt.}, 13:L651--L655, 1980.

\bibitem{Hod-81}
R.~V. {Hodges}, M.~J. {Coggiola}, and J.~R. {Peterson}.
\newblock Photodetachment cross sections for {H}e$^{-}$ $^{4}\!{P}$.
\newblock {\em Phys. Rev.~A}, 23(1):59--63, January 1981.

\bibitem{Peg-90}
D.~J. {Pegg}, J.~S. {Thompson}, J.~{Dellwo}, R.~N. {Compton}, and G.~D.
  {Alton}.
\newblock Partial cross sections for the photodetachment of metastable
  {H}e$^{-}$.
\newblock {\em Phys. Rev. Lett.}, 64:278--281, 1990.

\bibitem{Haz-81}
A.~U. {Hazi} and K.~{Reed}.
\newblock Theoretical photodetachment cross section for
  {H}e$^{-}$($^{4}\!{P}^{\mathrm{o}}$).
\newblock {\em Phys. Rev.~A}, 24(4):2269--2272, October 1981.

\bibitem{Sah-90}
H.~P. {Saha} and R.~N. {Compton}.
\newblock Theoretical studies of the photophysics of
  {H}e$^{-}$($1s2s2p$)$^{4}\!{P}^{\mathrm{o}}$.
\newblock {\em Phys. Rev. Lett.}, 64(13):1510--1513, March 1990.

\bibitem{Dou-90}
Maryvonne~Le {Dourneuf} and Shinichi {Watanabe}.
\newblock Grandparent model of the doubly excited {H}e$^{-\ast\ast}$ resonances
  from a hypersperical viewpoint.
\newblock {\em J. of Phys.~B: At., Mol. Opt.}, 23:3205--3224, 1990.

\bibitem{Xi-96}
Jinhua {Xi} and Charlotte {Froese Fischer}.
\newblock Cross section and angular distribution for the photodetachment of
  {H}e$^{-}$($1s2s2p\,^{4}\!{P}^{\mathrm{o}}$) below the {H}e($n=4$) threshold.
\newblock {\em Phys. Rev.~A}, 53(5):3169--3177, May 1996.

\bibitem{Wie-66-1}
W.~L. {Wiese}, M.~W. {Smith}, and B.~M. {Glennon}, editors.
\newblock {\em Atomic transition probabilities}, volume \textsc{I}, hydrogen
  through neon of {\em National standard reference data series}.
\newblock United States Department of Commerce, 1966.

\bibitem{Byl-97}
Miroslaw {Bylicki}.
\newblock Spectrum of doubly excited $^{4}${P}$^{\mbox{e}}$ resonances in
  {He}$^{-}$.
\newblock {\em J. of Phys.~B: At., Mol. Opt.}, 30:189--201, 1997.

\bibitem{The-95}
Spyros~I. {Themelis} and Cleanthes~A. {Nicolaides}.
\newblock Energies, widths and $l$-dependence of the {H}$^{-}$ $^{3}${P} and
  {He}$^{-}$ $^{4}${P} \textsc{teil} states.
\newblock {\em J. of Phys.~B: At., Mol. Opt.}, 28:L379--L385, 1995.

\bibitem{Meg-75}
W.~F. {Meggers}, C.~H. {Corliss}, and B.~F. {Scribner}.
\newblock NBS Monograph 145, 1975.

\bibitem{Kur-75}
R.~L. {Kurucz} and E.~{Peytremann}.
\newblock SAO Special Report 362, 1975.

\bibitem{The-94}
Spyros~I. {Themelis} and Cleanthes~A. {Nicolaides}.
\newblock Effect of interchannel coupling on the partial and total
  auto-ionization widths: {A}pplication to the $1s3s3p\,^{4}\!{P}^{\text{o}}$
  and $1s3p^{2}\,^{4}\!{P}$ states for ${Z}$=$2-5$,\,10.
\newblock {\em Phys. Rev.~A}, 49(3):1618--1622, March 1994.

\bibitem{Dav-90}
Brian~F. {Davis}.
\newblock Energy and auto-ionization width of the
  $1s3s3p\,^{4}\!{P}^{\mbox{o}}$ and the $1s3p3p\,^{4}\!{P}$ states in
  lithium-like ions.
\newblock {\em Phys. Rev.~A}, 41(11):5844--5855, June 1990.

\bibitem{Tho-96-1}
J.~{Th{\o}gersen}, M.~{Scheer}, L.~D. {Steele}, H.~K. {Haugen}, and W.~P.
  {Wijesundera}.
\newblock Two-photon detachment of negative ions via magnetic dipol
  transitions.
\newblock {\em Phys. Rev. Lett.}, 76(16):2870--2873, April 1996.

\bibitem{Del-92-1}
J.~{Dellwo}, Y.~{Liu}, C.~Y. {Tang}, D.~J. {Pegg}, and G.~D. {Alton}.
\newblock Photodetachment cross section for {L}i$^{-}$.
\newblock {\em Phys. Rev.~A}, 46(7):3924--3928, 1992.

\bibitem{Vos-91-1}
S.~H. {Vosko}, J.~A. {Chevary}, and I.~L. {Mayer}.
\newblock Predictions of stable {Y}b$^{-}$ in the {P}$^{\text{o}}_{1/2}$ state:
  {T}he importance of spin-orbit coupling.
\newblock {\em J. of Phys.~B: At., Mol. Opt.}, 24:L225--L231, 1991.

\bibitem{Vos-91-2}
S.~H. {Vosko}, J.~B. {Lagowski}, I.~L. {Mayer}, and J.~A. {Chevary}.
\newblock Theoretical study of even- and odd-parity states in {L}a$^{-}$ and
  {A}c$^{-}$: {E}vidence for the uniqueness of {L}a$^{-}$.
\newblock {\em Phys. Rev.~A}, 43:6389--6392, 1991.

\bibitem{Gri-92}
A.~A. {Gribakina}, G.~F. {Gribakin}, and V.~K. {Ivanov}.
\newblock The structure and photodetachment of the {Y}b$^{-}$ negative ion.
\newblock {\em Phys. Rev. Lett.}, 68:280--284, August 1992.

\bibitem{Din-94}
K.~{Dinov}, D.~R. {Beck}, and D.~{Datta}.
\newblock Electron affinities of six bound states of {C}e$^{-}$ formed by
  attachment of $6p$ and $5d$ electrons to {C}e.
\newblock {\em Phys. Rev.~A}, 50(2):1144--1148, August 1994.

\bibitem{Dat-94}
Debais~{Datta} nad Donald R.~{Beck}.
\newblock Electron affinities of opposite parity bound states on {T}h$^{-}$:
  Relativistic-configuration-interaction studies.
\newblock {\em Phys. Rev.~A}, 50(2):1107--1111, August 1994.

\bibitem{Din-95-1}
Konstantin {Dinov} and Donald~R. {Beck}.
\newblock Electron affinities of $6p$ electrons in {P}r$^{-}$.
\newblock {\em Phys. Rev.~A}, 51(2):1680--1682, February 1995.

\bibitem{Din-95-2}
Konstantin {Dinov} and Donald~R. {Beck}.
\newblock Electron affinity and hyperfine structure for {U}$^{-}$ and {U}
  \textsc{i} obtained from relativistic configuration-interaction calculations.
\newblock {\em Phys. Rev.~A}, 52(4):2632--2637, October 1995.

\bibitem{Din-96}
Konstantin {Dinov} and Danald~R. {Beck}.
\newblock Electron affinity of {P}a by $7p$ attachment and hyperfine constants
  for {P}a$^{-}$.
\newblock {\em Phys. Rev.~A}, 53(6):4031--4035, June 1996.

\end{thebibliography}

\bibliographystyle{unsrt}
\end{document}